\documentclass[useAMS,usedcolumn,usenatbib,usegraphicx]{mn2e}

\usepackage{epsfig,graphicx,natbib}
\usepackage{./reference/mycite}
\usepackage{amssymb}
\usepackage{gensymb}
\usepackage{amsfonts}
\usepackage{amsmath}
\usepackage{color}
\usepackage{lscape,longtable}
 \usepackage{times}
 
\begin{document}

\title[Cold Dust Emission from X-Ray AGN in S2CLS]{Cold Dust Emission from X-ray AGN in the SCUBA-2 Cosmology Legacy Survey: Dependence on Luminosity, Obscuration \& AGN Activity} 
\author[M. Banerji et al.]{ \parbox{\textwidth}
{Manda Banerji$^{1,2}$\thanks{E-mail: mbanerji@ast.cam.ac.uk}, R. G. McMahon$^{1,2}$, C. J. Willott$^{3}$, J. E. Geach$^{4}$, C. M. Harrison$^{5}$, S. Alaghband-Zadeh$^{1}$, D. M. Alexander$^{5}$, N. Bourne$^{6}$, K. E. K. Coppin$^{4}$, J. S. Dunlop$^{6}$, D. Farrah$^{7}$, M. Jarvis$^{8, 9}$, M. J. Micha{\l}owski$^{6}$, M. Page$^{10}$, D. J. B. Smith$^{4}$, A. M. Swinbank$^{5}$, M. Symeonidis$^{10}$, P. P. Van der Werf$^{11}$}
  \vspace*{6pt} \\
$^{1}$Institute of Astronomy, University of Cambridge, Madingley Road, Cambridge, CB3 0HA, UK.\\
$^{2}$Kavli Institute for Cosmology, University of Cambridge, Madingley Road, Cambridge, CB3 0HA, UK.\\
$^{3}$Herzberg Institute of Astrophysics, National Research Council, 5071 West Saanich Rd, Victoria, BC V9E 2E7, Canada. \\
$^{4}$Centre for Astrophysics Research, University of Hertfordshire, College Lane, Hatfield, Hertfordshire AL10 9AB, UK. \\
$^{5}$Centre for Extragalactic Astronomy, Department of Physics, Durham University, South Road, Durham DH1 3LE, UK. \\
$^{6}$Institute for Astronomy, University of Edinburgh, Royal Observatory, Blackford Hill, Edinburgh EH9 3HJ, UK. \\
$^{7}$ Department of Physics, Virginia Tech, Blacksburg, VA 24061, USA \\
$^{8}$Oxford Astrophysics, Department of Physics, Keble Road, Oxford OX1 3RH, UK. \\
$^{9}$Physics Department, University of the Western Cape, Bellville 7535, South Africa. \\
$^{10}$Mullard Space Science Laboratory, University College London, Holmbury St Mary Dorking, Surrey RH5 6NT, UK.\\
$^{11}$Leiden Observatory, Leiden University, PO Box 9513, NL-2300 RA Leiden, the Netherlands. \\
} 

\maketitle

\begin{abstract} 

We study the 850$\mu$m emission in X-ray selected active galactic nuclei (AGN) in the $\sim$2 deg$^2$ COSMOS field using new data from the SCUBA-2 Cosmology Legacy Survey. We find nineteen 850$\mu$m bright X-ray AGN in a ``high-sensitivity" region covering 0.89 deg$^2$ with flux densities of S$_{850}$=4-10 mJy. The 19 AGN span the full range in redshift and hard X-ray luminosity covered by the sample - $0.7 \lesssim z \lesssim 3.5$ and $43.2 \lesssim$ log$_{10}$(L$_{\rm{X}}) \lesssim 45$. We report a highly significant stacked 850$\mu$m detection of a hard X-ray flux-limited population of 699 $z>1$ X-ray AGN  - S$_{850}$=0.71$\pm$0.08mJy. We explore trends in the stacked 850$\mu$m flux densities with redshift, finding no evolution in the average cold dust emission over the redshift range probed. For Type 1 AGN, there is no significant correlation between the stacked 850$\mu$m flux and hard X-ray luminosity. However, in Type 2 AGN the stacked submillimeter flux is a factor of 2 higher at high luminosities. When averaging over all X-ray luminosities, no significant differences are found in the stacked submillimeter fluxes of Type 1 and Type 2 AGN as well as AGN separated on the basis of X-ray hardness ratios and optical-to-infrared colours. However, at log$_{10}$(L$_{2-10}$ / erg s$^{-1}>44.4$), dependences in average submillimeter flux on the optical-to-infrared colours become more pronounced. We argue that these high luminosity AGN represent a transition from a secular to a merger-driven evolutionary phase where the star formation rates and accretion luminosities are more tightly coupled. Stacked AGN 850$\mu$m fluxes are compared to the stacked fluxes of a mass-matched sample of $K$-band selected non-AGN galaxies. We find that at $10.5<$log$_{10}$(M$_*$/M$_\odot$)$<11.5$, the non-AGN 850$\mu$m fluxes are $1.5-2\times$ higher than in Type 2 AGN of equivalent mass. We suggest these differences are due to the presence of massive dusty, red starburst galaxies in the $K$-band selected non-AGN sample, which are not present in optically selected catalogues covering a smaller area. 

\end{abstract}


\section{INTRODUCTION}

Studying the connection between star formation in galaxies and accretion onto the supermassive black-hole is of prime importance in understanding galaxy formation (e.g. see \citealt{Alexander:12} for a recent review). The discovery that supermassive black-holes are ubiquitous with their mass connected tightly to the stellar bulge mass of galaxies \citep{Magorrian:98, Kormendy:13} has led to the realisation that these active galactic nuclei (AGN) play a fundamental role in regulating the growth of galaxies. Sophisticated galaxy formation simulations now incorporate feedback from the AGN, which is shown to be critical in order to explain the observed numbers of massive galaxies (e.g. \citealt{Springel:05, Croton:06, diMatteo:05}). In these simulations, massive galaxies are assembled through gas-rich mergers which induce luminous bursts of obscured star formation. Star formation and black-hole accretion are fuelled by a common gas supply and the star formation is initially dust-obscured. As the gas and dust are expelled from the galaxy e.g. via AGN-driven winds, the accreting black-hole emerges as an X-ray/UV luminous quasar. In such a picture of galaxy formation, luminous starburst galaxies, obscured and unobscured AGN are linked through a well-defined evolutionary sequence. Within unified AGN models \citep{Antonucci:93, Urry:95}, the obscuration of the central accreting source can also be explained purely as a result of orientation effects and the distinction between unobscured Type 1 AGN and obscured Type 2 AGN is attributed to the line of sight through which they are viewed.

Obscuration in AGN almost certainly results from a combination of orientation and evolution driven factors. One way to distinguish between these is to look for evidence for obscured AGN having excess star formation in their host galaxies relative to their unobscured counterparts. Several early studies of X-ray luminous AGN using the SCUBA bolometer on the James Clerk Maxwell Telescope (JCMT) showed this to be the case \citep{Page:01, Page:04, Stevens:05}. More recently, new data at far infrared wavelengths from the \textit{Herschel} Space Observatory, has been used to study the connection between star formation in galaxies and accretion onto the central AGN (e.g. \citealt{Bonfield:10, Lutz:08, Shao:10, Page:12, Harrison:12, Mullaney:12, Rovilos:12, Rosario:12, Santini:12, Azadi:14, Stanley:15}). \citet{Lutz:10} have also  conducted a study of the 870$\mu$m properties of $z\sim1$ AGN using LABOCA. There is now a growing consensus that moderate luminosity AGN tend to occupy star-forming host galaxies and that the star formation rate is independent of AGN X-ray luminosity in this regime. AGN activity in galaxies occurs on very different spatial as well as temporal scales relative to star formation, which means that the instantaneous star formation rate in galaxies is unlikely to be tightly coupled to the instantaneous accretion rate \citep{Hickox:14}. However, at high luminosities, major-merger driven evolution can lead to a tighter coupling between AGN activity and star formation. 

Studying the connection between star formation and AGN activity at high luminosities requires multi-wavelength data from the X-ray to submillimeter wavelengths over reasonably wide fields. The COSMOS field is among the most well-studied extragalactic fields with a plethora of multi-wavelength photometric and spectroscopic data. X-ray data is available from both the \textit{XMM-Newton} and \textit{Chandra} observatories and the field has been imaged at many other wavelengths, in particular with \textit{Spitzer} and \textit{Herschel}. Moreover, the wide area covered by many of these datasets now provides crucial information for the most luminous, high-redshift galaxies and AGN. Recently, the SCUBA-2 bolometer on the JCMT has provided a submillimeter map at 850$\mu$m over the entire 2 deg$^2$ field as part of the SCUBA-2 Cosmology Legacy Survey \citep{Geach:13}. The 850$\mu$m data offers several advantages over \textit{Herschel} observations over the same region. Firstly, the superior angular resolution and smaller beam size relative to \textit{Herschel} mean that source blending is less of an issue when inferring the average submillimeter properties of AGN (see Fig. \ref{fig:radio}). Secondly, the negative k-correction at these wavelengths at $z>1$ \citep{Blain:02}, ensures that we are sensitive to roughly constant dust luminosity for essentially all high redshift galaxies in an 850$\mu$m flux-limited sample. Furthermore, even in the \textit{Herschel}-SPIRE bands, there can be a significant contribution from AGN dust heating in the case of the most luminous quasars (e.g. \citealt{Rosario:12}). The submillimeter dust emission at 850$\mu$m is much less prone to such contamination from the AGN. Although it could potentially be more contaminated by synchrotron emission, this is unlikely to be an issue for the majority of AGN which are radio-quiet The dust emission at 850$\mu$m can therefore more safely be attributed to heating by star formation. At shorter wavelengths it becomes necessary to disentangle the relative contributions of starburst and AGN heating to the total infrared flux and linking the two phenomena is therefore more complicated, with potential biases arising to derived star formation rates due to different assumptions regarding the intrinsic AGN model.  

The aim of this paper is to study the 850$\mu$m properties of both unobscured (type 1) and obscured (type 2) AGN in the COSMOS field where statistically significant numbers of these populations now exist. We look for dependences of this submillimeter emission on the AGN host galaxy properties and compare to non-AGN galaxies. As such, our analysis is intended to provide the first SCUBA-2 Cosmology Legacy Survey 850$\mu$m view of the X-ray AGN population at high redshift. Throughout this paper we assume a flat concordance cosmology with H$_0$=70 km s$^{-1}$ Mpc$^{-1}$.

\section{DATA}

\subsection{SCUBA-2 850$\mu$m}

\label{sec:data}

\begin{figure*}
\begin{center}
\centering
\begin{tabular}{ccc}
\large{\textit{Herschel}-SPIRE 250$\mu$m} & \large{\textit{Herschel}-SPIRE 500$\mu$m} & \large{SCUBA-2 850$\mu$m} \\
\includegraphics[scale=0.3,angle=0]{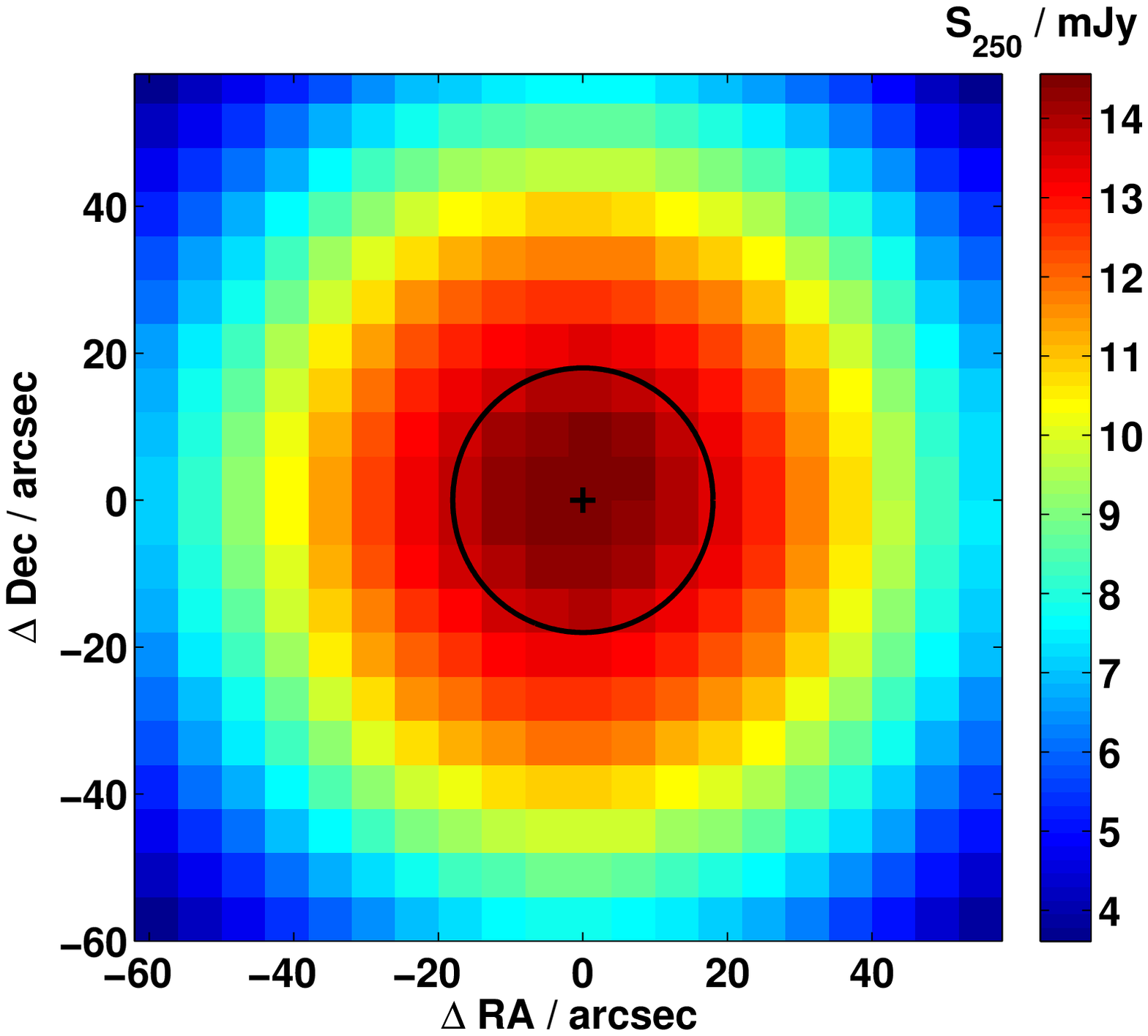} & \includegraphics[scale=0.3,angle=0]{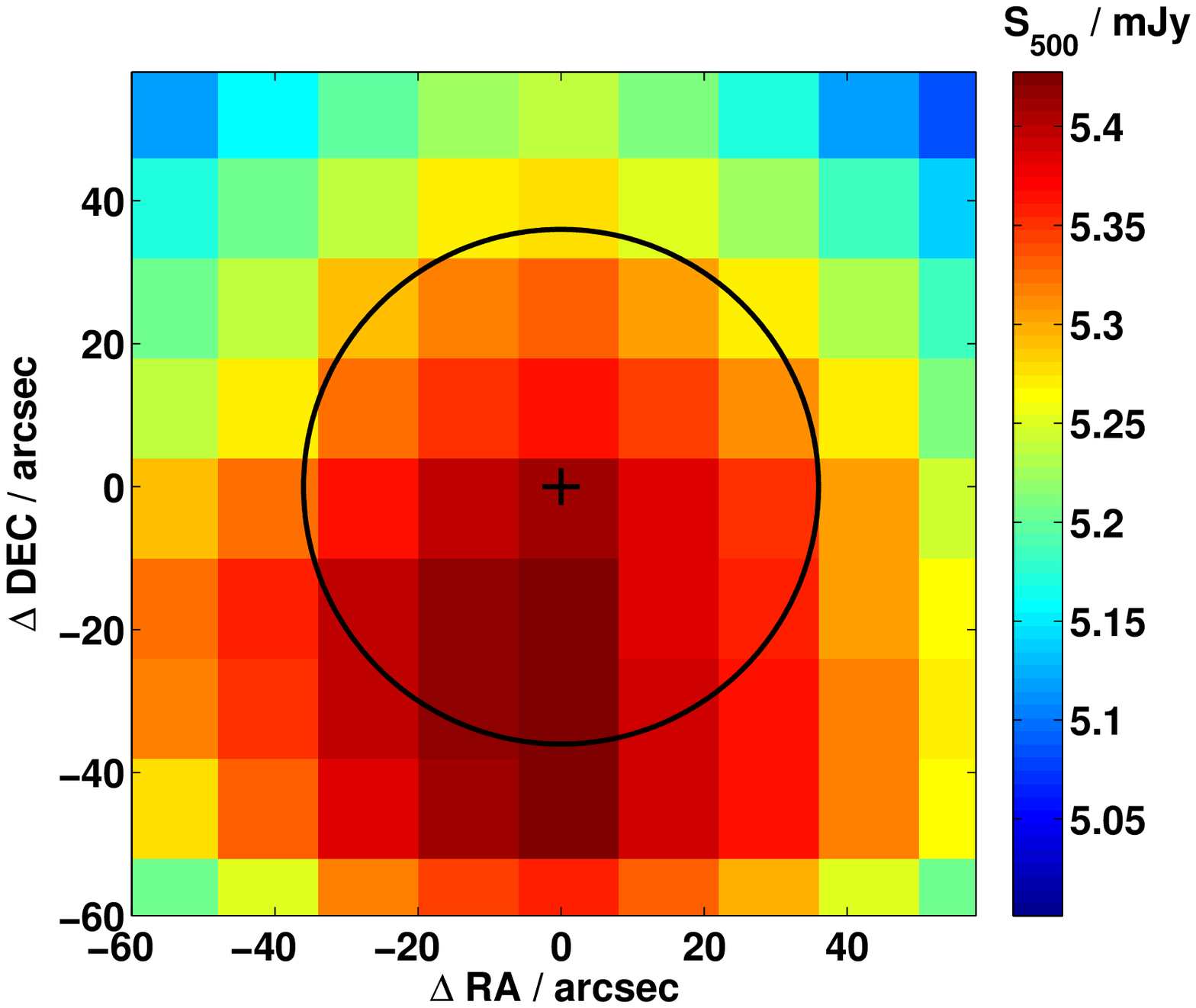} & \includegraphics[scale=0.3,angle=0]{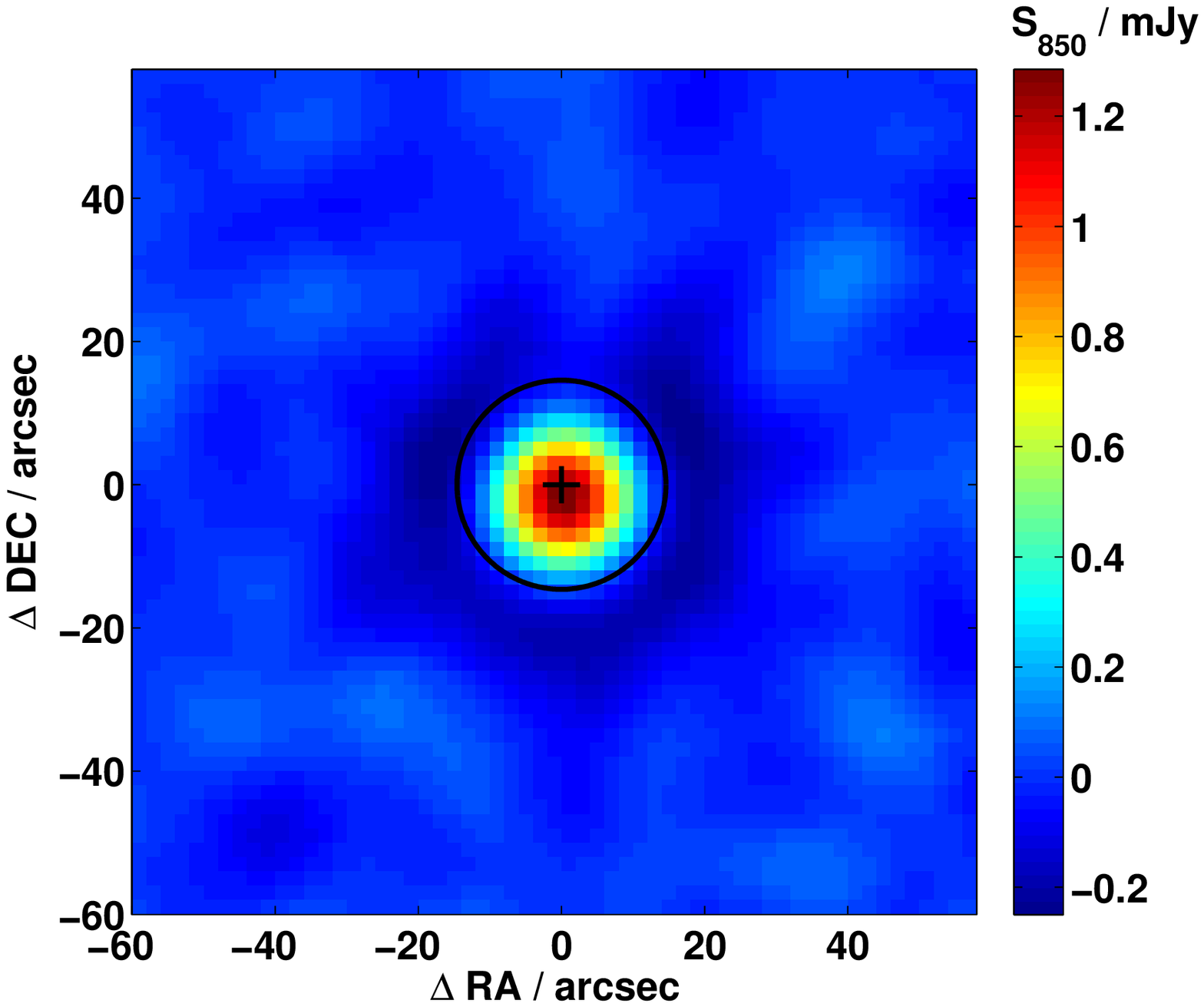} \\
\end{tabular}
\caption{Stacked 250$\mu$m (left), 500$\mu$m (middle) and 850$\mu$m (right) images of 2865 radio sources in the COSMOS field from \citet{Schinnerer:10}. The superior spatial resolution and angular scale of the SCUBA-2 data is apparent relative to \textit{Herschel}. For the 850$\mu$m stack, we confirm that the peak emission is clearly coming from the central pixels demonstrating that the astrometry of the maps is accurate at the sub-pixel level. The negative ``bowling" around the bright central source seen in the SCUBA-2 850$\mu$m image, is an artefact of the filtering procedure employed to generate the map.}
\label{fig:radio}
\end{center}
\end{figure*}

Our primary dataset used in this analysis is the SCUBA-2 850$\mu$m map in the COSMOS field obtained as part of the SCUBA-2 Cosmology Legacy Survey (S2CLS; \citealt{Geach:13}). The wide-field map covers a total area of $\sim$2 deg$^2$ imaged using a 4$\times$45\arcmin\@ PONG scanning pattern.  The basic principle of the data reduction is to extract astronomical signal
from the timestreams recorded by each bolometer in the SCUBA-2 array and map
them onto a two dimensional celestial projection. We have used the Dynamical
Iterative Map-Maker ({\sc dimm}) within the {\it Sub-Millimetre Common User
Reduction Facility} ({\sc smurf}; \citealt{Chapin:13}). We refer readers to
\citet{Chapin:13} for a detailed overview of {\sc smurf}, and detailed
data reduction steps will be given in Geach et al. (in preparation) but
describe the main steps here.

Flat-fields are applied to the time-streams using flat scans that bracket each
observation, and a polynomial baseline fit is subtracted from each bolometer's
time-stream. Each time-stream is cleaned for spikes and DC steps are removed
and gaps filled. After cleaning, the {\sc dimm} enters an iterative process
that aims to fit the data with a model comprising a common mode signal,
astronomical signal and noise. Next, a filtering step is performed in the
Fourier domain, which rejects data at frequencies corresponding to angular
scales $\theta>150''$ and $\theta< 2''$. The next step is to estimate the
astronomical signal, which is subtracted from the data. Finally, a noise
model is estimated for each bolometer by measuring the residual, which is then
used to weight the data during the mapping process in additional steps. The
iterative process above runs until convergence is met. In this case, we
execute a maximum of 20 iterations, or when the map tolerance reaches 0.05.

The final 850$\mu$m map in the COSMOS field has non-uniform depth with typical RMS values of $\sigma_{850}\sim$2 mJy in one half, $\sigma_{850}\sim$4 mJy in the other half and $\sigma_{850}\sim$1 mJy in the central region. An 850$\mu$m catalogue containing 360 sources has been produced from this map by searching for all sources above 3.5$\sigma$ in regions with $\sigma_{850}<2$ mJy (Geach et al. in preparation). This 850$\mu$m catalogue covers an area of 0.89 deg$^2$. Our search for AGN that are individually detected at 850$\mu$m covers this smaller area, higher sensitivity region. However, when studying the stacked properties of the AGN, we utilise the full area in order to get the largest AGN samples possible. As stacked fluxes are weighted by the noise (see Section \ref{sec:stack}), AGN in lower RMS regions are automatically down-weighted when calculating average 850$\mu$m fluxes.   

We begin by checking the astrometry of this 850$\mu$m map by stacking a sample of 2865 radio sources in the COSMOS field \citep{Schinnerer:10} in this map. The stacking procedure is described in more detail in Section \ref{sec:stack} and the stacked 850$\mu$m image for these radio sources is shown in Fig. \ref{fig:radio}. The peak 850$\mu$m emission has an offset of 1.4$\arcsec$ relative to the radio position. The SCUBA-2 850$\mu$m pixel scale is 2$\arcsec$. This confirms that the maps have the correct astrometry at the sub-pixel level and can therefore be used in our stacking analysis. We also show the stacked images of the same radio sources in the \textit{Herschel}-SPIRE 250 and 500$\mu$m bands in Fig. \ref{fig:radio}. The \textit{Herschel}-SPIRE data in the COSMOS field is taken from the \textit{Herschel} Multi-tiered Extragalactic Survey (HerMES; \citealt{Oliver:12}). Maps have been produced using the Level 2 data products from the ESA archive as detailed in \citet{Swinbank:14}. While the stacked 250$\mu$m emission appears to be well centred at the radio position, the 500$\mu$m emission shows an offset relative to the radio position. The superior spatial resolution and pixel scale of the SCUBA-2 data relative to \textit{Herschel}-SPIRE, are clearly apparent in this plot. The longer wavelength SCUBA-2 850$\mu$m observations are therefore extremely complementary to already published \textit{Herschel} observations of AGN in this field. 

\subsection{AGN Catalogue}

\label{sec:cat}

\begin{figure*}
\begin{center}
\begin{tabular}{cc}
\includegraphics[scale=0.4,angle=0]{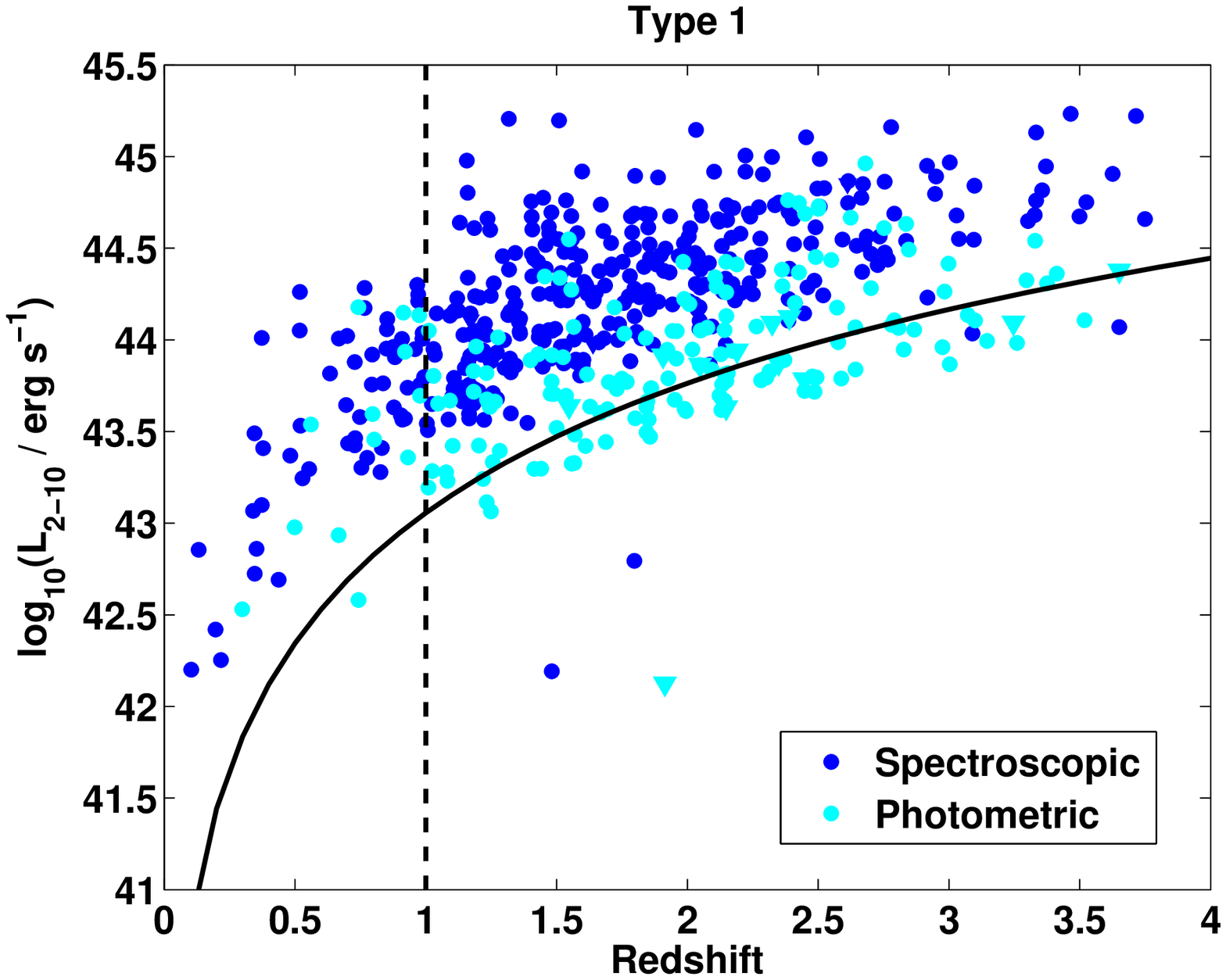} & \includegraphics[scale=0.4,angle=0]{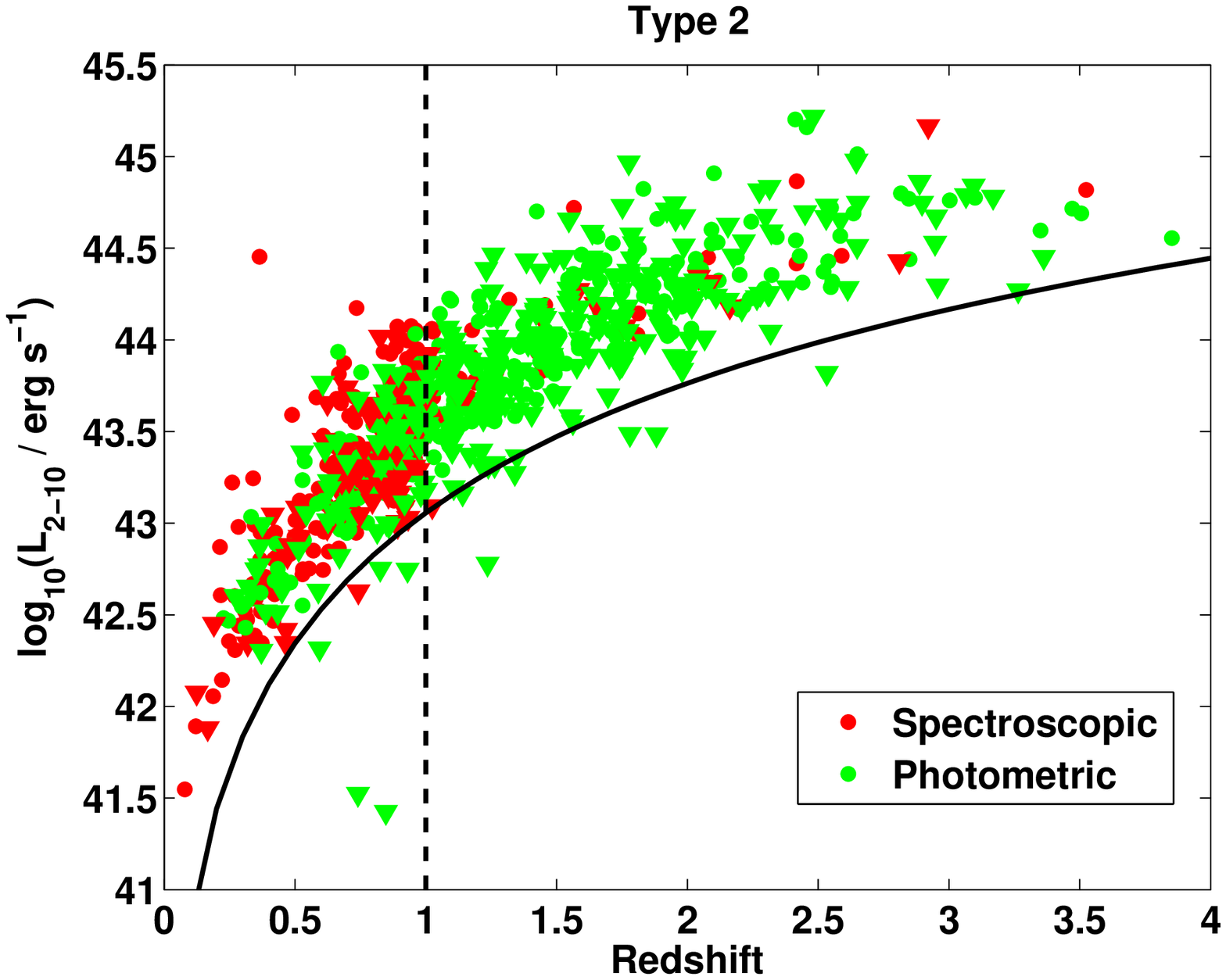} \\
\end{tabular}
\caption{Redshift versus observed hard (2-10 keV) band luminosity for Type 1 AGN (left) and Type 2 AGN (right). Both spectroscopic and photometric AGN are shown. The circles mark actual detections whereas the downward triangles represent upper limits in the case of the Type 2 AGN and hard-band luminosities estimated from the soft-band flux in the case of the Type 1 AGN. The solid lines denote the flux limit applied to the data to generate homogenous populations of the two populations for our stacking analysis. The vertical dashed lines correspond to $z>1$ which is the redshift cut applied before stacking such that the 850$\mu$m flux is invariant with dust luminosity as a function of redshift.}
\label{fig:LXz}
\end{center}
\end{figure*}

The aim of this paper is to study the submillimeter properties of a statistically significant and homogenous sample of AGN. The COSMOS field \citep{Scoville:07} provides the largest and most well characterised AGN sample overlapping the new SCUBA-2 CLS data. The AGN parent sample is a catalogue of \textit{XMM-Newton} detected X-ray point sources, which goes down to a flux limit of 9.3e-15 erg/s/cm$^2$ in the 2-10 keV band over 90\% of the area. The flux limit for the faintest sources detected in the catalogue are 5e-16 erg/s/cm$^2$ and 2.5e-15 erg/s/cm$^2$ in the 0.5-2 keV and 2-10 keV bands respectively. We only consider X-ray sources with robust multi-wavelength optical and infrared cross-identifications from \citet{Brusa:10}, which corresponds to $\sim$98\% of the X-ray sources. We make use of the latest spectroscopic and photometric redshifts for these AGN \citep{Bongiorno:12, Salvato:11} together with the corresponding spectroscopic and photometric classifications\footnote{http://www2011.mpe.mpg.de/XMMCosmos/xmm53\_release/}. The Type 1 and Type 2 samples specifically constitute AGN with broad emission lines in the case of the spectroscopic Type 1s and narrow emission lines in the case of the spectroscopic Type 2s. For the AGN with photometric redshifts, Type 1 AGN are considered to be those fit by templates 19-30 in \citet{Salvato:11} whereas Type 2 AGN are those fit by templates 1-18 or $>$100 (which correspond to galaxy templates). The photometric redshift accuracy is in general very good $\sigma_{\Delta z}/(1+z)$ = 0.014 for $i_{AB} < 22.5$ and 0.015 for $i_{AB}<24.5$ \citep{Salvato:09}. Photometry at mid infra-red wavelengths is from the \textit{Spitzer} S-COSMOS Survey \citep{Sanders:07}. There are a total of 1797 AGN with multi-wavelength cross-identifications in these catalogues. The observed hard-band (2-10 keV) X-ray luminosity versus redshift for both the Type 1 and Type 2 AGN can be seen in Fig. \ref{fig:LXz}. These X-ray luminosities have been calculated from the hard-band fluxes from \citet{Brusa:10} assuming a photon index of $\Gamma$=1.8. In the case of the Type 1 AGN that are detected in the soft (0.5-2 keV) band but not in the hard band, the soft X-ray fluxes from \citet{Brusa:10} are used to estimate the hard X-ray flux, once again assuming $\Gamma$=1.8. We note that no absorption correction has been applied to these luminosities as the X-ray data for the entire sample is in general of insufficient quality to allow full spectral fitting. The Type 1 AGN have negligible absorption and in the case of the Type 2 COSMOS AGN that are detected in the hard X-ray, \citet{Lusso:11} estimate the average shift in the hard X-ray luminosity introduced by an absorption correction to be $< \Delta \rm{log} (\rm{L}_{2-10})> = 0.04 \pm 0.01$. In Appendix A we compare the hard X-ray luminosities derived in this paper to those derived by \citet{Brightman:14} for a subset of the COSMOS AGN with \textit{Chandra} data and where complex absorption spectral models can be fit, demonstrating that there are no significant biases in the X-ray luminosities used in this paper. 

In the majority of the analysis presented in this paper we consider the average stacked submillimeter fluxes of AGN as a function of various properties. For these stacking studies, we construct flux-limited populations of AGN restricted to sources with S$>$2.5$\times 10^{-15}$ erg s$^{-1}$ cm$^{-2}$ in the hard (2-10 keV) band \citep{Lusso:11}. As previously mentioned, Type 1 AGN that are detected in the soft band but not in the hard band, have had their hard band fluxes estimated from the soft X-ray flux using a $\Gamma=1.8$ power-law SED. Type 2 AGN that are detected in the soft band only have been discarded for the majority of the stacking analysis due to their essentially unknown dust columns and therefore intrinsic luminosities. These soft band detected Type 2 AGN are however included when searching for AGN that are individually detected at 850$\mu$m in Section \ref{sec:850detect}. The hard-band X-ray flux limit is illustrated in Fig. \ref{fig:LXz} and corresponds to $L_{\rm{X}}\gtrsim10^{43}$ erg/s at $z>1$, where the X-ray luminosity should be completely dominated by AGN activity. We also restrict ourselves to only those AGN at $z>1$ for all the stacking analysis. At $z>1$, the 850$\mu$m flux is essentially invariant with redshift for a given dust luminosity and therefore star formation rate, allowing us to average the submillimeter fluxes of AGN over relatively wide redshift bins in order to improve the statistical significance of our results. The 850$\mu$m data does not however trace the peak of the dust SED at these redshifts and there could potentially be biases in how effectively this submillimeter flux is tracing star formation due to variations in the template SEDs over this rest-frame wavelength range. In Section \ref{sec:sfr} we will therefore explicitly check the dependence of 850$\mu$m flux on star formation rates. 

Finally, there is a concern that the 850$\mu$m flux could be contaminated by synchrotron emission for populations of radio-loud AGN. The fraction of radio loud AGN in X-ray selected AGN samples is expected to be small ($\lesssim$5\% for Type 1 AGN from \citealt{Hao:14}). Nevertheless, we match the X-ray AGN sample to the VLA radio catalogue \citep{Schinnerer:10} and remove all sources with S$_{1.4\rm{GHz}}>$0.5 mJy. This removes 10 Type 1 and 7 Type 2 AGN at $z>1$ from the sample, which represents $\sim$2\% of both the Type 1 and Type 2 samples\footnote{Including these radio-detected AGN in the 850$\mu$m stacking does not change the average 850$\mu$m flux derived for our sample in Section \ref{sec:stack} to within 0.1mJy}. The radio flux limit imposed corresponds to a synchrotron 850$\mu$m flux of $<$0.01 mJy assuming a synchrotron spectral index of 0.7. Therefore we can be confident that the 850$\mu$m stacks that we study are completely dominated by thermal dust emission. Our final $z>1$ AGN sample used in the stacking analysis, totals 699 AGN. These are comprised of 428 Type 1 AGN (314 spectroscopic, 114 photometric) and 271 Type 2 AGN (46 spectroscopic, 225 photometric).  

\citet{Bongiorno:12} have used spectral energy distribution (SED) fits to the multi wavelength COSMOS photometry, in order to estimate stellar masses, star formation rates and other host galaxy parameters for these Type 1 and Type 2 AGN. Both galaxy and AGN components are fit allowing us to study the dependence of 850$\mu$m flux densities on a wide range of AGN and host galaxy properties. These SED fits incorporate photometry at optical, near infrared as well as mid infrared wavelengths. Around 80 per cent of the sources are detected in the \textit{Spitzer} MIPS 24$\mu$m band. In addition, several of the AGN are also detected in the MIPS 70$\mu$m band as well as in the \textit{Herschel}-PACS 100$\mu$m and 160$\mu$m bands \citep{Bongiorno:12}. All of this multi-wavelength photometry is used in the SED fitting and both pure AGN and pure galaxy templates are allowed in the fits. The galaxy templates constitute a set of \citet{BC:03} stellar population synthesis models with a \citet{Chabrier:03} initial mass function and assuming exponentially declining star formation histories with varying amounts of dust reddening. The AGN template is the mean quasar SED from \citet{Richards:06}, again with varying amounts of reddening. This choice of AGN template is appropriate for the luminous XMM-COSMOS AGN investigated in this work. \citet{Bongiorno:12} have already discussed the accuracy of the stellar mass estimates for both the Type 1 and Type 2 AGN samples and when considering the properties of the AGN as a function of stellar mass, we only consider those AGN that have robust stellar masses as specified by the \textit{flags} parameter. As expected, $>99.5$ per cent of the Type 2 AGN have robust stellar mass estimates.

\subsubsection{Limitations of the X-ray AGN Catalogue}

In this analysis we consider the submillimeter properties of a hard X-ray flux limited catalogue of AGN. However, before proceeding it is important to highlight the selection biases and incompletenesses of an AGN catalogue selected in this way. \citet{Donley:12} have used a mid infrared IRAC selection for AGN candidates in the COSMOS field demonstrating that only $\sim$38\% of these AGN candidates have X-ray counterparts due to the relatively shallow flux limit of the \textit{XMM}-COSMOS hard X-ray data. Hence, we may potentially be missing a significant population of heavily obscured to mildly Compton thick AGN in this analysis. Recently \citet{Lacy:15} have also used mid infra-red selected AGN samples to demonstrate that X-ray surveys begin to become incomplete at $z\gtrsim1.6$ where a deficit in the X-ray luminosity function is seen relative to the mid infra-red luminosity function even in the highest luminosity bins. We refer readers to these papers for a detailed consideration of the incompleteness in hard X-ray selected samples of AGN.

\subsection{Non-AGN Galaxy Catalogue}

Finally, as a control sample to compare the AGN properties to, we also utilise a galaxy sample from the UltraVISTA survey Data Release 1 \citep{McCracken:12} over the same region. We work with the $K$-band flux limited sample from \citet{Muzzin:13}, which essentially represents a mass-limited sample of galaxies. Corresponding photometric redshifts, stellar masses and other SED-fitting parameters are also taken from \citet{Muzzin:13}. An advantage of using a galaxy catalogue that is selected in the redder $K$-band, is that these catalogues are potentially more sensitive to dusty galaxies that would be detected at 850$\mu$m compared to galaxy catalogues constructed using data at optical wavelengths. We note that several previous studies that we compare our results to in this paper, have used non-AGN galaxy samples selected using flux-limited surveys at bluer (optical) wavelengths. 

\section{850$\mu\rm{\lowercase{m}}$ DETECTED AGN}

\label{sec:850detect}

We begin by looking for AGN that are detected at $>$3.5$\sigma$ in the SCUBA-2 CLS 850$\mu$m catalogues. As stated above, only regions with $\sigma_{850}<2$mJy were used in the construction of the 850$\mu$m catalogue, which therefore does not cover the entire field. A matching radius of 7.5$^{\prime \prime}$ is used to associate 850$\mu$m catalogued sources with our AGN sample. The AGN positions in all cases correspond to the optical positions from \citet{Brusa:10}, which have much smaller uncertainties than the \textit{XMM-Newton} X-ray positions. Given the errors on the optical positions are likely to be negligible, we follow \citet{Ivison:07} and assume that for a signal-to-noise ratio (SNR) of 3.5, the positional uncertainty, $\sigma_{\rm{pos}}$=0.6$\times$FWHM/SNR. For a FWHM of 14.5$^{\prime \prime}$ at 850$\mu$m, $\sigma_{\rm{pos}}$ is therefore 2.5$\arcsec$ and the 7.5$\arcsec$ matching radius corresponds to 3$\sigma_{\rm{pos}}$ and is therefore expected to encompass $>$99.7\% of true counterparts. 

\begin{figure}
\begin{center}
\includegraphics[scale=0.45,angle=0]{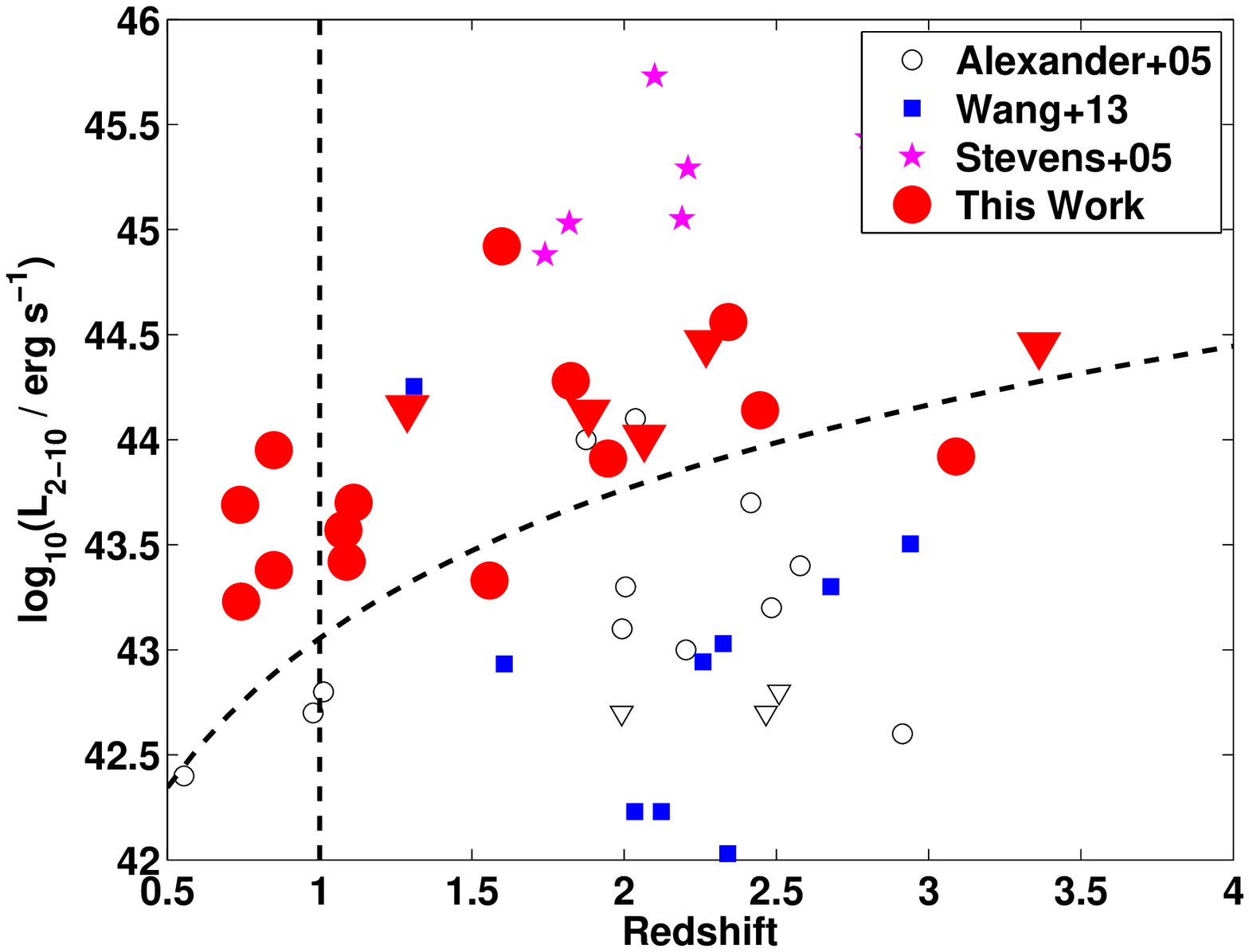} 
\caption{Redshift versus observed hard X-ray (2-10 keV) luminosity for the X-ray AGN that are individually detected at submillimeter wavelengths in this work and compared to individually detected submillimeter bright X-ray AGN from \citet{Alexander:05} and \citet{Wang:13} and submillimeter detected quasars from \citet{Stevens:05}. Downward triangles denote upper limits. The hard X-ray flux limit and $z>1$ redshift limit applied prior to our stacking analysis, are denoted by the dashed lines. Both \citet{Alexander:05} and \citet{Wang:13} probe fainter X-ray luminosities compared to our work, while the X-ray AGN from \citet{Stevens:05} that were individually followed up, probe very high X-ray luminosities. Our study is useful in bridging the gap between these different datasets.}
\label{fig:LXz_ind}
\end{center}
\end{figure}

There are a total of 19 850$\mu$m sources that lie within 7.5$\arcsec$ of an X-ray AGN and there are no instances of multiple submillimeter sources matched to the same X-ray AGN. All 19 matches are presented in Table \ref{tab:IDs} and we estimate the corrected Poisson probability, $p_{\rm{corr}}$ of these being chance associations based on the background density of all X-ray AGN that have fluxes brighter than the target being considered. Specifically:

\begin{equation}
\begin{split}
p_{\rm{raw}}=1-\rm{exp}\left(\pi r^2 N(>S)\right) \\
p_{\rm{det}}=1-\rm{exp}\left(\pi r_{\rm{max}}^2 N(>S_{\rm{lim}})\right) \\
p_{\rm{corr}}=1-\rm{exp}\left(p_{\rm{raw}}\left(1+ln\left(\frac{p_{\rm{raw}}}{p_{\rm{det}}}\right)\right)\right) \\
\end{split}
\label{eq:p}
\end{equation}

\noindent where $r$ is the separation between the submillimeter source and the X-ray AGN, $r_{\rm{max}}$ is the maximum search radius and N($>$S) and N($>$S$_{\rm{lim}}$) represent the background surface density of AGN brighter than the source being considered and brighter than the flux limit of the survey respectively. We estimate N($>$S) using the source counts in the soft and hard X-ray bands from \citet{Cappelluti:09} and assume S$_{\rm{lim}}$=5e-16 erg/s/cm$^2$ and 2.5e-15 erg/s/cm$^2$ in the 0.5-2 keV and 2-10 keV bands respectively. Given the low surface density of X-ray AGN and submillimeter sources, all 19 matches have very small probabilities ($\lesssim$ 1\%) of being chance associations (see Table \ref{tab:IDs}). The 19 sources are comprised of 6 spectroscopically confirmed broad-line AGN, 1 spectroscopically confirmed narrow-line AGN, 3 AGN with photometric redshifts that are best fit by Type 1 AGN templates and 9 AGN with photometric redshifts that are best fit by Type 2 AGN or galaxy templates. Thus out of a sample of 360 850$\mu$m bright sources, we find that $\sim$5\% are associated with X-ray luminous AGN. The \textit{XMM}-COSMOS data used in this work corresponds to relatively bright AGN X-ray flux limits as can be seen in Fig. \ref{fig:LXz}. \citet{Wang:13} identify the X-ray counterparts to ALMA detected submillimeter bright galaxies in the Extended Chandra Deep Field South (E-CDFS). \textit{Chandra} observations are used to go down to X-ray luminosities of $\sim$7$\times$10$^{42}$ erg/s and the authors find a larger AGN fraction of $\sim$17\% among the submillimeter galaxies. However as can be seen in their figure 10, this AGN fraction decreases as a function of increasing X-ray luminosity. Down to similar X-ray luminosities as \citet{Wang:13}, \citet{Symeonidis:14} find an AGN fraction of 18\% among infrared luminous galaxies in the CDF-N and CDF-S fields at $z<1.5$. \citet{Alexander:05} have used ultra-deep X-ray observations in the CDF-N to study the AGN fraction in $S_{850} \gtrsim 4$ mJy sub-mm bright galaxies at $z>1$, finding that the majority host AGN activity. However, the X-ray data from \textit{Chandra} used in that work, is an order of magnitude fainter than our \textit{XMM}-COSMOS sample. In Fig \ref{fig:LXz_ind} we plot the redshift versus hard (2-10 keV) X-ray luminosity for the submillimeter detected AGN in this work as well as \citet{Alexander:05} and \citet{Wang:13}. We also show here the submillimeter detected X-ray absorbed quasars from \citet{Stevens:05}, which were selected using X-ray surveys with \textit{ROSAT}, \textit{XMM-Newton} and \textit{Chandra} and followed up at submillimeter wavelengths using SCUBA. The \citet{Wang:13} observed 0.5-8 keV luminosities have been converted to 2-10 keV luminosities assuming the intrinsic photon indices derived for each object in that paper whereas conversions from 0.5-2 keV fluxes in \citet{Stevens:05} assume $\Gamma=1.8$.  Our work can immediately be place in the context of these previous studies. While both \citet{Alexander:05} and \citet{Wang:13} probe significantly fainter in terms of X-ray luminosity, our sample fills in the luminosity gap between these studies and the individual follow-up observations of X-ray bright quasars by \citet{Stevens:05}. The X-ray brightest sub-mm AGN in our sample is brighter than any of the X-ray AGN detected in the sub-mm by \citet{Alexander:05} and \citet{Wang:13}. It corresponds to XID18 and is an extremely red AGN with evidence for outflowing gas with velocities of $\sim$300 km/s \citep{Brusa:14}. This class of objects will be discussed later in Section \ref{sec:obscuration}. 

\textit{Herschel} fluxes for all 850$\mu$m detected sources are extracted from the catalogues of \citet{Swinbank:14} by matching the optical positions of the AGN in our catalogue to the 24$\mu$m positions in the \textit{Herschel} catalogues, using a matching radius of 3$\arcsec$. The \textit{Herschel} catalogues have already been corrected for blending using the 24$\mu$m positional priors \citep{Swinbank:14}. These fluxes are presented in Appendix B together with other derived properties for these submillimeter bright X-ray AGN. As the \textit{Herschel} fluxes have been extracted by \citet{Swinbank:14} using the 24$\mu$m positional priors, all 24$\mu$m detected sources have measured \textit{Herschel} fluxes whereas sources not detected at 24$\mu$m are not in the \textit{Herschel} catalogues. 

\newpage
\pagestyle{empty}
\begin{landscape}
\begin{table*}
\begin{center}
\caption{Summary of X-ray AGN with potential 850$\mu$m bright identifications in the COSMOS field. The corrected Poisson probability of a chance association is quoted as calculated from the soft X-ray flux and the hard X-ray flux respectively. Sources with photometric redshifts are marked with a $pz$. Sources with hard X-ray luminosities marked with an asterisk are Type 1 AGN that are only detected in the soft band and where the hard band luminosity has been calculated assuming $\Gamma=1.8$.}
\label{tab:IDs}
\begin{tabular}{lcccccccccc}
XID & RA & Dec & Redshift & log$_{10}($L$_{2-10}$) & Separation & S$_{0.5-2\rm{keV}}$ & p$_{0.5-2 \rm{keV}}$ & S$_{2-10\rm{keV}}$ & p$_{2-10 \rm{keV}}$ & S$_{850}$ \\
& & & & erg s$^{-1}$ & $\arcsec$ & erg s$^{-1}$ cm$^{-2}$ & & erg s$^{-1}$ cm$^{-2}$ & & mJy \\
\hline
 13    &   150.00924    &   2.2755119  &   0.850 & 43.95 & 3.42 &  2.11e-14 & 0.0006 &  2.9e-14 & 0.0009 & 4.5$\pm$1.1 \\
18    &   150.13304 &      2.3032850  &    1.598 & 44.92 & 5.92 &  1.80e-14  & 0.002 & 6.04e-14 & 0.0009 &  4.4$\pm$1.2 \\
139$^{pz}$   &    150.04184   &    2.6294839   & 0.739 & 43.69 &  3.30 & 9.52e-15 &  0.001 & 2.22e-14 & 0.001 & 7.4$\pm$1.7 \\
160   & 150.15839   &    2.1396031   &   1.825    & 44.28 &   1.48 &  2.39e-15  & 0.001 & 1.02e-14 & 0.0008 & 8.1$\pm$1.2 \\
246$^{pz}$    &   150.05100    &   2.4938550    &    2.342 &  44.56 &    3.65 &  8.81e-16  & 0.008 & 1.09e-14 & 0.003 &      6.4$\pm$1.7 \\
 250   &    150.06455   &    2.3290511  &    2.446 &   44.14$^{*}$ & 3.04 &  2.41e-15 &  0.004 & $<$3.78e-15 & -- &       4.0$\pm$1.1    \\
 270$^{pz}$    &   150.10612   &    2.0144781   & 1.883  &   $<$44.13 &  1.83 &  1.17e-15 &  0.003 &  $<$6.64e-15 & -- & 9.8$\pm$1.7 \\
278$^{pz}$  &     150.09308   &    2.1014100   &   2.269 &  $<$44.46  & 2.61 &  1.46e-15 &   0.004 & $<$9.39e-15 & -- & 5.0$\pm$1.4 \\
 353  &      150.08438   &    2.2904881  &    1.112 & 43.70 & 2.04 &  2.34e-15 &  0.002 & 8.57e-15 & 0.002 &  4.7$\pm$1.0 \\
 402$^{pz}$ &       150.25226    &   2.2619131  & 1.078 &  43.57 &     2.86 &  2.22e-15  & 0.003 & 6.82e-15 & 0.003 &       5.7$\pm$1.5 \\
 415$^{pz}$ &       149.96981   &    2.1834869 & 1.558  &  43.33$^{*}$ &   2.90 &  1.04e-15 &  0.005 & $<$8.31e-15 & -- &  5.5$\pm$1.4 \\
 469$^{pz}$ &       150.09685    &   2.0215000   & 3.362 &   $<$44.45 & 2.22 &  1.08e-15 &  0.004 & $<$3.73e-15 & -- &   6.6$\pm$1.7 \\
10675   &    150.19262  &     2.2198489   &   3.090 &  43.92$^{*}$ & 0.28 &  8.45e-16  & 0.0001 & $<$1.73e-15 & -- & 5.7$\pm$1.2 \\
 10809$^{pz}$ &       150.20758   &    2.3816111  & 1.288 & $<$44.15 &    1.79 & 1.12e-15  &  0.003 & $<$1.7e-14 & -- &   6.5$\pm$1.5 \\ 
 30182$^{pz}$ &       150.12541   &    2.6978931 & 0.742 & 43.23 & 6.00 &  6.26e-15 &  0.005 & 7.52e-15 & 0.009 &  6.9$\pm$1.7 \\
53922 &       150.09454    &   2.7029131  &   0.850   & 43.38 & 1.81 &  2.25e-15 &  0.002 & 7.74e-15 & 0.001 &  8.6$\pm$1.8 \\
54440$^{pz}$    &   150.06308   &    1.9446778   & 1.947 & 43.91    &   4.39 &  1.13e-15 &  0.009 & $<$3.74e-15 & -- &   7.7$\pm$1.7 \\
60070$^{pz}$    &   150.22719   &    2.2324781  & 2.066 &  $<$44.01 & 1.93 &  7.33e-16 &  0.003 & $<$4.11e-15 & -- &  5.8$\pm$1.4 \\
60490$^{pz}$  & 150.10544   & 2.1852681   & 1.089  & 43.42 & 5.33 & $<$7.57e-16  & -- & 4.76e-15 & 0.011 &  6.9$\pm$1.1 \\          
\hline
\end{tabular} 
\end{center} 
\end{table*}
\end{landscape}
\newpage

\section{AVERAGE 850$\mu\rm{\lowercase{m}}$ EMISSION FROM STACKED SUB-SAMPLES}

\label{sec:stack}

\begin{figure}
\begin{center}
\includegraphics[scale=0.5,angle=0]{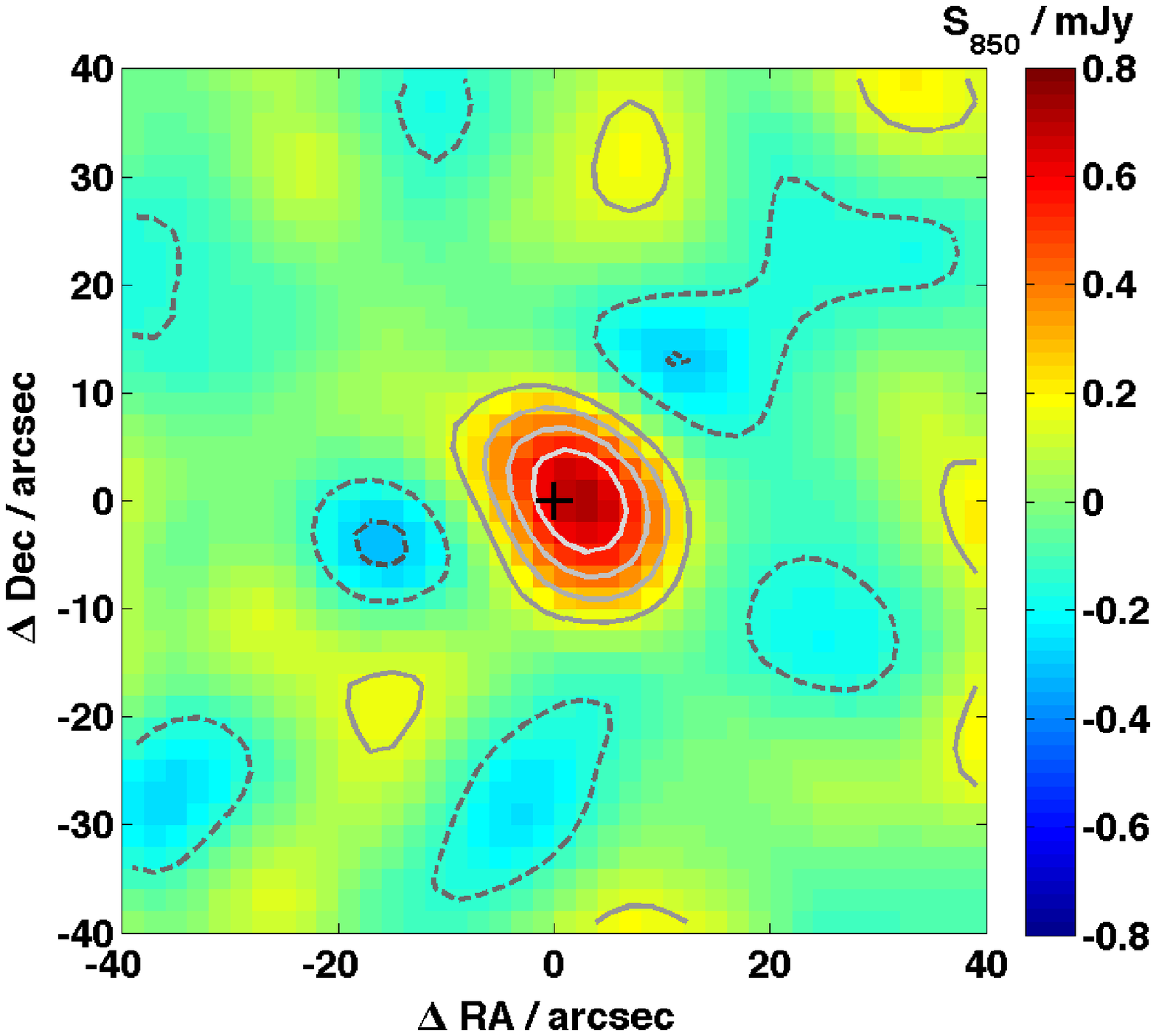}
\includegraphics[scale=0.5,angle=0]{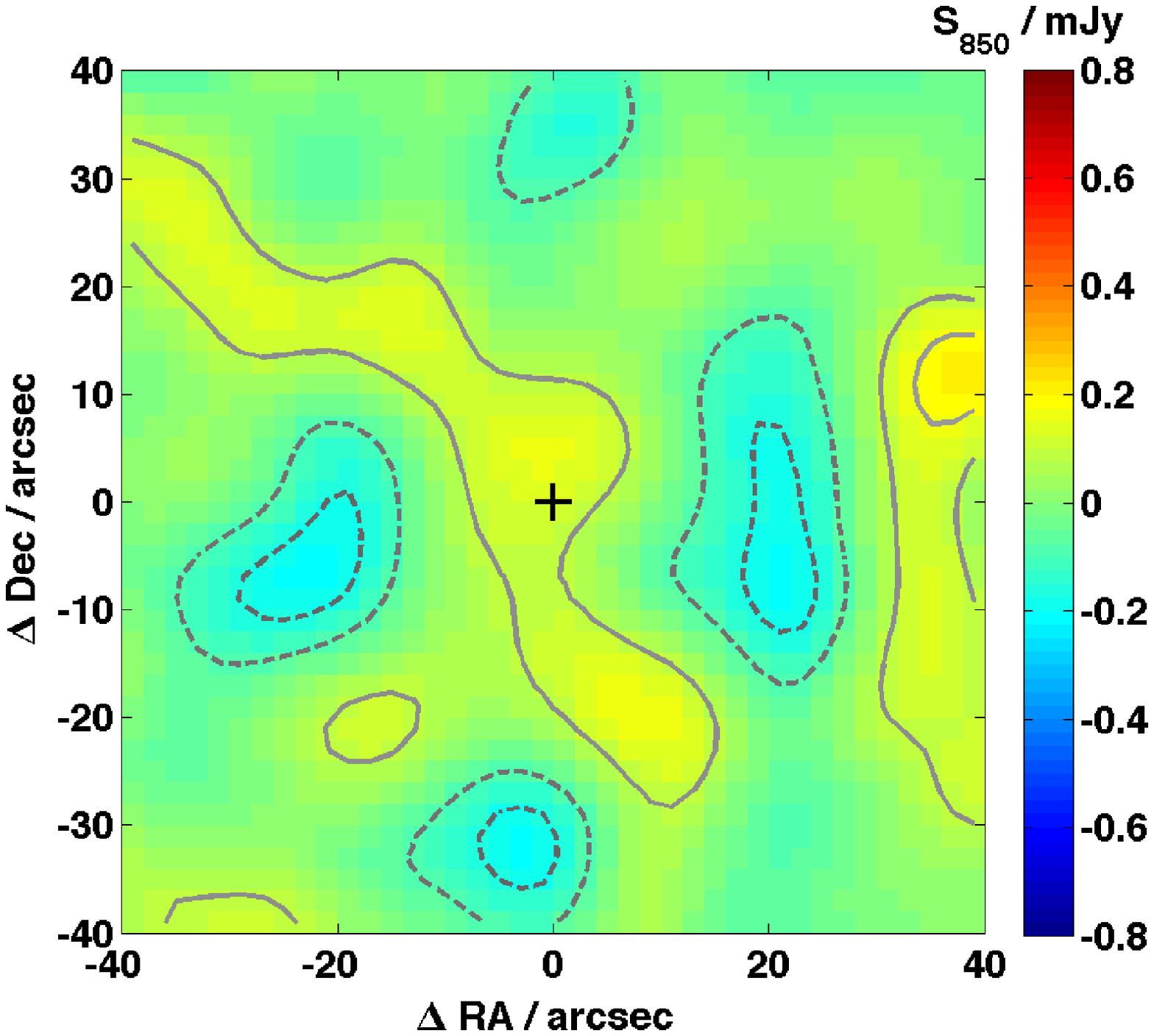}
\caption{80$\times$80$^{\prime \prime}$ weighted, stacked 850$\mu$m images centred on the positions of the 699 $z>1$ X-ray AGN in our flux limited sample (top) and centred on exactly the same number of random points (bottom). The AGN stack includes 428 Type 1 AGN and 271 Type 2 AGN. The colour scaling is the same in both images. The contours are at 1-5$\sigma$ and the solid contours correspond to positive fluxes whereas the dashed contours correspond to negative fluxes. We clearly observe a statistically significant 850$\mu$m detection from the AGN relative to the random points.}
\label{fig:stack}
\end{center}
\end{figure}

The majority of the X-ray AGN are not individually detected in the 850$\mu$m catalogue, which goes down to a 3.5$\sigma$ flux limit of $\sim$3.5 mJy in the highest sensitivity regions. To understand the average submillimeter properties of the AGN we therefore have to study stacked 850$\mu$m maps. We begin by constructing an inverse variance weighted stacked image at 850$\mu$m for all 699 $z>1$ AGN (428 Type 1 and 271 Type 2) that will be used from hereon in our stacking analysis. Specifically:

\begin{equation}
S_{ij}=\frac{\Sigma_{k=1}^{N} P_{ij}\times f_{ij} ^{k}/ (\sigma_{ij}^{k} \times \sigma_{ij}^{k})}{\Sigma_{k=1}^{N}P_{i,j}/(\sigma_{ij}^k \times \sigma_{ij}^k)}
\label{eq:stack}
\end{equation}

\noindent where $S_{ij}$ represents the stacked 850$\mu$m image, $P_{ij}$ is the point response function of SCUBA-2 at 850$\mu$m generated by stacking all $>$10$\sigma$ sources in the map, $f_{ij}$ are the 40$\times$40 pixel (80$\times$80${\arcsec}$) flux images of each of the $N$ sources going into the stacks and $\sigma_{ij}$ are the corresponding RMS images. Stacked images generated in this way can be seen in Fig. \ref{fig:stack} for both the AGN sample and a set of random points in the map, where any random points within 7.5$\arcsec$ of a catalogued 850$\mu$m source, have been removed. Fig. \ref{fig:stack} clearly demonstrates the presence of a statistically significant 850$\mu$m signal from the $z>1$ X-ray AGN relative to the random points. The typical RMS noise in the maps is well above the confusion limit. As stated in Section \ref{sec:data}, the noise is non-uniform over the full 2 deg$^2$ map. Out of the 699 $z>1$ sources being stacked, 80\% lie in regions with RMS noise $<$3.5 mJy. From the 850$\mu$m source counts presented in \citet{Casey:13}, we estimate that the integral source count at this RMS noise level is only $\sim$630 deg$^{-2}$. Assuming a beam FWHM of 14.5$\arcsec$ for SCUBA-2 at 850$\mu$m, the beam density is $\sim$78,000 deg$^{-2}$. Hence only 1 in 120 beams are expected to be affected by confusion at this RMS noise. In the lowest RMS regions in our map ($\sigma$=1mJy), the integral source counts reach $\sim$3000 deg$^{-2}$ and 1 in 25 beams are affected by confusion, which is still a small effect.

Before investigating the dependence of the 850$\mu$m emission on the AGN properties, we first describe our methodology for calculating the stacked 850$\mu$m fluxes. Throughout this analysis the stacked 850$\mu$m flux is measured as the median or mean value of the central pixel in 100 bootstrapped inverse variance weighted stacked images (see Eq. \ref{eq:stack}). We have already checked in Section \ref{sec:data} that the astrometry of the SCUBA-2 map is accurate to within a pixel so this central pixel flux provides the most conservative and unbiased estimate of the average flux. The central pixel flux is measured to be 0.71$\pm$0.08 mJy for all 699 AGN. The median redshift of the sample is $z=1.66$ and the median hard X-ray luminosity is log$_{10}$(L$_{\rm{2-10}})=44.2$. We emphasise that these X-ray luminosities correspond to the most luminous AGN in the ``quasar" regime. For comparison, the average 870$\mu$m flux of the \citet{Lutz:10} AGN is 0.49$\pm$0.04mJy. The \citet{Lutz:10} sample has both a lower median redshift ($z=1.17$) and a lower median hard X-ray luminosity - log$_{10}$(L$_{\rm{2-10}})=43.6$ calculated from the median hard X-ray flux in that paper and using our assumed value of $\Gamma=1.8$ as described in Section \ref{sec:cat}. Given these differences in the median properties of the two samples as well as the correction from 870$\mu$m to 850$\mu$m for a typical single temperature greybody SED and the statistical errors in both measurements, our results are broadly consistent with those of \citet{Lutz:10}. Excluding the individually detected 850$\mu$m X-ray AGN described in Section \ref{sec:850detect} from the stacks, we find that the average flux falls to 0.60$\pm$0.07 mJy, which is still a $>$8$\sigma$ detection. 

We have checked that the median and mean stacked fluxes are very similar in both cases. The error on the flux is calculated using 100 bootstrap samples. We also calculate stacked fluxes using the publicly available code {\sc simstack} \citep{Viero:13}. The code is designed to correctly account for clustering of sources within the scale of the SCUBA-2 beam which is assumed to be 14.5$^{\prime \prime}$. The method essentially involves performing a regression of the true flux map with a `hits' map, where the `hits' map is created by counting the number of sources to be stacked that contribute to each pixel in the map. We stack the 699 flux-limited $z>1$ AGN with the $z<1$ AGN. If the photometric redshifts are accurate, there should be no clustering between the high and low redshift AGN populations. The stacked 850$\mu$m flux for the 699 $z>1$ AGN is 0.82$\pm$0.02 mJy from {\sc simstack} which is slightly higher than our central pixel flux measurement. The higher {\sc simstack} flux is likely due to the fact that the peak 850$\mu$m emission in Fig. \ref{fig:stack} is offset from the central pixel. The {\sc simstack} errors reflect the regression errors and do not account for any errors due to sample variance. Our bootstrap errors are therefore more conservative. The {\sc simstack} results demonstrate that the central pixel flux represents a conservative estimate and is not biased up due to clustering. For the rest of the analysis, we therefore use the central pixel flux to measure the average submillimeter emission for all the sub-samples we construct. As a final check, given that some regions of the 850$\mu$m wide S2CLS COSMOS maps are significantly noisier than others (Section \ref{sec:data}), we also restrict our stacks only to the regions with $\sigma_{850}<$2mJy to mitigate against the effects of flux boosting. There are 341 $z>1$ X-ray AGN in our flux-limited sample in these high sensitivity regions and the mean flux is S$_{850}$=0.69$\pm$0.10 mJy, consistent with the results derived using the entire 850$\mu$m wide map. We will therefore proceed with using the 850$\mu$m data over the full COSMOS wide area for all the stacking analysis that follows. 

Before continuing, it is important to highlight the assumptions inherent in our work. Throughout this paper, we assume that the 850$\mu$m flux is tracing the total amount of cool dust emission in the AGN host galaxies and that this cool dust is primarily heated by star formation and has a negligible contribution from the AGN. Given that the 850$\mu$m flux is invariant with redshift at $z>1$, we can average these fluxes over relatively broad redshift bins without bias. Note that this is not the case for \textit{Herschel} fluxes for example, which trace the peak of the dust SED at $z\sim1-3$, and are expected to be varying more strongly with redshift. As we do not have large enough AGN samples to be able to split our AGN into narrow redshift bins, we restrict our analysis to the 850$\mu$m properties only rather than constructing full SEDs from the \textit{Herschel} and 850$\mu$m data. The 850$\mu$m emission can also only be used as a proxy for the star formation rate provided the dust temperature is not varying significantly between AGN sub-samples, and also assuming that the dust heating at these wavelengths is dominated by recent star formation rather than being dominated by cold dust heated by old stars and the interstellar medium (e.g. \citealt{Bourne:13}). We therefore begin by testing how the 850$\mu$m flux varies with star formation rate. 

\subsection{Star Formation Rate}

\label{sec:sfr}

\citet{Bongiorno:12} have estimated the star formation rates for the COSMOS AGN using SED-fitting to the available multi-wavelength photometry \citep{Brusa:10} - SFR$_{\rm{SED}}$ hereafter. These can be used to check the dependence of the 850$\mu$m flux density on star formation rate. As described in Section \ref{sec:cat}, the SED based star-formation rates include photometry in the \textit{Spitzer} IRAC as well as MIPS 24$\mu$m bands in most cases. In some cases, the SED fits also include longer wavelength photometry at MIPS 70$\mu$m as well as at \textit{Herschel} 100 and 160$\mu$m. We select only the Type 2 $z>1$ AGN with robust star formation rate estimates as specified by the \textit{flags} parameters in \citet{Bongiorno:12}. We note that soft X-ray detected Type 2 AGN that are not detected in the hard band are also included in this analysis to improve the statistics. Here we are primarily concerned with checking whether the 850$\mu$m fluxes correlate with other independently derived star formation rate constraints and knowledge of the true dust columns and therefore intrinsic X-ray luminosities of the AGN is not relevant. In luminous, unobscured Type 1 AGN, the big blue bump typically seen in quasar spectra may mimic the UV bump from massive stars so star formation rates based on SED fitting are no longer reliable. Hence Type 1 AGN are not considered in this test.  

We split the Type 2 AGN into two SFR$_{\rm{SED}}$ bins at SFR$>$10M$_\odot$yr$^{-1}$ and SFR$<$10M$_\odot$yr$^{-1}$. The median redshifts are $z=1.36$ and $z=1.77$ for the low and high SFR$_{\rm{SED}}$ sub-samples respectively. The median star formation rates are 1.5M$_\odot$yr$^{-1}$ and 33M$_\odot$yr$^{-1}$ while the mean star formation rates are 1.1M$_\odot$yr$^{-1}$ and 46M$_\odot$yr$^{-1}$ in the two bins. The average 850$\mu$m fluxes for these sub-samples are 0.3$\pm$0.1 mJy and 1.0$\pm$0.2 mJy respectively and are therefore clearly higher in the higher SFR bin. The excess 850$\mu$m flux in the high SFR bin is significant at the $\sim$3$\sigma$ level. Several previous studies have instead used the 250$\mu$m fluxes as a proxy for SFR. We therefore also calculate average \textit{Herschel} 250$\mu$m fluxes for the Type 2 AGN in these two SFR bins although we caution that the bins encompass AGN over a relatively broad range of redshifts, which could be adding scatter to the 250$\mu$m fluxes. The 250$\mu$m fluxes are calculated by injecting the AGN in each SFR bin in turn into the \textit{Herschel} SPIRE 250$\mu$m map from \citet{Swinbank:14}, which has had all 24$\mu$m detected sources removed. The maps are then stacked at the AGN position and the average flux is once again measured as the value at the central pixel. Note that we have also checked the astrometry of the 250$\mu$m map in Section \ref{sec:data}. The 250$\mu$m fluxes in these two SFR bins are 2.0$\pm$0.3 mJy and 5.3$\pm$0.6 mJy respectively. In order to check whether the 250$\mu$m and 850$\mu$m stacks are consistent with each other, we fit a single temperature greybody with $\beta$=2.0 to these data points in the two SFR bins. The dust temperatures are 24$\pm$6K and 27$\pm$5K respectively, consistent with a single temperature in both SFR bins. From the integral under the greybody, we can then calculate the far infrared luminosity (between 60--300$\mu$m to eliminate any contributions from the AGN) and average star formation rate for each stack. These are log$_{10}$(L$_{\rm{FIR}}$/L$_\odot$)=10.9$\pm$0.3 corresponding to SFR=12$\pm^{11}_{6}$M$_\odot$yr$^{-1}$ for the low SFR stack and log$_{10}$(L$_{\rm{FIR}}$/L$_\odot$)=11.5$\pm$0.2 corresponding to SFR=60$\pm^{30}_{20}$M$_\odot$yr$^{-1}$ for the high SFR stack. In both cases the star formation rates derived from the \textit{Herschel} and 850$\mu$m data are higher than the mean star formation rates from \citet{Bongiorno:12} although the error bars and scatter in both average SFR measurements are large. In Appendix B where we have shown best-fit far infrared SEDs for our X-ray AGN that are individually detected at 850$\mu$m, we also show that the far infrared luminosity and star formation rate as measured from these best-fit SEDs, correlates well with both the 250 and 850$\mu$m fluxes (Fig. \ref{fig:SLIR}). In Appendix B, the dust mass from full SED fitting has also been plotted against both the 250 and 850$\mu$m fluxes and in general shows a weaker correlation with the infra-red and sub-mm flux densities compared to the far infra-red luminosity. We will therefore proceed by assuming that the stacked 850$\mu$m flux in our $z>1$ AGN is giving us a direct estimate of the average star formation rate, with the caveat that for galaxies with low levels of star formation, there may also be a contribution to the dust heating at 850$\mu$m from old stars and the interstellar medium. 

\subsection{Redshift}

\label{sec:redshift}

\begin{figure}
\begin{center}
\includegraphics[scale=0.45,angle=0]{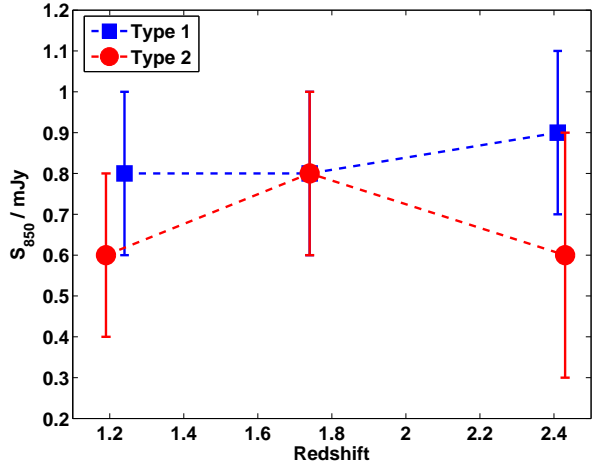} 
\caption{Average 850$\mu$m flux versus redshift for both Type 1 and Type 2 AGN.}
\label{fig:S850_z}
\end{center}
\end{figure}

The size of our current sample necessitates stacking over relatively broad redshift bins. Evolutionary effects could therefore influence any derived results and introduce differences in stacked 850$\mu$m fluxes between different sub-samples. Before looking for trends with other AGN properties we therefore first consider the redshift evolution of the 850$\mu$m fluxes over the full redshift range considered in our stacked analyses. In order to do this, we split the $z>1$ Type 1 and Type 2 AGN into three redshift bins and consider their stacked 850$\mu$m fluxes. The results are illustrated in Fig. \ref{fig:S850_z}. Both the Type 1 and Type 2 AGN submillimeter fluxes show little evolution with redshift over the redshift range probed in this work. In all three redshift bins, the Type 1 and Type 2 AGN 850$\mu$m fluxes are consistent with each other given the error bars. When considering trends in AGN submillimeter fluxes with luminosity, obscuration and AGN activity in the following sections, we will therefore assume that the 850$\mu$m fluxes do not evolve with redshift over the redshift range considered.

\subsection{X-ray Luminosity}

\label{sec:Xray}

We now explore trends in these 850$\mu$m fluxes with the hard X-ray luminosity, which can be used as a proxy for the energy emitted by the central accreting power source. Several recent studies have found a reasonably flat relationship between AGN luminosity and star formation rate (e.g. \citealt{Harrison:12, Rosario:12, Azadi:14, Stanley:15}). At modest X-ray luminosities where large samples of X-ray AGN with far infrared and submillimeter data now exist, these results are now robust. However, the situation is less clear at very high X-ray luminosities. While \citet{Page:12} have suggested that the far infra-red fluxes and therefore star formation rates of AGN with high X-ray luminosities, are suppressed relative to AGN of lower luminosity, \citet{Lutz:10} and \citet{Rosario:12} instead find an increase in star formation rate at high X-ray luminosities, with the results being more pronounced for low redshift AGN in \citet{Rosario:12}. The results from recent hydrodynamical simulations suggest that there is a large scatter in the L$_{\rm{X}}$-SFR relation over the dynamic range probed by current surveys \citep{Sijacki:14}, which may explain some of the discrepancies between previous studies. 

\begin{table*}
\begin{center}
\caption{Stacked 850$\mu$m fluxes for the Type 1 and Type 2 AGN in three bins in hard X-ray luminosity. Detections above 3$\sigma$ are shown in bold and the number of sources in each bin is given in brackets.}
\label{tab:Xray_table}
\begin{tabular}{lcc}
\hline
& Type 1 AGN & Type 2 AGN \\
\hline
43.5$<$log$_{10}$(L$_{2-10}$/ erg s$^{-1}$)$<$44.0 & 0.7$\pm$0.3 (N=102) & 0.5$\pm$0.2 (N=114) \\
44.0$<$log$_{10}$(L$_{2-10}$/ erg s$^{-1}$)$<$44.4 & \textbf{1.0$\pm$0.2} (N=168) & \textbf{0.6$\pm$0.2} (N=94) \\
log$_{10}$(L$_{2-10}$/ erg s$^{-1}$)$>$44.4 & \textbf{0.7$\pm$0.2} (N=146) & \textbf{1.3$\pm$0.3} (N=50) \\
\hline
\end{tabular}
\end{center}
\end{table*} 

\begin{figure}
\begin{center}
\includegraphics[scale=0.45,angle=0]{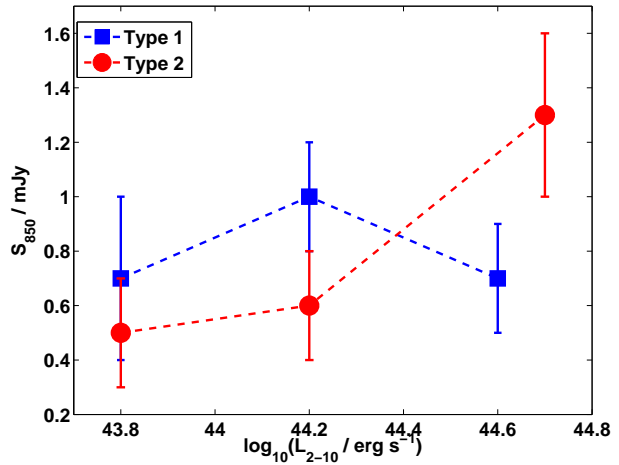} 
\caption{Average 850$\mu$m flux versus hard X-ray luminosity for both Type 1 and Type 2 AGN.}
\label{fig:L_S850}
\end{center}
\end{figure}

The SCUBA-2 photometry presented in this paper now allows an independent investigation of the relation between star formation and AGN luminosity with the advantage that the submillimeter flux is less contaminated by AGN emission than the \textit{Herschel} fluxes in the case of the most luminous quasars. As illustrated in Fig. \ref{fig:LXz}, at $z>1$, the COSMOS field is only complete at relatively high X-ray luminosities. We select Type 1 and Type 2 AGN in three X-ray luminosity bins  - 43.5$<\rm{log}_{10}(L_{\rm{2-10}})<44$,  44.0$<\rm{log}_{10}(L_{\rm{2-10}})<44.4$ and $\rm{log}_{10}(L_{\rm{2-10}})>44.4$. The AGN that are individually detected at 850$\mu$m (see Section \ref{sec:850detect}) are also included in the stack. There are only 7 X-ray AGN out of the 19 that satisfy both the hard X-ray flux limit and the $z>1$ redshift cut and we have checked that inclusion of these AGN does not change the stacked fluxes quoted within the error bars. Note that the lowest X-ray luminosity bin is incomplete at $z \gtrsim 2$ so the results here should be interpreted with caution. The average 850$\mu$m fluxes are presented in Table \ref{tab:Xray_table} and illustrated in Fig. \ref{fig:L_S850}. While the Type 1 submillimeter flux appears to be fairly independent of X-ray luminosity, there is some evidence that the most X-ray luminous Type 2 AGN have higher submillimeter fluxes. The Type 2 AGN at  $\rm{log}_{10}(L_{\rm{2-10}})>44.4$ have 850$\mu$m fluxes that are $\sim$2$\sigma$ higher than in the previous X-ray bin although the numbers of Type 2 AGN at such high X-ray luminosities are small so the results could be affected by small number statistics. 

In the highest X-ray luminosity bin, the AGN are expected to have a relatively narrow Eddington ratio distribution based on the models of \citet{Aird:13}. These AGN therefore likely correspond to high-mass black-holes hosted in massive galaxies. If AGN were preferentially found in star-forming galaxies on the main sequence, these massive galaxies would be expected to have higher star formation rates. The high X-ray luminosity bins may therefore be dominated by a larger fraction of quiescent hosts for the Type 1 AGN, which would naturally explain the lower 850$\mu$m fluxes. X-ray luminous Type 1 AGN could also represent a skewed population of highly accreting black-holes seen during a brief phase of black-hole growth. As such, their host galaxies may be under-massive relative to their black-hole masses, and this could once again explain the lower dust luminosities and therefore star formation rates in these sources. 

The Type 2 AGN show a more marked increase in average submillimeter flux with X-ray luminosity with the most X-ray luminous AGN having higher star formation rates. As can be seen in Fig. \ref{fig:LXz}, for a flux-limited sample, the median redshift increases with increasing X-ray luminosity. However, we have shown in Section \ref{sec:redshift} that the 850$\mu$m flux does not evolve significantly with redshift in the Type 2 AGN population. Therefore the trend observed in Fig. \ref{fig:L_S850} is unlikely to be driven by redshift. Evidence of higher star formation rates in more X-ray luminous AGN was also found by \citet{Lutz:10} who argue for two modes of star formation in the population - secular and merger-driven. The most X-ray luminous Type 2 AGN could therefore represent a transition to a merger-driven scenario where star formation and black hole accretion are more tightly coupled. Fig. \ref{fig:L_S850} also suggests that there may be systematic differences in the host galaxy properties of the most X-ray luminous Type 1 and Type 2 AGN with the luminous Type 2 AGN occupying more highly star-forming galaxies.  Alternatively, it is possible that the Type 2 AGN have higher 850$\mu$m fluxes simply as a result of a higher dust mass relative to their Type 1 counterparts of comparable luminosity. 

Overall, it is difficult to establish a direct causal connection between AGN activity and star formation when studying average properties of AGN host galaxies over a wide range in luminosity, redshift and host galaxy type. The situation is further complicated by the fact that the timescales for star formation and AGN activity are very different so AGN variability can wipe out any underlying correlations \citep{Hickox:14, Gabor:13}. Indeed the AGN that are individually detected at $>$3.5$\sigma$ at 850$\mu$m (Table \ref{tab:IDs}) have a range of hard X-ray luminosities and would populate all three X-ray luminosity bins considered in this analysis. However, at least in the Type 2 AGN population there appears to be evidence for a transition to more highly star-forming hosts and/or higher dust mass at the highest X-ray luminosities.  

\subsection{Obscuration}

\label{sec:obscuration}

AGN classifications into Type 1s and Type 2s as described in Section \ref{sec:cat}, are one way to potentially separate unobscured and obscured AGN. However, other obscuration measures have also previously been used in the literature, many of these tracing obscuration on potentially very different physical scales. A key aim of this work is to understand if any of these obscuration measures are correlated with the 850$\mu$m flux and therefore star formation rates of AGN. Having established trends in the 850$\mu$m flux with both redshift and X-ray luminosity in the preceding sections, we will aim to disentangle these effects from the effects of obscuration on the cold dust emission. 

\subsubsection{Type 1 versus Type 2 AGN}

\label{sec:type12}

There are 428 $z>1$ Type 1 AGN and 271 $z>1$ Type 2 AGN in our sample as classified based on their spectroscopic properties/best-fit SEDs (see Section \ref{sec:cat}). The median redshifts are 1.78 and 1.45 and the median hard X-ray luminosities are log$_{10}$(L$_{\rm{2-10}}$)=44.2 and 44.0 respectively. In Appendix C we also show both the X-ray luminosity and redshift distributions for the Type 1 and Type 2 populations. The Type 2 population clearly shows a paucity of sources at the highest redshifts. However in Section \ref{sec:redshift} we have already established that redshift trends are unlikely to be significant within the population. The stacked 850$\mu$m flux densities for the Type 1s and Type 2s are 0.8$\pm$0.1 mJy and 0.6$\pm$0.1 mJy respectively. In Section \ref{sec:Xray}, we have seen that the high luminosity Type 2 AGN have higher submillimeter fluxes but at lower luminosities, the Type 1 AGN 850$\mu$m fluxes are slightly higher. Given that the X-ray luminosity distribution for the Type 2 AGN sample peaks at lower luminosities than the Type 1 AGN, this helps explain the small difference in the overall stacked fluxes observed here.

\subsubsection{X-ray Hardness Ratio}

The X-ray hardness ratio (HR) is used to characterise the colour index associated with the X-ray spectrum of an AGN and therefore the level of absorption seen in the X-ray. The HR is defined as $(H+S)/(H-S)$ where $H$ represents the counts in the \textit{XMM-Newton} hard X-ray band and $S$ represents the counts in the \textit{XMM-Newton} soft X-ray band. An HR value of $-0.2$ has been used by \citet{Brusa:10} to separate obscured and unobscured AGN and following this, we split our AGN sample into two HR bins. Only those AGN that are detected in \textit{both} the soft and hard X-ray bands are used for this analysis. Our sample comprises 129 HR$\ge -0.2$ obscured AGN (34 Type 1 and 95 Type 2) and 346 HR$<-$0.2 unobscured AGN (263 Type 1 and 83 Type 2). The average 850$\mu$m fluxes are very similar for the obscured and unobscured samples as shown in Table \ref{tab:obs} - 0.7$\pm$0.3 mJy at HR$\ge -0.2$ and 0.7$\pm$0.1 mJy at HR$<-0.2$. In Appendix C we also show the redshift and luminosity distributions for the two samples, which are very similar. Various previous studies have considered the dependence of cool dust emission from AGN on X-ray obscuration estimates parametrized by the neutral hydrogen column density (e.g. \citealt{Stevens:05, Shao:10, Lutz:10, Rosario:12, Merloni:14}). In most of these studies, no significant correlation is seen between cool dust emission and the dust column towards the central accreting black-hole. Thus the dust responsible for the X-ray absorption is probably not directly associated with star formation processes. Indeed most X-ray studies place warm absorbers on much smaller spatial scales than typical star-forming regions in galaxies.

\begin{table*}
\begin{center}
\caption{Summary of stacked 850$\mu$m fluxes for subsamples of AGN selected according to various obscuration parameters. Detections above 3$\sigma$ are shown in bold.}
\label{tab:obs}
\begin{tabular}{llcccc}
\hline \hline
Parameter Investigated & Sample & N & $z$ & log$_{10}($L$_{2-10}$/erg s$^{-1}$) & S$_{850}$ (mJy)  \\
\hline
\hline
Spectroscopic/Photometric Classification & Type 1 & 428 & 1.78 & 44.2 & \textbf{0.8$\pm$0.1} \\
& Type 2 & 271 & 1.45 & 44.0 & \textbf{0.6$\pm$0.1} \\
\\
Hardness Ratio & Type 1 \& 2 $HR>-0.2$ & 129 & 1.60 & 44.2 & 0.7$\pm$0.3 \\
& Type 1 \& 2 $HR<-0.2$ & 346 & 1.61 & 44.3 & \textbf{0.7$\pm$0.1} \\
\\
Optical/Infrared Colour & Type 1 \& 2, (r-4.5)$>$6.1 (Vega) & 303 & 1.56 & 44.1 & \textbf{0.7$\pm$0.1} \\
& Type 1 \& 2, (r-4.5)$<$6.1 (Vega) & 391 & 1.74 & 44.2 & \textbf{0.7$\pm$0.1} \\
& Type 1 \& 2, (r-4.5)$<$6.1 (Vega) & 124 & 2.15 & 44.7 & \textbf{0.7$\pm$0.2} \\
& log$_{10}($L$_{2-10})>$44.4 & & & & \\ 
& Type 1 \& 2, (r-4.5)$>$6.1 (Vega) & 60 & 2.31 & 44.7 & \textbf{1.2$\pm$0.3} \\
& log$_{10}($L$_{2-10})>$44.4 & & & & \\ 
& Type 1 \& 2, (r-K)$>$5 (Vega), log$_{10}$(X/0)$>$1 & 149 & 1.59 & 44.2 & \textbf{0.7$\pm$0.2} \\
& Type 1 \& 2, (r-K)$>$5 (Vega), log$_{10}$(X/0)$>$1 & 31 & 1.96 & 44.7 & \textbf{1.6$\pm$0.4} \\
& log$_{10}($L$_{2-10})>$44.4 & & & & \\ 
\\
SED Fit Extinction & Type 1 \& 2 $E(B-V)_{\rm{AGN}}>0.3$ & 348 & 1.50 & 44.1 & \textbf{0.7$\pm$0.1} \\
& Type 1 \& 2 $E(B-V)_{\rm{AGN}}<0.3$ & 351 & 1.80 & 44.2 & \textbf{0.7$\pm$0.1} \\
& Type 1 \& 2 $E(B-V)_{\rm{GAL}}>0.3$ & 298 & 1.63 & 44.2 & \textbf{0.9$\pm$0.1} \\
& Type 1 \& 2 $E(B-V)_{\rm{GAL}}<0.3$ & 366 & 1.60 & 44.1 & \textbf{0.7$\pm$0.1} \\
\hline
\end{tabular}
\end{center}
\end{table*}

\subsubsection{Optical-to-infrared colours}

Optical to infrared flux ratios have been proposed as a way of identifying luminous obscured quasars that are not in X-ray surveys (e.g. \citealt{Hickox:07}) and recent results suggest that far infrared fluxes and therefore star formation rates could be correlated with such colour measures of obscuration \citep{Chen:15}. We therefore split our AGN sample into two (r-4.5$\mu$m) colour bins using the threshold of (r-4.5$\mu$m)=6.1 (Vega) used in both \citet{Hickox:07} and \citet{Chen:15}. The red sample has 303 AGN (65 Type 1 and 238 Type 2) at a median redshift of 1.56, whereas the blue sample has 391 AGN (359 Type 1 and 32 Type 2) at a median redshift of 1.74. The stacked 850$\mu$m fluxes are presented in Table \ref{tab:obs} and are very similar for the two populations - 0.7$\pm$0.1 mJy. Our result, that red and blue AGN have very similar submillimeter properties, is apparently at odds with that of \citet{Chen:15}. However, these authors have investigated differences in red and blue AGN at the highest quasar luminosities. \citet{Lutz:10} also find that trends with obscuration start to become more significant at high X-ray luminosities and as illustrated in Fig. \ref{fig:L_S850}, differences in Type 1 and Type 2 AGN submillimeter fluxes only become significant at high luminosities. We therefore only select those AGN with log$_{10}$(L$_{2-10})>$44.4 and again split these into two samples based on the $(r-4.5\mu$m) colours. The red sample has 60 AGN (18 Type 1 and 42 Type 2) and the blue sample has 124 AGN (116 Type 1 and 8 Type 2). The 850$\mu$m fluxes in the two bins are once again given in Table \ref{tab:obs}. An excess in the 850$\mu$m flux is now observed in the red AGN sample although this is only significant at the $\sim$1.3$\sigma$ level given the sample size.    

Another interesting population of red, dusty AGN have also been identified in the COSMOS field by \citet{Brusa:10}. These authors use the $(r-K)$ colour and X-ray to optical flux ratio to select obscured AGN. Follow-up observations of this population demonstrate that they are hosted in starburst / main-sequence star-forming galaxies \citep{Bongiorno:14} with strong evidence for powerful outflows affecting the ionised gas \citep{Brusa:14, Cresci:14}, consistent with these dusty quasars being caught during a brief evolutionary phase when both star formation and black hole accretion are synchronously occurring. The most X-ray luminous red AGN in the COSMOS field, XID2028, was independently selected in a wide-field search for reddened, high-redshift quasars in the UKIDSS Large Area Survey \citep{Banerji:12}. The population of hyper-luminous reddened quasars also shows evidence for very high dust luminosities \citep{Banerji:14} and large reservoirs of molecular gas \citep{Feruglio:14}, suggesting that the dust obscuration is, at least partially associated with star formation in the quasar host. This population of moderately obscured or reddened AGN appears to dominate the AGN luminosity function even at very high luminosities \citep{Banerji:15,Lacy:15}. These are therefore candidates where we might expect excess submillimeter emission relative to the rest of the AGN population.  

The brightest red AGN XID2028 lies in a region of poor sensitivity in our 850$\mu$m wide map, with $\sigma_{850} \sim 3.4$ mJy and is therefore not present in the 850$\mu$m catalogue. XID18, another obscured quasar from \citet{Brusa:14}, is associated with an 850$\mu$m bright source in Table \ref{tab:IDs}. We once again employ a stacking analysis in order to test whether such red obscured quasars are associated with high cold dust luminosities and therefore high star formation rates. Specifically, we follow \citet{Brusa:10} and isolate a population of AGN with (r-K)$>$5 (Vega) and log$_{10}$(X/O)$>$1, where (X/O) represents the hard X-ray to optical $r$-band flux ratio and the monochromatic $r$-band flux is estimated as described in \citet{Brusa:10}. Our sample comprises 149 AGN (14 Type 1 and 135 Type 2) at a median redshift of 1.59. The average 850$\mu$m flux is 0.7$\pm$0.2 mJy and there is no strong evidence for an excess in the submillimeter fluxes compared to the various other obscured populations that we have been investigating above. However, once again isolating only the X-ray luminous sub-set of this red population (log$_{10}$(L$_{2-10})>$44.4), we now find that the stacked 850$\mu$m flux of $1.6\pm0.4$ mJy to be more than 2$\times$ higher at a significance of $\sim$2$\sigma$, once again suggesting that an increase in cold dust emission with obscuration only becomes significant at high luminosities. While we have put forward the interpretation that the cold dust emission is directly tracing star formation in these high luminosity, obscured AGN samples, it is also possible that the more highly obscured X-ray luminous AGN simply have higher dust mass, which is responsible for driving the observed 850$\mu$m emission trends. However, as dust mass in galaxies is closely tied to their gas content, large dust masses generally imply large gas reservoirs and therefore presumably also higher star formation rates.   

\subsubsection{E(B-V) from SED Fitting}

In \citet{Bongiorno:12}, extinctions have been derived for the COSMOS AGN sample from SED fitting to the extensive multi wavelength photometry. Separate extinction values were fit for the AGN and galaxy components. As expected the Type 2 AGN are dominated by high extinctions towards the AGN component whereas the Type 1 sources are fit by relatively modest extinctions to the AGN component. We begin by separating the $z>1$ AGN sample based on these AGN component extinction values. Two samples are created with $E(B-V)_{\rm{AGN}}<0.3$ and $E(B-V)_{\rm{AGN}}>0.3$ respectively. The low extinction sample comprises 349 Type 1 AGN and only 2 Type 2 AGN and has a median redshift of 1.80 and a mean $E(B-V)_{\rm{AGN}}=0.04$. The high extinction sample has 269 Type 2 AGN and 79 Type 1 AGN with a median redshift of 1.50 and a mean $E(B-V)_{\rm{AGN}}=1.3$. The average 850$\mu$m fluxes are 0.7$\pm$0.1 mJy in both the low and high extinction bins respectively and there is therefore no evidence for a dependence of star formation rate on these AGN component extinction values. 

We now consider the extinction values fit to the galaxy component instead, using a threshold of $E(B-V)_{\rm{GAL}}$=0.3 to separate the AGN into two samples. The reliability of SED-based extinction estimates for AGN host galaxies has not been explored in detail thus far and should clearly be interpreted with caution, particularly for Type 1 AGN. Before using these extinction values we therefore look for systematic biases in the two populations. Firstly, we confirm using the rest-frame monochromatic UV-luminosities from \citet{Lusso:10} for the Type 1 AGN, that there are no systematic differences in the UV-luminosity between the two extinction bins (log$_{10}$(L$_{\rm{UV}}/\rm{erg s}^{-1} \rm{cm}^{-2} \rm{Hz}^{-1})$=29.7 in both bins). Next we calculate the rest-frame 5.8$\mu$m luminosity directly from the infrared photometry from \citet{Brusa:10} and look for differences in the mid infrared luminosity between the two extinction bins. A systematic difference in the rest-frame 5.8$\mu$m luminosity between host galaxy extinction bins would indicate that AGN split by their host galaxy extinction have different mid infra-red SEDs, which could reflect differences in covering factor and viewing angle. Once again, we find no notable difference in the rest-frame 5.8$\mu$m luminosity as a function of extinction, with average values of log$_{10}$(L$_{5.8}$/erg s$^{-1}$)=45.1 in both bins. 

Having confirmed that the Type 1 AGN split by host extinction do not show systematic differences in either their UV or IR luminosities, we now proceed to look at the average 850$\mu$m flux in these two bins. The average fluxes are quoted in Table \ref{tab:obs} and are 0.7$\pm$0.1 mJy and 0.9$\pm$0.1 mJy in the low and high extinction bins respectively. The mean E(B-V)$_{\rm{GAL}}$ values in these bins are 0.06 and 0.6 respectively. The excess 850$\mu$m flux in the high extinction bin is only significant at the 1.4$\sigma$ level. The median stellar masses and hard X-ray luminosities of the AGN in these two extinction bins are very similar - log$_{10}$(M$_*$/M$_\odot$)=10.80 and 10.85 respectively and log$_{10}$(L$_{\rm{2-10}}/\rm{erg s}^{-1})$=44.14 and 44.16 respectively. The luminosity and redshift distributions of the various E(B-V) selected samples are also shown in Appendix C and demonstrate that the samples overlap significantly in both X-ray luminosity and redshift space.

Overall, we find that trends in the 850$\mu$m flux with different measures of obscuration are weak when the AGN sample is considered as a whole. However, we do find that these trends become more significant at the highest luminosities where a transition from secular to merger-driven co-evolution may be occurring. Much larger samples over wider fields are required to confirm these results and investigate the overlap between obscured populations selected using different classification methods. Starting with mid infrared rather than X-ray selected parent AGN samples would also help in getting better statistics on luminous, obscured AGN where high levels of obscuration cause the AGN to drop out of flux-limited X-ray samples. These studies are now becoming possible with very large mid infrared surveys such as the \textit{WISE} All-Sky Survey \citep{Wright:10}.

\subsection{AGN Versus Non-AGN}

\label{sec:gal}

\begin{figure}
\begin{center}
\includegraphics[scale=0.4,angle=0]{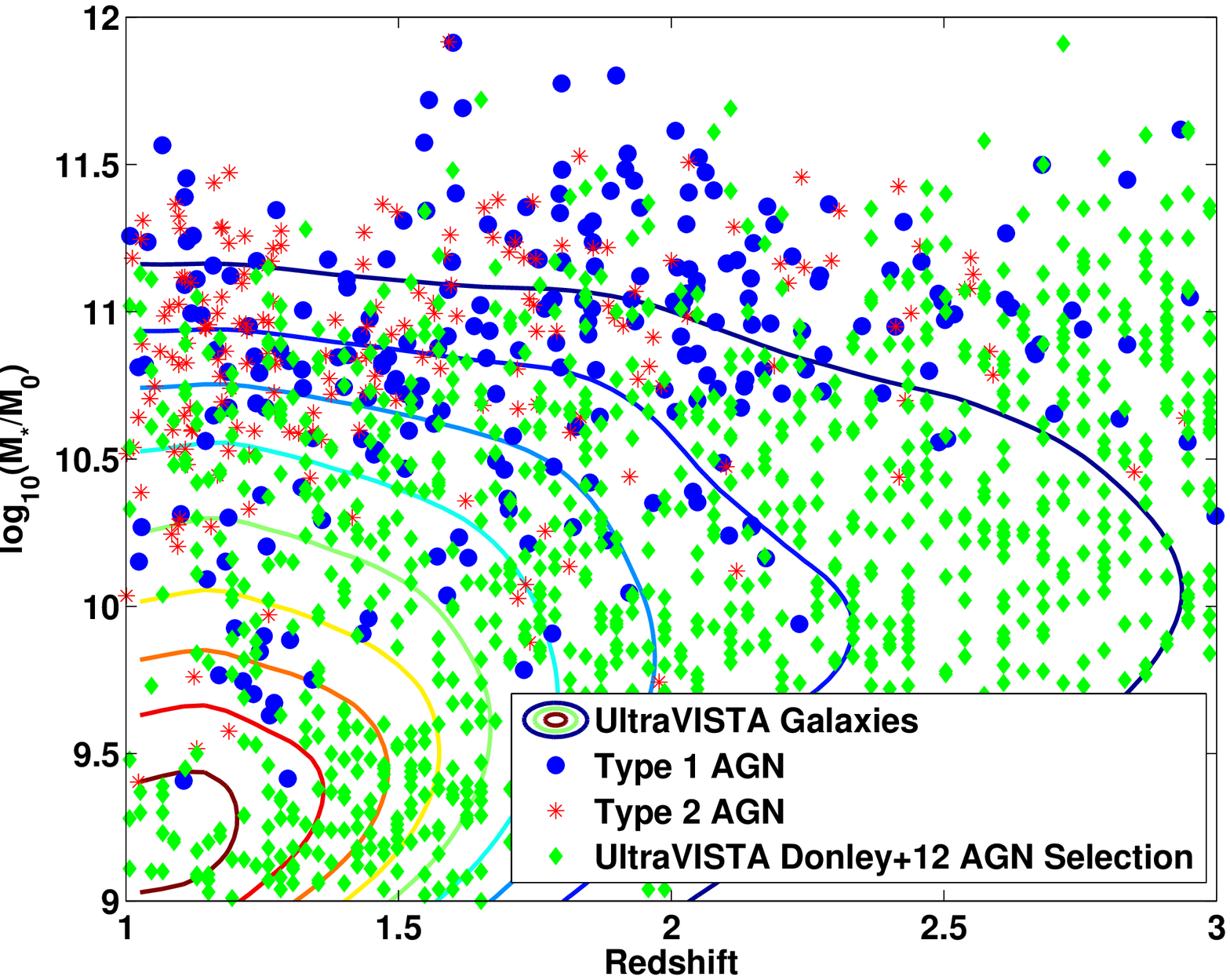} 
\caption{Redshift versus stellar mass of $\sim$114,000 UltraVISTA galaxies at $z>1$ from \citet{Muzzin:13} compared to the Type 1 and Type 2 AGN \citep{Bongiorno:12}. In the case of the UltraVISTA galaxies, the coloured contours represent the density of points on the plot with red contours corresponding to the highest density. UltraVISTA galaxies that satisfy the \citet{Donley:12} IRAC AGN selection criteria are also shown and occupy the full redshift-mass plane. On the other hand, the X-ray AGN host galaxies correspond to the most massive end of the UltraVISTA galaxy stellar mass distribution at $z>1$.}
\label{fig:Mstar_z}
\end{center}
\end{figure} 

Finally, we would like to understand if there is evidence for cool dust emission being preferentially associated with AGN activity or whether all galaxies of similar mass and at similar redshift, show similar dust emission properties. Stellar masses are available for both the Type 1 and Type 2 AGN samples from spectral energy distribution fitting \citep{Bongiorno:12}. The stellar masses as a function of redshift are shown in Fig. \ref{fig:Mstar_z}. As described in Section \ref{sec:gal}, we also have a control sample of galaxies from the UltraVISTA survey with photometric redshifts and stellar masses from \citet{Muzzin:13}. The stellar mass versus redshift distribution for $\sim$114,000 galaxies selected from this sample is also shown in Fig. \ref{fig:Mstar_z}. In both cases, the stellar masses have been computed using the \citet{BC:03} stellar population synthesis models with a \citet{Chabrier:03} initial mass function and assuming exponentially declining star formation histories so the two stellar mass estimates should be comparable. Matching the UltraVISTA galaxy catalogue to the \citet{Bongiorno:12} catalogue, we find that for the Type 2 AGN sample, where the emission is dominated by the galaxy SED and the contribution from the AGN template is negligible, the stellar masses in the two catalogues agree to within 0.1 dex with a median difference of 0.08 dex in the masses. In the analysis that follows, we separate sources into stellar mass bins of width 0.5 dex so for our purposes, differences in stellar masses between \citet{Bongiorno:12} and \citet{Muzzin:13} are negligible. 

From Fig. \ref{fig:Mstar_z}, it is apparent that the AGN host galaxy masses are at the top end of the distribution of masses in the UltraVISTA galaxy population. We therefore begin by selecting both samples of galaxies and AGN to be in the stellar mass range 10.0$<$log$_{10}(M_*/M_\odot)<$11.5 and require the AGN stellar masses to have \textit{flag 0} in the catalogue of \citet{Bongiorno:12}, which indicates that these stellar mass estimates are robust. In the case of the Type 1 AGN, these \textit{flag} 0 sources correspond to those where the galaxy contribution at 1$\mu$m is $>$10\% so the mass of the galaxy can effectively be constrained \citep{Merloni:10, Bongiorno:12}. We also restrict our analysis to only those AGN with errors of $<$0.1 dex on their stellar masses. In order to remove any contamination from AGN in the non-AGN galaxy sample, we cross-match the UltraVISTA galaxies to the \textit{XMM-Newton} point-source catalogue of \citet{Cappelluti:09} and remove any galaxies associated with an X-ray point source within a matching radius of 8$^{\prime \prime}$. Even within the mass range selected for our AGN versus non-AGN galaxy comparison study, the mass distribution for the AGN is more strongly skewed towards higher masses than the corresponding galaxy distribution. As it is well known that more massive galaxies have higher star formation rates, this can introduce biases into our analysis. Starting with the non-AGN, we therefore match both the mass and redshift distributions of the galaxies to that of the AGN by randomly sampling from the galaxies in $\delta$log$(M_*)$ bins of width 0.1 and $\delta z$ bins of width 0.1. The number of AGN to non-AGN in each mass bin is fixed to be the same and we therefore ensure that the median stellar mass of the non-AGN galaxy sample is the same as that of the AGN sample - log$_{10}(M_*/M_\odot)=$10.9 in both cases. The median redshifts are 1.66 for the AGN and 1.68 for the non-AGN. A potential concern is that highly obscured AGN that are not detected in the X-ray, still remain in the non-AGN galaxy sample. However, visual inspection of the best-fit galaxy SEDs from \citet{Muzzin:13} confirms that the majority of these galaxies do not have power-law type AGN dominated SEDs in the \textit{Spitzer} IRAC bands. From our sample of 782 mass-matched non-AGN galaxies, we find that $<$5\% satisfy the IRAC AGN selection criteria presented by \citet{Donley:12}. As illustrated in Fig. \ref{fig:Mstar_z}, UltraVISTA galaxies that satisfy the \citet{Donley:12} AGN selection criteria, occupy the entire redshift-mass plane and do not dominate at high masses. In this high-mass regime, the UltraVISTA galaxies are therefore unlikely to suffer from heavy contamination from highly obscured AGN. 

\begin{table*}
\begin{center}
\caption{Stacked 850$\mu$m fluxes for the Type 1 and Type 2 AGN and a sample of non-AGN galaxies selected from UltraVISTA in three stellar mass bins. Detections above 3$\sigma$ are shown in bold and the number of sources in each bin is given in brackets.}
\label{tab:mass_table}
\begin{tabular}{lccc}
\hline
& Type 1 AGN & Type 2 AGN & Non-AGN UltraVISTA Galaxies \\
\hline
10.0$<$log$_{10}$(M$_*$/M$_\odot$)$<$10.5 & 0.6$\pm$0.3 mJy (N=35) & 0.3$\pm$0.4 mJy (N=22) & \textbf{0.6$\pm$0.2 mJy} (N=98) \\
10.5$<$log$_{10}$(M$_*$/M$_\odot$)$<$11.0 & \textbf{1.0$\pm$0.3 mJy} (N=97) & 0.4$\pm$0.2 mJy (N=91) & \textbf{0.8$\pm$0.1 mJy} (N=329) \\
11.0$<$log$_{10}$(M$_*$/M$_\odot$)$<$11.5 & \textbf{1.0$\pm$0.2 mJy} (N=91) & \textbf{0.9$\pm$0.2 mJy} (N=69) & \textbf{1.4$\pm$0.1 mJy} (N=355) \\
\hline
\end{tabular}
\end{center}
\end{table*} 

\begin{figure}
\begin{center}
\includegraphics[scale=0.45,angle=0]{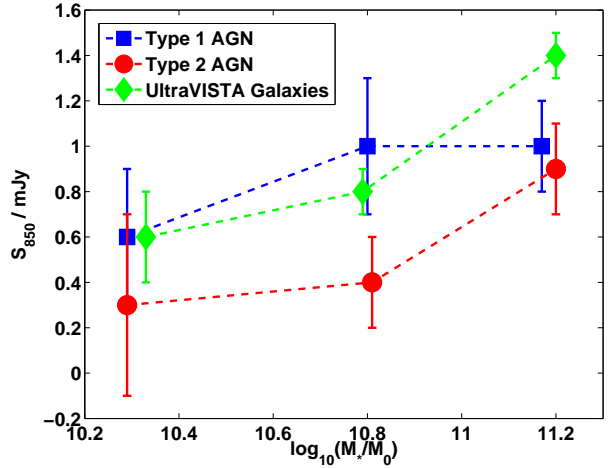} 
\caption{Average 850$\mu$m flux versus stellar mass for both Type 1 and Type 2 AGN and compared to a sample of non-AGN galaxies selected from the UltraVISTA survey.}
\label{fig:M_S850}
\end{center}
\end{figure}

With the mass and redshift distributions of the inactive galaxies and AGN now matched, we proceed to calculating their average 850$\mu$m fluxes. The non-AGN galaxy sample has a higher 850$\mu$m flux compared to the AGN - 1.01$\pm$0.09 mJy versus 0.72$\pm$0.12 mJy. The median and mean 850$\mu$m fluxes for both the AGN and non-AGN are also consistent with each other confirming that the stacked properties are not being dominated by a few outliers. To investigate this excess in more detail, we now split our galaxy and AGN sample into three stellar mass bins and consider the Type 1 and Type 2 AGN populations independently. The average submillimeter fluxes in the three stellar mass bins are presented in Table \ref{tab:mass_table} and illustrated in Fig. \ref{fig:M_S850}. While the Type 2 AGN 850$\mu$m fluxes increase as a function of stellar mass, the Type 1 AGN fluxes remain fairly constant. The redshift and luminosity distributions of the AGN in all three stellar mass bins can be seen in Appendix C. In the case of both the Type 1 and Type 2 AGN, the highest stellar mass bins peak at higher X-ray luminosities consistent with more massive galaxies having more powerful accreting black holes. Combined with the results in Fig. \ref{fig:L_S850}, this suggests that the trends in submillimeter emission with mass are equivalent to the trends with X-ray luminosity, with Type 1 AGN showing little evolution and high luminosity/high-mass Type 2 AGN being more submillimeter bright. Interestingly however, the galaxy stacked fluxes are in general comparable to or higher than those seen in the AGN population. For Type 2 AGN this appears to be the case over pretty much the entire mass range probed. Non-AGN hosts therefore appear to be more highly star-forming than the X-ray AGN.

These results are somewhat at odds with similar analyses conducted using \textit{Herschel} data where evidence for excess star formation is seen in AGN host galaxies relative to non-AGN host galaxies of similar mass (e.g. \citealt{Santini:12}). The excess disappears when considering star-forming galaxies only leading the authors to conclude that AGN are preferentially hosted in star-forming galaxies with the same gas supply likely fuelling both processes. However, we note that in contrast to these investigations, \citet{Bongiorno:12} find that AGN are preferentially hosted in redder, more quiescent galaxies. \citet{Bongiorno:12} have already pointed out that their SED-fitting method, by correcting for dust extinction to both the AGN and host galaxy components and disentangling the UV-luminous AGN emission from that from star formation, leads to lower star formation rate estimates for AGN hosts relative to \citet{Santini:12}. The AGN template used by \citet{Santini:12} is a Seyfert template. While this is clearly appropriate for low-luminosity AGN, for the high-luminosity XMM-COSMOS AGN investigated here, it may not account for all the UV emission from the AGN, therefore leading to an over-estimate of the star formation rate. 

The higher average 850$\mu$m flux of the non-AGN galaxy sample is primarily driven by the most massive galaxies with log$_{10}(M_*/M_\odot)>$11.0 as can be seen in Fig. \ref{fig:M_S850}. These most massive galaxies have extremely red SEDs with very little flux at rest-frame UV wavelengths. Many of these ultra-red galaxies would not be selected in flux-limited catalogues generated in bluer bands such as the $r$ or $i$-band and could be one reason for the different conclusions from our analysis and that of e.g. \citet{Santini:12}. Interpreting the higher 850$\mu$m fluxes of the \citet{Muzzin:13} massive galaxies as signifying that they have higher star formation rates relative to AGN, requires us to assume that the average dust properties are similar in both populations. Inspection of the best-fit non-AGN galaxy SEDs at log$_{10}(M_*/M_\odot)>$11.0 from \citet{Muzzin:13} however confirms that these galaxies are almost always fit by highly star-forming galaxy templates with significant extinction (median A$_{\rm{V}}=0.9$), lending credence to the idea that a significant fraction of the massive galaxies found by \citet{Muzzin:13} at these redshifts may indeed correspond to starburst galaxies with very high star formation rates and therefore submillimeter fluxes. 

We have shown that AGN host galaxies appear to have lower 850$\mu$m fluxes compared to a mass-matched sample of non-AGN galaxies. It is important to highlight that photometric redshifts as well as stellar masses for both the AGN and non-AGN samples are clearly model dependent and affected by the choice of templates used in the SED fitting. Stellar mass estimates are also likely to be less reliable for Type 1 AGN where the host galaxy is less visible. However, as can be seen in Fig. \ref{fig:M_S850}, it is the difference between the submillimeter fluxes of Type 2 AGN hosts and non-AGN that is more significant. Different model choices in SED fitting could perhaps partially explain the different conclusions reached regarding the nature of AGN host galaxies in the literature. However, another key issue is that many previous studies have selected non-AGN galaxy samples from optical survey data, which are less sensitive to star-forming galaxies with significant dust extinctions. High-mass, high-redshift, red populations of non-AGN galaxies that are matched in mass to the most luminous AGN are now selectable using new deep infrared surveys such as UltraVISTA that cover reasonably wide fields. Due to their rarity, these massive red galaxies are only present in wide-field extragalactic surveys such as COSMOS and due to the red colours of this population, they are often not present in optical flux-limited surveys. These galaxies are also the most difficult to obtain spectra of. Understanding their true nature - i.e. whether they correspond to massive dusty starbursts  - is now vital in order to be able to relate them to the AGN host galaxies which have equivalently high stellar masses. 

\section{CONCLUSIONS}

We have studied the 850$\mu$m properties of X-ray selected AGN in the COSMOS field using new data from the SCUBA-2 Cosmology Legacy Survey. The 850$\mu$m data covering $\sim$2 deg$^2$ presents an opportunity to study the cold dust emission from a statistically significant number of luminous AGN as a function of properties such as luminosity, stellar mass and obscuration. We begin by searching for AGN that are detected at $>$3.5$\sigma$ in our 850$\mu$m data in a ``high sensitivity" ($\sigma_{850}<$2 mJy) region covering 0.89 deg$^2$. There are 360 submillimeter sources over this area and we identify that 19 of these are X-ray luminous AGN. These submillimeter galaxies have S$_{850}$=4-10 mJy. The 19 submillimeter detected X-ray AGN comprise 6 spectroscopically confirmed broad-line AGN, 1 spectroscopically confirmed narrow-line AGN, 3 photometric redshift AGN that are best fit by Type 1 AGN templates and 9 photometric redshift AGN that are best fit by Type 2 or galaxy templates. In terms of X-ray luminosity, these AGN fill the luminosity gap between the very X-ray luminous quasars that have been individually followed up at submillimeter wavelengths (e.g. \citealt{Stevens:05}) and sub-mm detected X-ray AGN from much smaller area, deeper X-ray surveys (e.g. \citealt{Alexander:05, Wang:13}). 

We construct inverse-variance weighted stacked 850$\mu$m images of different sub-samples of $z>1$ AGN in the X-ray luminosity range 43$\lesssim$log$_{10}$(L$_{\rm{X}}$)$\lesssim$45, utilising a hard X-ray (2-10 keV) flux-limited population of 699 radio quiet AGN (428 Type 1 and 271 Type 2). The average 850$\mu$m fluxes are measured from a 100 such bootstrap images and for the entire AGN population, S$_{850}$=0.71$\pm$0.08 mJy. The 850$\mu$m flux directly traces the dust luminosity at $z>1$ due to the effects of the negative k-correction so the 850$\mu$m fluxes can be averaged over relatively wide redshift bins at these high redshifts. Under the assumption of an isothermal SED, the 850$\mu$m flux can therefore be used as a proxy for star formation rate. Using independent estimates of the star formation rate in the COSMOS AGN sample estimated via SED fitting to the optical through infrared photometry, we demonstrate that AGN with higher star formation rates do, on average, have higher 850$\mu$m fluxes. In particular, AGN with SFR$>$10M$_\odot$yr$^{-1}$ have 850$\mu$m fluxes that are $\sim$3.3$\times$ higher than those in AGN with SFR$<$10M$_\odot$yr$^{-1}$. We then study the dependence of this 850$\mu$m flux on various AGN properties. In particular, we reach the following conclusions:

\begin{itemize}

\item{The 850$\mu$m fluxes of both Type 1 and Type 2 AGN show little evolution with redshift and the Type 1 and Type 2 AGN have average submillimeter fluxes that are consistent with each other over the entire redshift range probed in this study.}

\item{We study the dependence of 850$\mu$m emission and therefore star formation on AGN X-ray luminosity. We find that the 850$\mu$m fluxes are relatively constant for $z>1$ Type 1 AGN at $\rm{log}_{10}(L_{\rm{2-10}})>43.5$. However, in the case of the $z>1$ Type 2 AGN, we find that the 850$\mu$m flux increases with X-ray luminosity. This may suggest a transition from secular to merger-driven evolution in the Type 2 AGN population at the highest X-ray luminosities ($\rm{log}_{10}(L_{\rm{2-10}})>44.4$) probed in this work, which could be responsible for the tighter coupling seen between cold dust emission (tracing star formation) and X-ray luminosity (tracing black hole accretion) at these luminosities. Our results also indicate that high-luminosity Type 1 and Type 2 AGN have systematic differences in their observed 850$\mu$m flux densities, which could reflect 
systematic differences in their host galaxy properties with the Type 2 AGN occupying more highly star-forming galaxies. Alternatively, the higher 850$\mu$m flux densities in high luminosity Type 2 AGN could simply arise as a result of these AGN having higher dust mass relative to Type 1 AGN of comparable luminosity.} 

\item{Having explored trends with redshift and luminosity, we study the dependence of submillimeter flux on obscuration, relying on various different obscuration measures. There is no significant dependence of the 850$\mu$m flux on X-ray hardness ratio or the (r-4.5$\mu$m) colour when averaging over all X-ray luminosities. When selecting the most X-ray luminous AGN ($\rm{log}_{10}(L_{\rm{2-10}})>44.4$), we find that the red AGN have submillimeter fluxes that are $\sim$1.7$\times$ higher than the blue AGN, although given the sample size, the result is only significant at the $\sim$1.3$\sigma$ level.} 

\item{Following \citet{Brusa:10}, we select a population of red $(r-K)$ AGN with high X-ray to optical flux ratios. Once again these AGN have very similar submillimeter properties to the rest of the AGN population when averaging over all X-ray luminosities. However in the highest X-ray luminosity bin ($\rm{log}_{10}(L_{\rm{2-10}})>44.4$), these red AGN have submillimeter fluxes that are more than a factor of 2 higher than the average. Despite the small samples in this single field, the results are still significant at the $\sim$2$\sigma$ level and consistent with a more marked dependence of submillimeter emission on obscuration for the most X-ray luminous AGN.}

\item{Finally, the $z>1$ AGN are compared to a $K$-band selected population of galaxies showing no evidence for AGN activity and that are carefully matched in stellar mass and redshift to the AGN sample. On account of the $K$-band selection, this non-AGN galaxy sample includes populations of extremely red galaxies at the high-mass end, many of which have very little flux at rest-frame UV wavelengths and would not be selected in optical flux-limited surveys. Type 2 AGN show an increase in submillimeter flux with stellar mass which is equivalent to the increase seen as a function of X-ray luminosity. Over almost the entire mass range probed however, non-AGN have higher 850$\mu$m fluxes compared to the Type 2 AGN and in the highest stellar mass bin - $11.0<$log$_{10}<11.5$ - the non-AGN also have higher submillimeter fluxes compared to the Type 1 AGN. The best-fit SED templates for the massive non-AGN galaxies indicate that they are dusty galaxies with high star formation rates and a median A$_{\rm{V}}$ of 0.9. Such massive, red non-AGN galaxies have eluded discovery in previous surveys that have either covered smaller areas or been flux-limited in the bluer optical bands. Due to their red colours, many of these galaxies also remain spectroscopically unconfirmed. Understanding their true nature is important to be able to relate them to their AGN counterparts of equivalent mass.}

\end{itemize}

We have demonstrated the unique sensitivity and resolution of the new SCUBA-2 850$\mu$m data provided by the SCUBA-2 Cosmology Legacy Survey, which has allowed us to probe rest-frame wavelengths corresponding to the Rayleigh Jeans tail in the SEDs of luminous X-ray AGN in the high redshift Universe. At these wavelengths the 850$\mu$m emission provides a unique probe of the cold dust luminosity and dust mass in AGN samples. Although the sample size used in this work has limited the significance of some of our results, nevertheless we find some evidence for a stronger coupling between AGN luminosity, dust luminosity and obscuration at high X-ray luminosities, where we may be seeing a transition from a secular to a merger-driven regime in galaxy evolution. Larger samples of AGN selected at multiple wavelengths are clearly required to establish more significant trends in the star formation properties of AGN host galaxies across a wide range in luminosity, redshift and mass. This will become possible when similar analyses are extended to cover all the fields imaged by the S2CLS.  

 \section*{ACKNOWLEDGEMENTS}
 
We thank the referee for a constructive review that helped improve the current analysis. We thank James Aird, Paul Hewett and Adam Muzzin for useful discussions, Angela Bongiorno for kindly providing the errors on the AGN stellar mass estimates and Ian Smail for helpful comments on the manuscript. MB, RGM and KEKC acknowledge support from STFC. JEG thanks the Royal Society for a University Research Fellowship. NB and JSD acknowledge the support of the European Research Council via the award of an Advanced Grant, and the contribution of the EC FP7 SPACE project ASTRODEEP (Ref.No: 312725). DMA and CMH acknowledge support from STFC grant ST/I001573/1.

This work was carried out based on observations obtained using SCUBA-2 on the James Clerk Maxwell Telescope. The James Clerk Maxwell Telescope has historically been operated by the Joint Astronomy Centre on behalf of the Science and Technology Facilities Council of the United Kingdom, the National Research Council of Canada and the Netherlands Organisation for Scientific Research. Additional funds for the construction of SCUBA-2 were provided by the Canada Foundation for Innovation. 

\bibliography{}

\begin{thebibliography}{}

\bibitem[\protect\citeauthoryear{{Aird}, {Coil}, {Georgakakis}, {Nandra},
  {Barro} \& {P{\'e}rez-Gonz{\'a}lez}}{{Aird} et~al.}{2015}]{Aird:15}
{Aird} J.,  {Coil} A.~L.,  {Georgakakis} A.,  {Nandra} K.,  {Barro} G.,
  {P{\'e}rez-Gonz{\'a}lez} P.~G.,  2015, \mnras, 451, 1892

\bibitem[\protect\citeauthoryear{{Aird}, {Coil}, {Moustakas}, {Diamond-Stanic},
  {Blanton}, {Cool}, {Eisenstein}, {Wong} \& {Zhu}}{{Aird}
  et~al.}{2013}]{Aird:13}
{Aird} J.,  {Coil} A.~L.,  {Moustakas} J.,  {Diamond-Stanic} A.~M.,  {Blanton}
  M.~R.,  {Cool} R.~J.,  {Eisenstein} D.~J.,  {Wong} K.~C.,    {Zhu} G.,  2013,
  \apj, 775, 41

\bibitem[\protect\citeauthoryear{{Alexander}, {Bauer}, {Chapman}, {Smail},
  {Blain}, {Brandt} \& {Ivison}}{{Alexander} et~al.}{2005}]{Alexander:05}
{Alexander} D.~M.,  {Bauer} F.~E.,  {Chapman} S.~C.,  {Smail} I.,  {Blain}
  A.~W.,  {Brandt} W.~N.,    {Ivison} R.~J.,  2005, \apj, 632, 736

\bibitem[\protect\citeauthoryear{{Alexander} \& {Hickox}}{{Alexander} \&
  {Hickox}}{2012}]{Alexander:12}
{Alexander} D.~M.,  {Hickox} R.~C.,  2012, 56, 93

\bibitem[\protect\citeauthoryear{{Antonucci}}{{Antonucci}}{1993}]{Antonucci:93}
{Antonucci} R.,  1993, \araa, 31, 473

\bibitem[\protect\citeauthoryear{{Azadi}, {Aird}, {Coil}, {Moustakas},
  {Mendez}, {Blanton}, {Cool}, {Eisenstein}, {Wong} \& {Zhu}}{{Azadi}
  et~al.}{2014}]{Azadi:14}
{Azadi} M.,  {Aird} J.,  {Coil} A.,  {Moustakas} J.,  {Mendez} A.,  {Blanton}
  M.,  {Cool} R.,  {Eisenstein} D.,  {Wong} K.,    {Zhu} G.,  2014, ArXiv
  e-prints

\bibitem[\protect\citeauthoryear{{Banerji}, {Alaghband-Zadeh}, {Hewett} \&
  {McMahon}}{{Banerji} et~al.}{2015}]{Banerji:15}
{Banerji} M.,  {Alaghband-Zadeh} S.,  {Hewett} P.~C.,    {McMahon} R.~G.,
  2015, \mnras, 447, 3368

\bibitem[\protect\citeauthoryear{{Banerji}, {Fabian} \& {McMahon}}{{Banerji}
  et~al.}{2014}]{Banerji:14}
{Banerji} M.,  {Fabian} A.~C.,    {McMahon} R.~G.,  2014, \mnras, 439, L51

\bibitem[\protect\citeauthoryear{{Banerji}, {McMahon}, {Hewett},
  {Alaghband-Zadeh}, {Gonzalez-Solares}, {Venemans} \& {Hawthorn}}{{Banerji}
  et~al.}{2012}]{Banerji:12}
{Banerji} M.,  {McMahon} R.~G.,  {Hewett} P.~C.,  {Alaghband-Zadeh} S.,
  {Gonzalez-Solares} E.,  {Venemans} B.~P.,    {Hawthorn} M.~J.,  2012, \mnras,
  427, 2275

\bibitem[\protect\citeauthoryear{{Blain}, {Smail}, {Ivison}, {Kneib} \&
  {Frayer}}{{Blain} et~al.}{2002}]{Blain:02}
{Blain} A.~W.,  {Smail} I.,  {Ivison} R.~J.,  {Kneib} J.-P.,    {Frayer} D.~T.,
   2002, 369, 111

\bibitem[\protect\citeauthoryear{{Bonfield} et~al.,}{{Bonfield}
  et~al.}{2011}]{Bonfield:10}
{Bonfield} D.~G.,  et~al., 2011, \mnras, 416, 13

\bibitem[\protect\citeauthoryear{{Bongiorno} et~al.,}{{Bongiorno}
  et~al.}{2012}]{Bongiorno:12}
{Bongiorno} A.,  et~al., 2012, \mnras, 427, 3103

\bibitem[\protect\citeauthoryear{{Bongiorno} et~al.,}{{Bongiorno}
  et~al.}{2014}]{Bongiorno:14}
{Bongiorno} A.,  et~al., 2014, \mnras, 443, 2077

\bibitem[\protect\citeauthoryear{{Bourne} et~al.,}{{Bourne}
  et~al.}{2013}]{Bourne:13}
{Bourne} N.,  et~al., 2013, \mnras, 436, 479

\bibitem[\protect\citeauthoryear{{Brightman}, {Nandra}, {Salvato}, {Hsu},
  {Aird} \& {Rangel}}{{Brightman} et~al.}{2014}]{Brightman:14}
{Brightman} M.,  {Nandra} K.,  {Salvato} M.,  {Hsu} L.-T.,  {Aird} J.,
  {Rangel} C.,  2014, \mnras, 443, 1999

\bibitem[\protect\citeauthoryear{{Brusa} et~al.,}{{Brusa}
  et~al.}{2010}]{Brusa:10}
{Brusa} M.,  et~al., 2010, \apj, 716, 348

\bibitem[\protect\citeauthoryear{{Brusa} et~al.,}{{Brusa}
  et~al.}{2015}]{Brusa:14}
{Brusa} M.,  et~al., 2015, \mnras, 446, 2394

\bibitem[\protect\citeauthoryear{{Bruzual} \& {Charlot}}{{Bruzual} \&
  {Charlot}}{2003}]{BC:03}
{Bruzual} G.,  {Charlot} S.,  2003, \mnras, 344, 1000

\bibitem[\protect\citeauthoryear{{Cappelluti} et~al.,}{{Cappelluti}
  et~al.}{2009}]{Cappelluti:09}
{Cappelluti} N.,  et~al., 2009, \aap, 497, 635

\bibitem[\protect\citeauthoryear{{Casey}}{{Casey}}{2012}]{Casey:12}
{Casey} C.~M.,  2012, \mnras, 425, 3094

\bibitem[\protect\citeauthoryear{{Casey}, {Chen}, {Cowie}, {Barger}, {Capak},
  {Ilbert}, {Koss}, {Lee}, {Le Floc'h}, {Sanders} \& {Williams}}{{Casey}
  et~al.}{2013}]{Casey:13}
{Casey} C.~M.,  {Chen} C.-C.,  {Cowie} L.~L.,  {Barger} A.~J.,  {Capak} P.,
  {Ilbert} O.,  {Koss} M.,  {Lee} N.,  {Le Floc'h} E.,  {Sanders} D.~B.,
  {Williams} J.~P.,  2013, \mnras, 436, 1919

\bibitem[\protect\citeauthoryear{{Chabrier}}{{Chabrier}}{2003}]{Chabrier:03}
{Chabrier} G.,  2003, \pasp, 115, 763

\bibitem[\protect\citeauthoryear{{Chapin}, {Berry}, {Gibb}, {Jenness}, {Scott},
  {Tilanus}, {Economou} \& {Holland}}{{Chapin} et~al.}{2013}]{Chapin:13}
{Chapin} E.~L.,  {Berry} D.~S.,  {Gibb} A.~G.,  {Jenness} T.,  {Scott} D.,
  {Tilanus} R.~P.~J.,  {Economou} F.,    {Holland} W.~S.,  2013, \mnras, 430,
  2545

\bibitem[\protect\citeauthoryear{{Chen} et~al.,}{{Chen}
  et~al.}{2015}]{Chen:15}
{Chen} C.-T.~J.,  et~al., 2015, ArXiv e-prints

\bibitem[\protect\citeauthoryear{{Cresci} et~al.,}{{Cresci}
  et~al.}{2015}]{Cresci:14}
{Cresci} G.,  et~al., 2015, \apj, 799, 82

\bibitem[\protect\citeauthoryear{{Croton} et~al.,}{{Croton}
  et~al.}{2006}]{Croton:06}
{Croton} D.~J.,  et~al., 2006, \mnras, 365, 11

\bibitem[\protect\citeauthoryear{{Di Matteo}, {Springel} \& {Hernquist}}{{Di
  Matteo} et~al.}{2005}]{diMatteo:05}
{Di Matteo} T.,  {Springel} V.,    {Hernquist} L.,  2005, \nat, 433, 604

\bibitem[\protect\citeauthoryear{{Donley} et~al.,}{{Donley}
  et~al.}{2012}]{Donley:12}
{Donley} J.~L.,  et~al., 2012, \apj, 748, 142

\bibitem[\protect\citeauthoryear{{Feruglio}, {Bongiorno}, {Fiore}, {Krips},
  {Brusa}, {Daddi}, {Gavignaud}, {Maiolino}, {Piconcelli}, {Sargent}, {Vignali}
  \& {Zappacosta}}{{Feruglio} et~al.}{2014}]{Feruglio:14}
{Feruglio} C.,  {Bongiorno} A.,  {Fiore} F.,  {Krips} M.,  {Brusa} M.,  {Daddi}
  E.,  {Gavignaud} I.,  {Maiolino} R.,  {Piconcelli} E.,  {Sargent} M.,
  {Vignali} C.,    {Zappacosta} L.,  2014, \aap, 565, A91

\bibitem[\protect\citeauthoryear{{Gabor} \& {Bournaud}}{{Gabor} \&
  {Bournaud}}{2013}]{Gabor:13}
{Gabor} J.~M.,  {Bournaud} F.,  2013, \mnras, 434, 606

\bibitem[\protect\citeauthoryear{{Geach} et~al.,}{{Geach}
  et~al.}{2013}]{Geach:13}
{Geach} J.~E.,  et~al., 2013, \mnras, 432, 53

\bibitem[\protect\citeauthoryear{{Hao} et~al.,}{{Hao}  et~al.}{2014}]{Hao:14}
{Hao} H.,  et~al., 2014, \mnras, 438, 1288

\bibitem[\protect\citeauthoryear{{Harrison} et~al.,}{{Harrison}
  et~al.}{2012}]{Harrison:12}
{Harrison} C.~M.,  et~al., 2012, \apjl, 760, L15

\bibitem[\protect\citeauthoryear{{Hickox}, {Jones}, {Forman}, {Murray},
  {Brodwin}, {Brown}, {Eisenhardt}, {Stern}, {Kochanek}, {Eisenstein}, {Cool},
  {Jannuzi}, {Dey}, {Brand}, {Gorjian} \& {Caldwell}}{{Hickox}
  et~al.}{2007}]{Hickox:07}
{Hickox} R.~C.,  {Jones} C.,  {Forman} W.~R.,  {Murray} S.~S.,  {Brodwin} M.,
  {Brown} M.~J.~I.,  {Eisenhardt} P.~R.,  {Stern} D.,  {Kochanek} C.~S.,
  {Eisenstein} D.,  {Cool} R.~J.,  {Jannuzi} B.~T.,  {Dey} A.,  {Brand} K.,
  {Gorjian} V.,    {Caldwell} N.,  2007, \apj, 671, 1365

\bibitem[\protect\citeauthoryear{{Hickox}, {Mullaney}, {Alexander}, {Chen},
  {Civano}, {Goulding} \& {Hainline}}{{Hickox} et~al.}{2014}]{Hickox:14}
{Hickox} R.~C.,  {Mullaney} J.~R.,  {Alexander} D.~M.,  {Chen} C.-T.~J.,
  {Civano} F.~M.,  {Goulding} A.~D.,    {Hainline} K.~N.,  2014, \apj, 782, 9

\bibitem[\protect\citeauthoryear{{Ivison} et~al.,}{{Ivison}
  et~al.}{2007}]{Ivison:07}
{Ivison} R.~J.,  et~al., 2007, \mnras, 380, 199

\bibitem[\protect\citeauthoryear{{Kormendy} \& {Ho}}{{Kormendy} \&
  {Ho}}{2013}]{Kormendy:13}
{Kormendy} J.,  {Ho} L.~C.,  2013, \araa, 51, 511

\bibitem[\protect\citeauthoryear{{Lacy}, {Ridgway}, {Sajina}, {Petric},
  {Gates}, {Urrutia} \& {Storrie-Lombardi}}{{Lacy} et~al.}{2015}]{Lacy:15}
{Lacy} M.,  {Ridgway} S.~E.,  {Sajina} A.,  {Petric} A.~O.,  {Gates} E.~L.,
  {Urrutia} T.,    {Storrie-Lombardi} L.~J.,  2015, \apj, 802, 102

\bibitem[\protect\citeauthoryear{{Lusso} et~al.,}{{Lusso}
  et~al.}{2010}]{Lusso:10}
{Lusso} E.,  et~al., 2010, \aap, 512, A34

\bibitem[\protect\citeauthoryear{{Lusso} et~al.,}{{Lusso}
  et~al.}{2011}]{Lusso:11}
{Lusso} E.,  et~al., 2011, \aap, 534, A110

\bibitem[\protect\citeauthoryear{{Lutz} et~al.,}{{Lutz}
  et~al.}{2010}]{Lutz:10}
{Lutz} D.,  et~al., 2010, \apj, 712, 1287

\bibitem[\protect\citeauthoryear{{Lutz}, {Sturm}, {Tacconi}, {Valiante},
  {Schweitzer}, {Netzer}, {Maiolino}, {Andreani}, {Shemmer} \&
  {Veilleux}}{{Lutz} et~al.}{2008}]{Lutz:08}
{Lutz} D.,  {Sturm} E.,  {Tacconi} L.~J.,  {Valiante} E.,  {Schweitzer} M.,
  {Netzer} H.,  {Maiolino} R.,  {Andreani} P.,  {Shemmer} O.,    {Veilleux} S.,
   2008, \apj, 684, 853

\bibitem[\protect\citeauthoryear{{Magorrian} et~al.,}{{Magorrian}
  et~al.}{1998}]{Magorrian:98}
{Magorrian} J.,  et~al., 1998, \aj, 115, 2285

\bibitem[\protect\citeauthoryear{{McCracken} et~al.,}{{McCracken}
  et~al.}{2012}]{McCracken:12}
{McCracken} H.~J.,  et~al., 2012, \aap, 544, A156

\bibitem[\protect\citeauthoryear{{Merloni} et~al.,}{{Merloni}
  et~al.}{2010}]{Merloni:10}
{Merloni} A.,  et~al., 2010, \apj, 708, 137

\bibitem[\protect\citeauthoryear{{Merloni} et~al.,}{{Merloni}
  et~al.}{2014}]{Merloni:14}
{Merloni} A.,  et~al., 2014, \mnras, 437, 3550

\bibitem[\protect\citeauthoryear{{Mullaney} et~al.,}{{Mullaney}
  et~al.}{2012}]{Mullaney:12}
{Mullaney} J.~R.,  et~al., 2012, \mnras, 419, 95

\bibitem[\protect\citeauthoryear{{Muzzin}, {Marchesini}, {Stefanon}, {Franx},
  {Milvang-Jensen}, {Dunlop}, {Fynbo}, {Brammer}, {Labb{\'e}} \& {van
  Dokkum}}{{Muzzin} et~al.}{2013}]{Muzzin:13}
{Muzzin} A.,  {Marchesini} D.,  {Stefanon} M.,  {Franx} M.,  {Milvang-Jensen}
  B.,  {Dunlop} J.~S.,  {Fynbo} J.~P.~U.,  {Brammer} G.,  {Labb{\'e}} I.,
  {van Dokkum} P.,  2013, \apjs, 206, 8

\bibitem[\protect\citeauthoryear{{Oliver} et~al.,}{{Oliver}
  et~al.}{2012}]{Oliver:12}
{Oliver} S.~J.,  et~al., 2012, \mnras, 424, 1614

\bibitem[\protect\citeauthoryear{{Page} et~al.,}{{Page}
  et~al.}{2012}]{Page:12}
{Page} M.~J.,  et~al., 2012, \nat, 485, 213

\bibitem[\protect\citeauthoryear{{Page}, {Stevens}, {Ivison} \&
  {Carrera}}{{Page} et~al.}{2004}]{Page:04}
{Page} M.~J.,  {Stevens} J.~A.,  {Ivison} R.~J.,    {Carrera} F.~J.,  2004,
  \apjl, 611, L85

\bibitem[\protect\citeauthoryear{{Page}, {Stevens}, {Mittaz} \&
  {Carrera}}{{Page} et~al.}{2001}]{Page:01}
{Page} M.~J.,  {Stevens} J.~A.,  {Mittaz} J.~P.~D.,    {Carrera} F.~J.,  2001,
  Science, 294, 2516

\bibitem[\protect\citeauthoryear{{Richards} et~al.,}{{Richards}
  et~al.}{2006}]{Richards:06}
{Richards} G.~T.,  et~al., 2006, \aj, 131, 2766

\bibitem[\protect\citeauthoryear{{Rosario} et~al.,}{{Rosario}
  et~al.}{2012}]{Rosario:12}
{Rosario} D.~J.,  et~al., 2012, \aap, 545, A45

\bibitem[\protect\citeauthoryear{{Rovilos} et~al.,}{{Rovilos}
  et~al.}{2012}]{Rovilos:12}
{Rovilos} E.,  et~al., 2012, \aap, 546, A58

\bibitem[\protect\citeauthoryear{{Salvato} et~al.,}{{Salvato}
  et~al.}{2009}]{Salvato:09}
{Salvato} M.,  et~al., 2009, \apj, 690, 1250

\bibitem[\protect\citeauthoryear{{Salvato} et~al.,}{{Salvato}
  et~al.}{2011}]{Salvato:11}
{Salvato} M.,  et~al., 2011, \apj, 742, 61

\bibitem[\protect\citeauthoryear{{Sanders} et~al.,}{{Sanders}
  et~al.}{2007}]{Sanders:07}
{Sanders} D.~B.,  et~al., 2007, \apjs, 172, 86

\bibitem[\protect\citeauthoryear{{Santini} et~al.,}{{Santini}
  et~al.}{2012}]{Santini:12}
{Santini} P.,  et~al., 2012, \aap, 540, A109

\bibitem[\protect\citeauthoryear{{Schinnerer}, {Sargent}, {Bondi}, {Smol{\v
  c}i{\'c}}, {Datta}, {Carilli}, {Bertoldi}, {Blain}, {Ciliegi}, {Koekemoer} \&
  {Scoville}}{{Schinnerer} et~al.}{2010}]{Schinnerer:10}
{Schinnerer} E.,  {Sargent} M.~T.,  {Bondi} M.,  {Smol{\v c}i{\'c}} V.,
  {Datta} A.,  {Carilli} C.~L.,  {Bertoldi} F.,  {Blain} A.,  {Ciliegi} P.,
  {Koekemoer} A.,    {Scoville} N.~Z.,  2010, \apjs, 188, 384

\bibitem[\protect\citeauthoryear{{Scoville} et~al.,}{{Scoville}
  et~al.}{2007}]{Scoville:07}
{Scoville} N.,  et~al., 2007, \apjs, 172, 1

\bibitem[\protect\citeauthoryear{{Shao} et~al.,}{{Shao}
  et~al.}{2010}]{Shao:10}
{Shao} L.,  et~al., 2010, \aap, 518, L26

\bibitem[\protect\citeauthoryear{{Sijacki}, {Vogelsberger}, {Genel},
  {Springel}, {Torrey}, {Snyder}, {Nelson} \& {Hernquist}}{{Sijacki}
  et~al.}{2014}]{Sijacki:14}
{Sijacki} D.,  {Vogelsberger} M.,  {Genel} S.,  {Springel} V.,  {Torrey} P.,
  {Snyder} G.,  {Nelson} D.,    {Hernquist} L.,  2014, arXiv:1408.6842

\bibitem[\protect\citeauthoryear{{Springel}, {Di Matteo} \&
  {Hernquist}}{{Springel} et~al.}{2005}]{Springel:05}
{Springel} V.,  {Di Matteo} T.,    {Hernquist} L.,  2005, \apjl, 620, L79

\bibitem[\protect\citeauthoryear{{Stanley}, {Harrison}, {Alexander},
  {Swinbank}, {Aird}, {Del Moro}, {Hickox} \& {Mullaney}}{{Stanley}
  et~al.}{2015}]{Stanley:15}
{Stanley} F.,  {Harrison} C.~M.,  {Alexander} D.~M.,  {Swinbank} A.~M.,  {Aird}
  J.~A.,  {Del Moro} A.,  {Hickox} R.~C.,    {Mullaney} J.~R.,  2015,
  arXiv:1502.07756

\bibitem[\protect\citeauthoryear{{Stevens}, {Page}, {Ivison}, {Carrera},
  {Mittaz}, {Smail} \& {McHardy}}{{Stevens} et~al.}{2005}]{Stevens:05}
{Stevens} J.~A.,  {Page} M.~J.,  {Ivison} R.~J.,  {Carrera} F.~J.,  {Mittaz}
  J.~P.~D.,  {Smail} I.,    {McHardy} I.~M.,  2005, \mnras, 360, 610

\bibitem[\protect\citeauthoryear{{Swinbank} et~al.,}{{Swinbank}
  et~al.}{2014}]{Swinbank:14}
{Swinbank} A.~M.,  et~al., 2014, \mnras, 438, 1267

\bibitem[\protect\citeauthoryear{{Symeonidis} et~al.,}{{Symeonidis}
  et~al.}{2014}]{Symeonidis:14}
{Symeonidis} M.,  et~al., 2014, \mnras, 443, 3728

\bibitem[\protect\citeauthoryear{{Urry} \& {Padovani}}{{Urry} \&
  {Padovani}}{1995}]{Urry:95}
{Urry} C.~M.,  {Padovani} P.,  1995, \pasp, 107, 803

\bibitem[\protect\citeauthoryear{{Viero} et~al.,}{{Viero}
  et~al.}{2013}]{Viero:13}
{Viero} M.~P.,  et~al., 2013, \apj, 779, 32

\bibitem[\protect\citeauthoryear{{Wang} et~al.,}{{Wang}
  et~al.}{2013}]{Wang:13}
{Wang} S.~X.,  et~al., 2013, \apj, 778, 179

\bibitem[\protect\citeauthoryear{{Wright} et~al.,}{{Wright}
  et~al.}{2010}]{Wright:10}
{Wright} E.~L.,  et~al., 2010, \aj, 140, 1868

\end{thebibliography}

\appendix

\section{X-ray Luminosities \& Absorption Corrections}

\begin{figure}
\begin{center}
\includegraphics[scale=0.45]{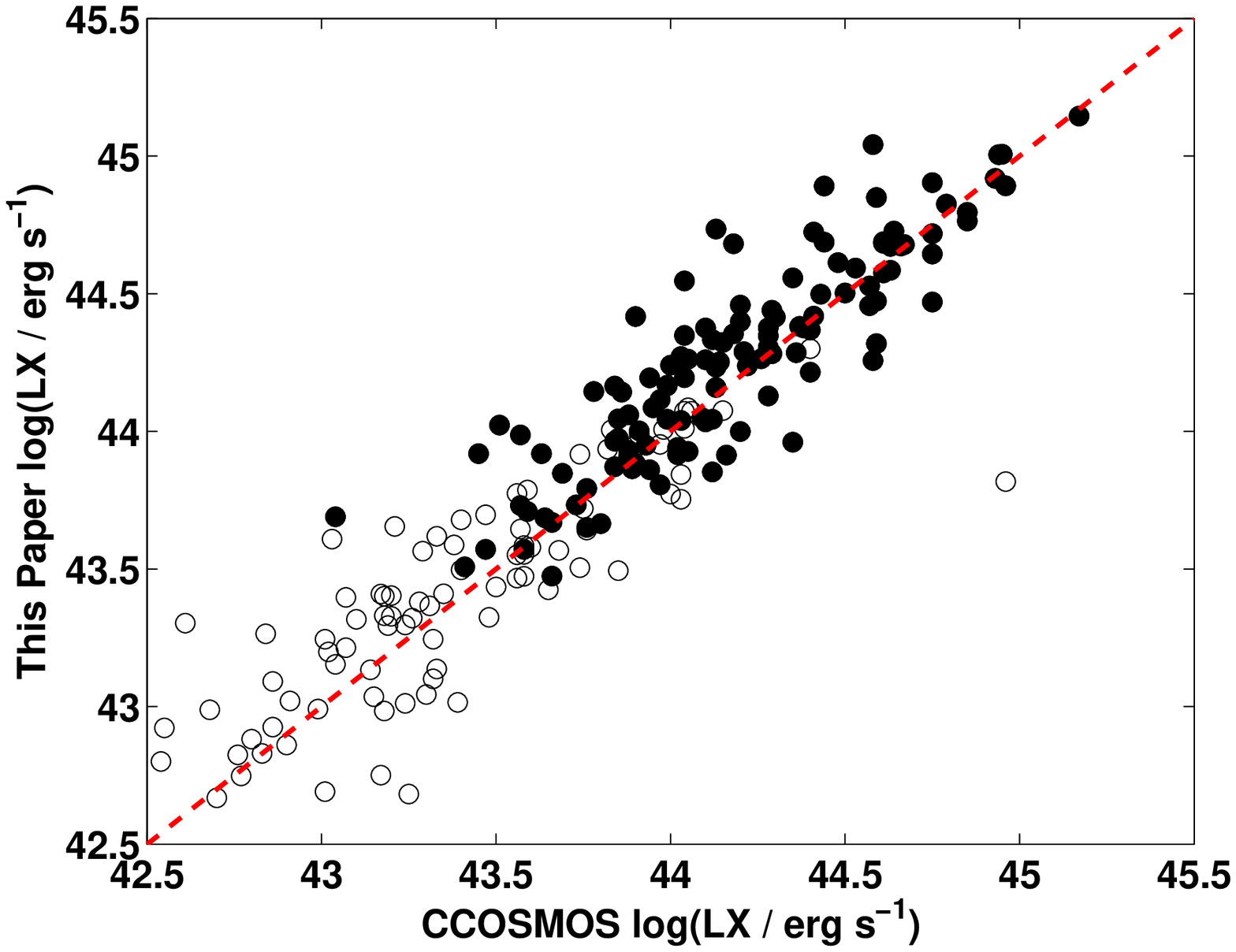}
\caption{Hard X-ray luminosities derived in this paper (Section \ref{sec:cat}) versus those derived using \textit{Chandra} data by \citet{Brightman:14} and including the effects of absorption. Filled circles correspond to AGN at $z>1$ which are used for all the stacking analysis considered in this work. No systematic bias is seen in our X-ray luminosities compared to the absorption corrected X-ray luminosities of \citet{Brightman:14}, confirming that absorption has a negligible impact on the derived luminosities for the majority of our sample.}
\label{fig:ccosmos}
\end{center}
\end{figure}

When deriving hard X-ray luminosities for our AGN sample in Section \ref{sec:cat}, we have neglected the effects of absorption. Significant absorption in a large fraction of our sample would systematically bias our X-ray luminosity estimates. As illustrated in figure 5 of \citet{Aird:15} however, hard X-ray luminosities are only likely to be significantly biased for absorbing columns of $N_H > 10^{23}$ cm$^{-2}$. A subset of the COSMOS AGN considered in this work also have deep X-ray observations obtained using the \textit{Chandra} observatory and where the data is of sufficient quality to enable detailed spectral fitting including different absorption components. \citet{Brightman:14} have conducted exactly such a study and derived absorption corrected hard X-ray luminosities as well as column densities for the X-ray AGN. We therefore match our sample of X-ray AGN to the \citet{Brightman:14} sample. There are 217 AGN in common between the two catalogues and where the redshifts in both catalogues also agree. In Fig. \ref{fig:ccosmos} we compare the hard X-ray luminosities derived in this paper (without absorption correction) to the hard X-ray luminosities from \citet{Brightman:14}. As can be seen, most of the AGN lie on the $x=y$ relation and the median difference between our X-ray luminosities and those of \citet{Brightman:14} is 0.06 dex. When considering the dependence of submillimeter flux on hard X-ray luminosity, we stack sub-samples in bins of 0.4-0.5 dex in X-ray luminosity. Hence we conclude that any biases in our X-ray luminosities due to a lack of absorption correction, are unlikely to significantly affect the trends observed in this paper.

\section{Properties of 850$\mu$\lowercase{m} Detected X-ray AGN}

\begin{figure*}
\begin{center}
\begin{tabular}{ccc} 
\large{XID13} & \large{XID139} & \large{XID160} \\
\includegraphics[scale=0.3]{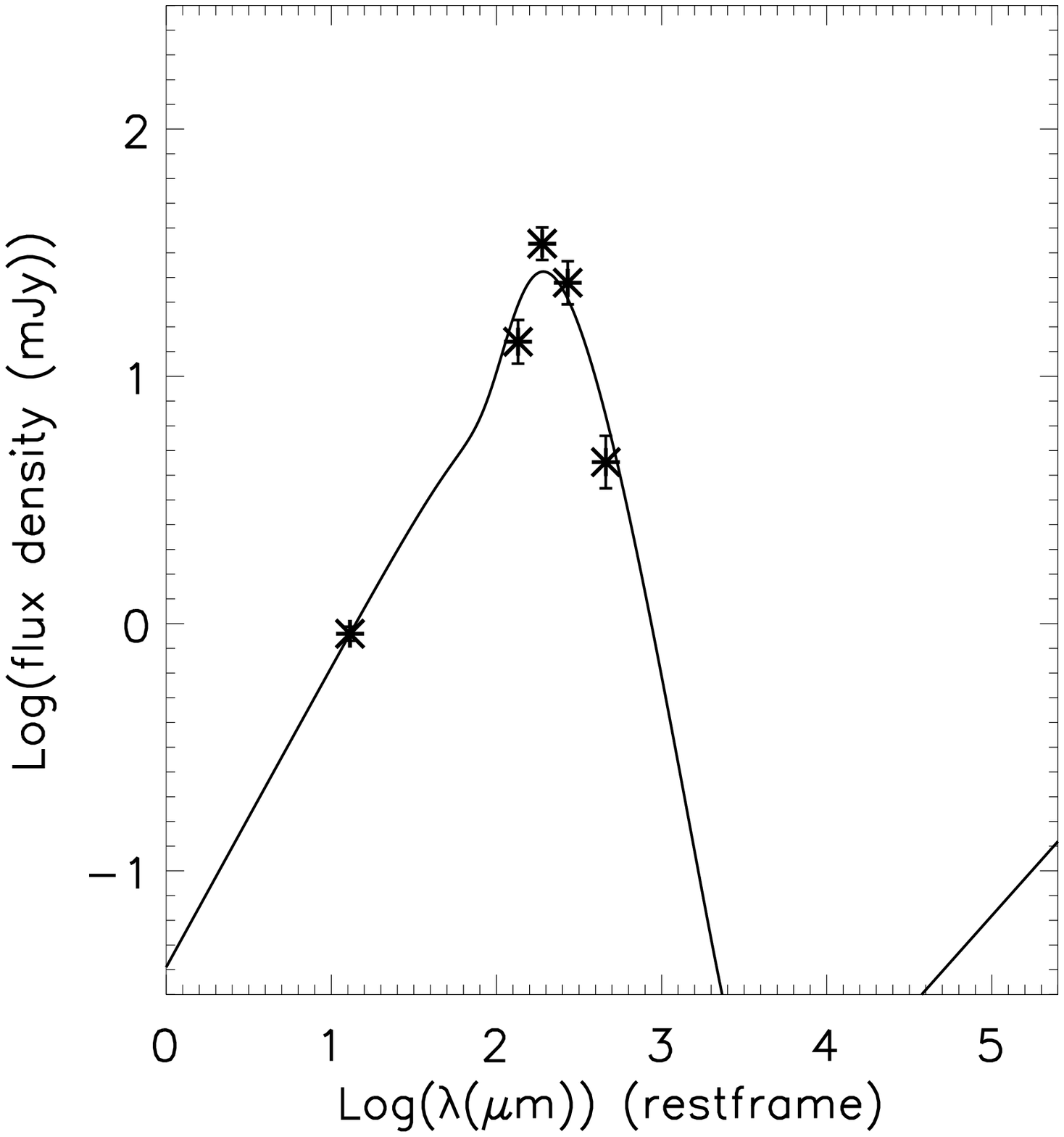} & \includegraphics[scale=0.3]{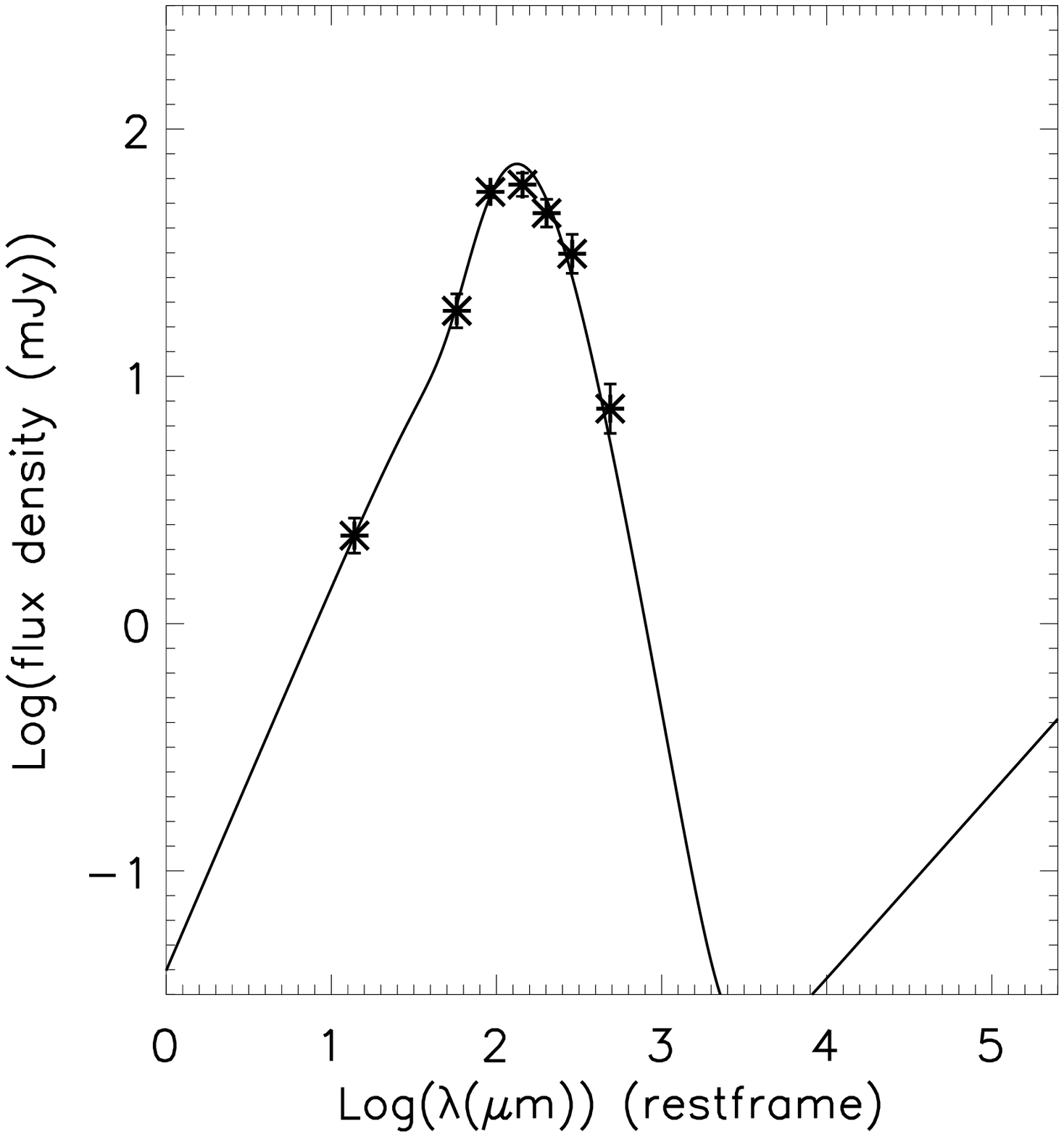} & \includegraphics[scale=0.3]{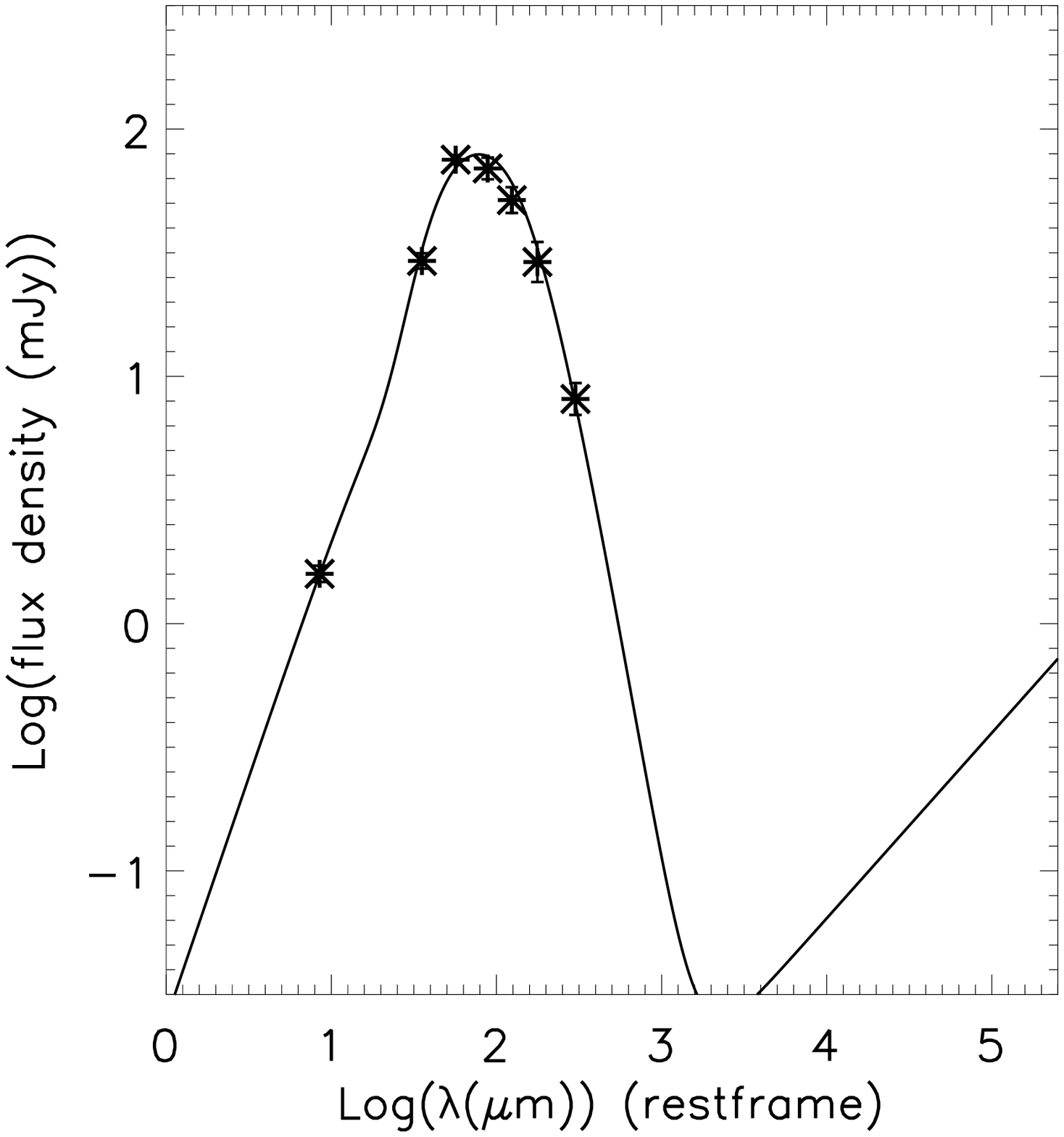} \\
\large{XID270} & \large{XID278} & \large{XID353} \\
\includegraphics[scale=0.3]{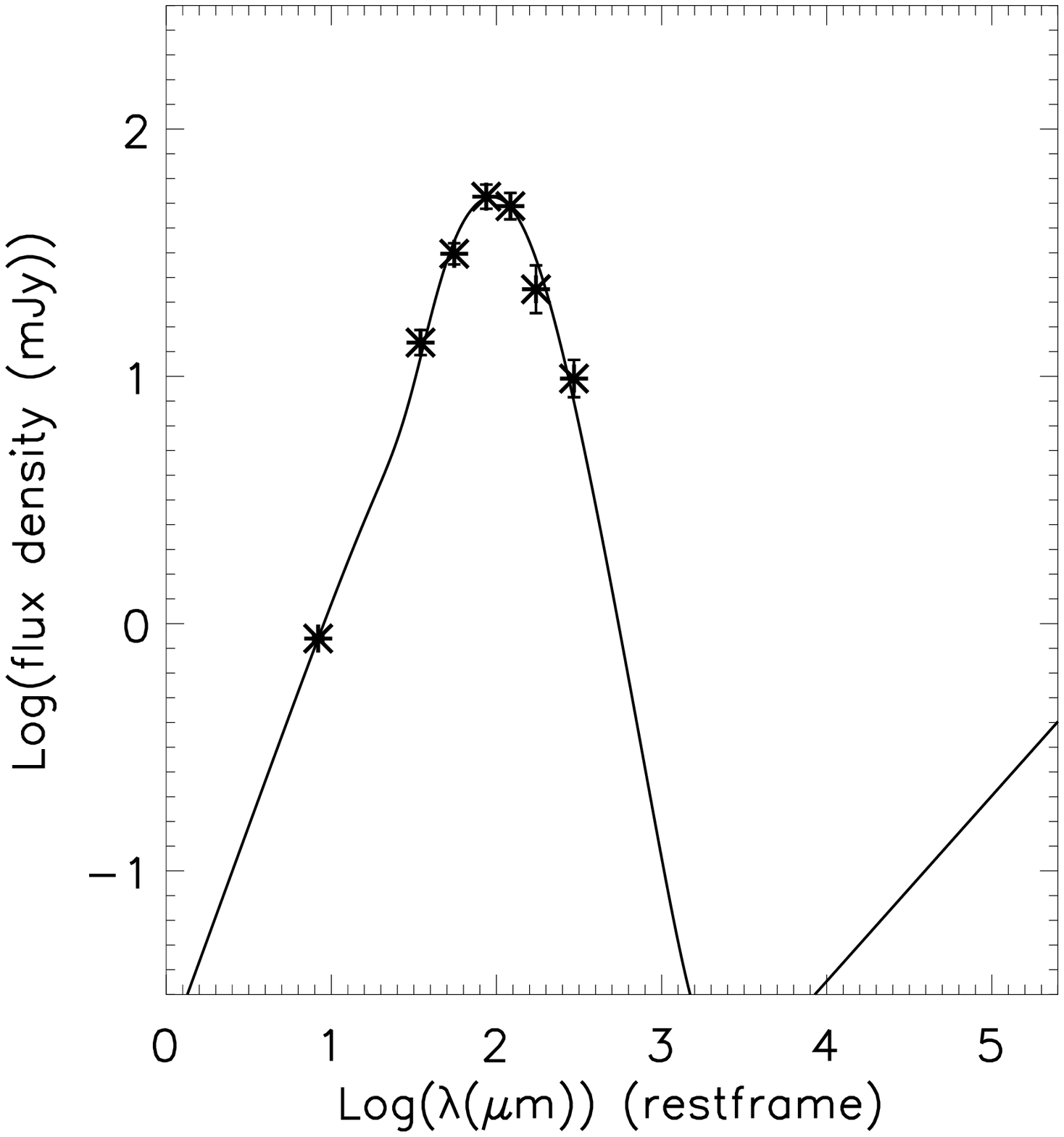} & \includegraphics[scale=0.3]{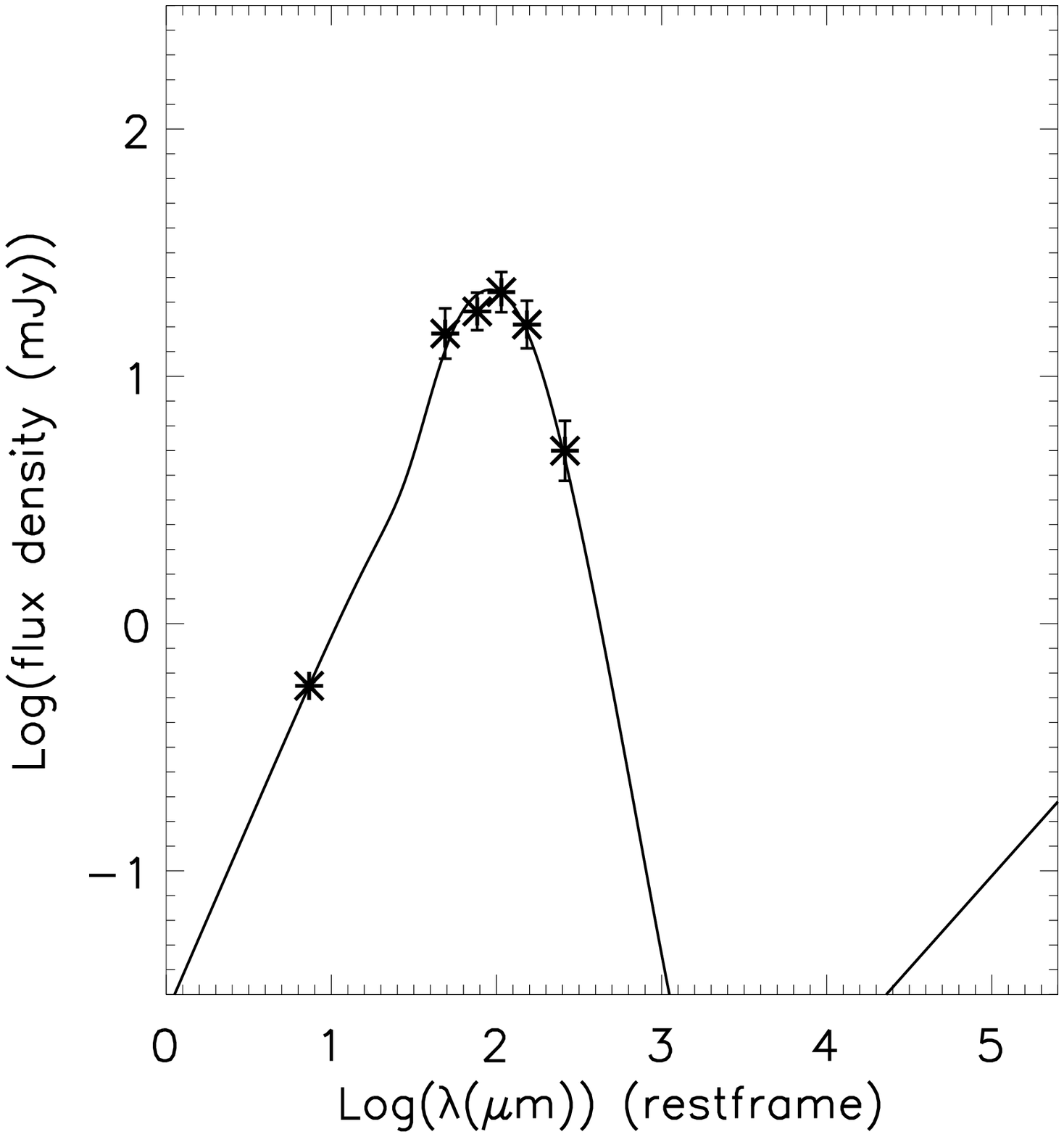} & \includegraphics[scale=0.3]{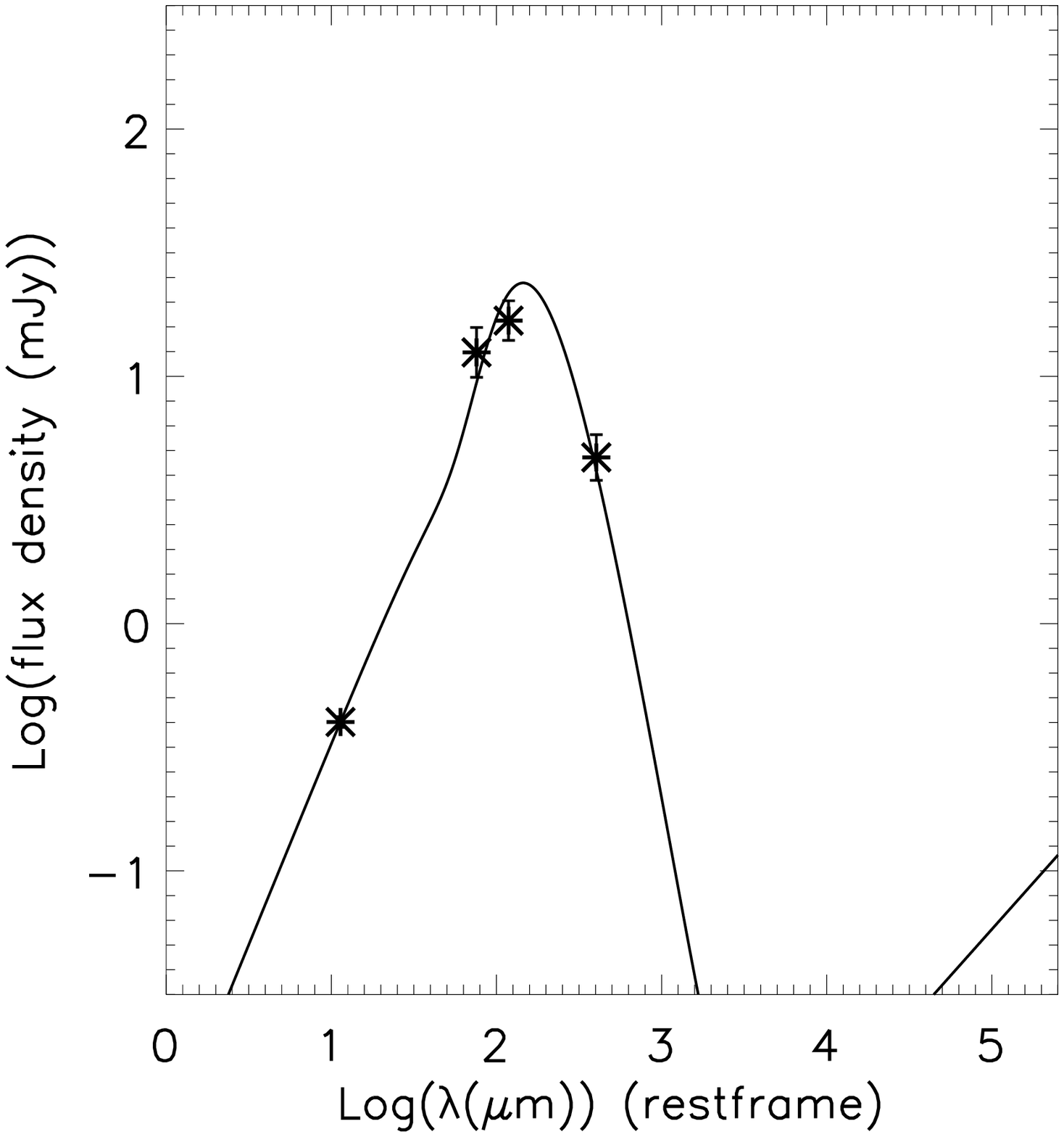} \\
\large{XID53922} & \large{XID54440} & \large{XID60070} \\
\includegraphics[scale=0.3]{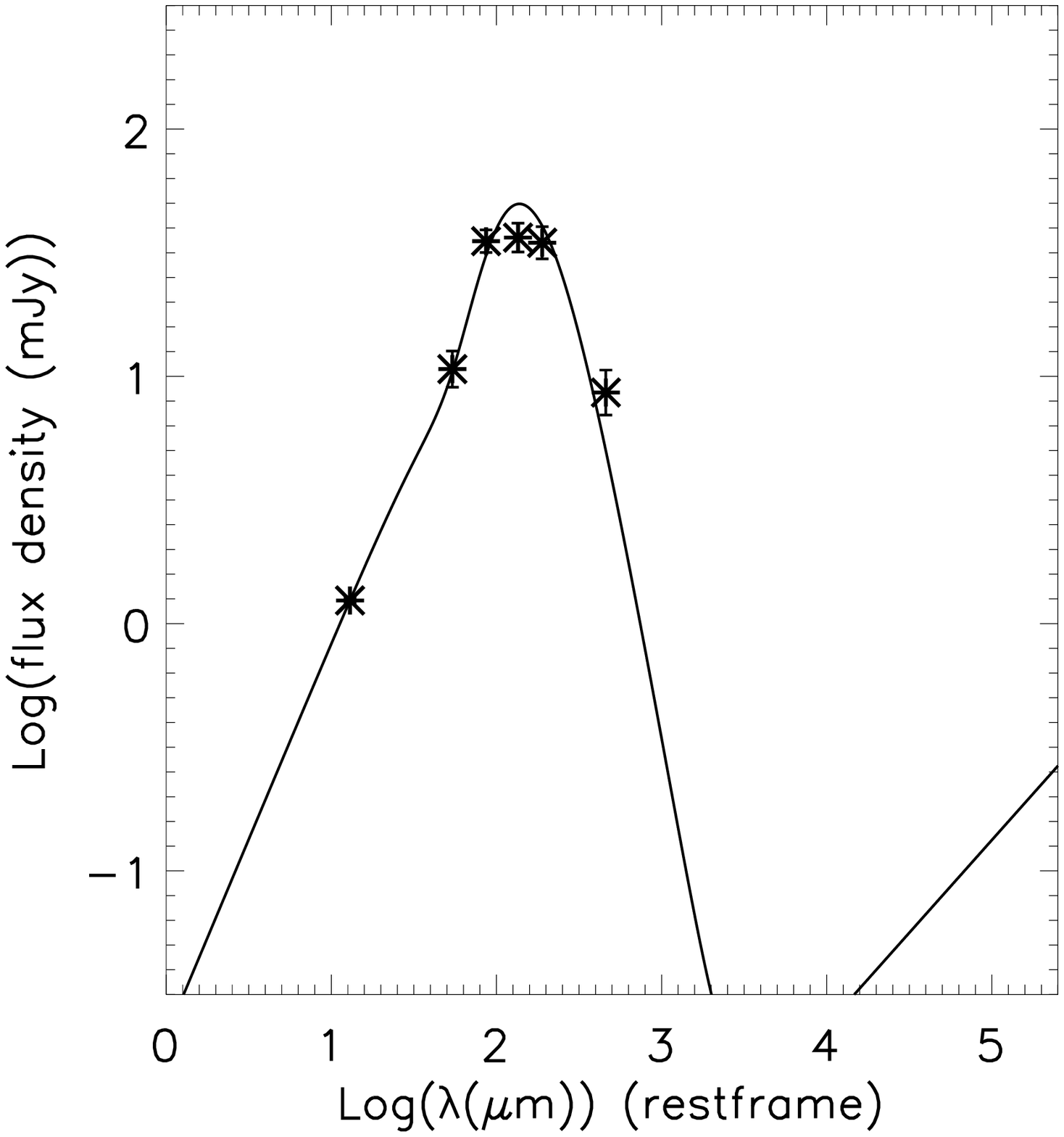} & \includegraphics[scale=0.3]{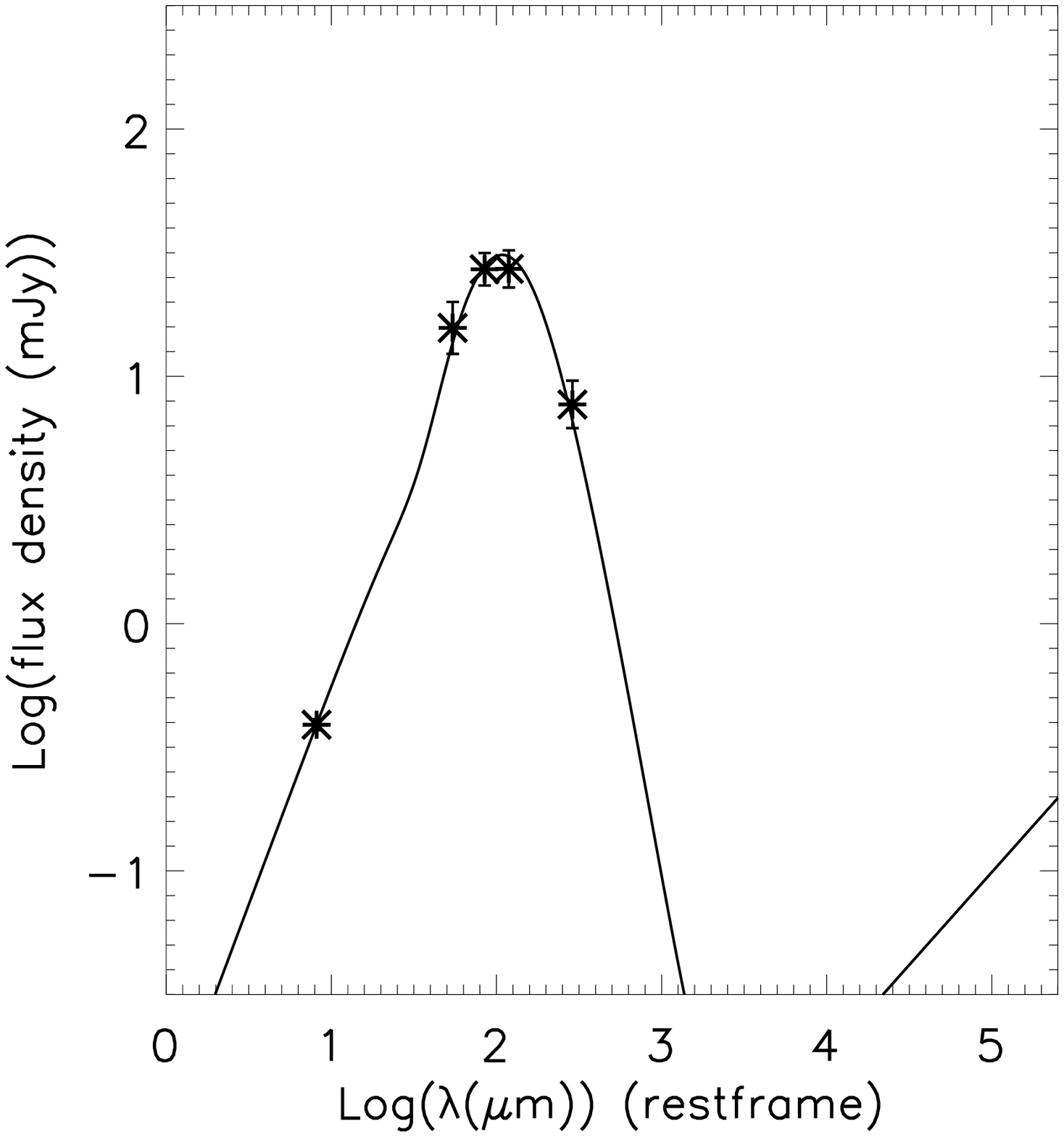} & \includegraphics[scale=0.3]{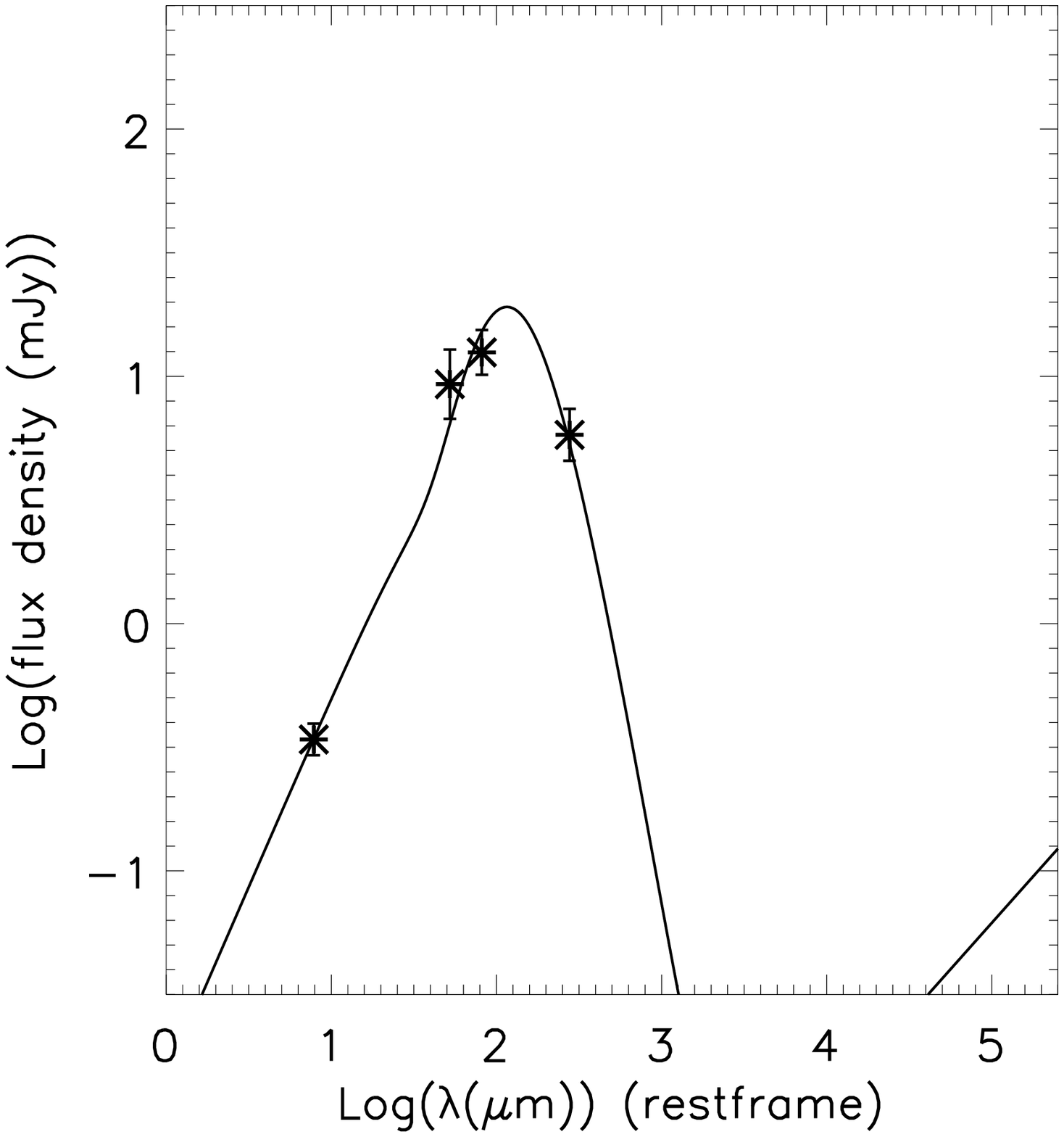} \\
\end{tabular}
\caption{Best-fit SEDs for the 9 AGN in Table \ref{tab:fluxes} with at least two detections in the \textit{Herschel} bands.}
\label{fig:SED}
\end{center}
\end{figure*}

\begin{figure*}
\begin{center}
\begin{tabular}{cc} 
\includegraphics[scale=0.45]{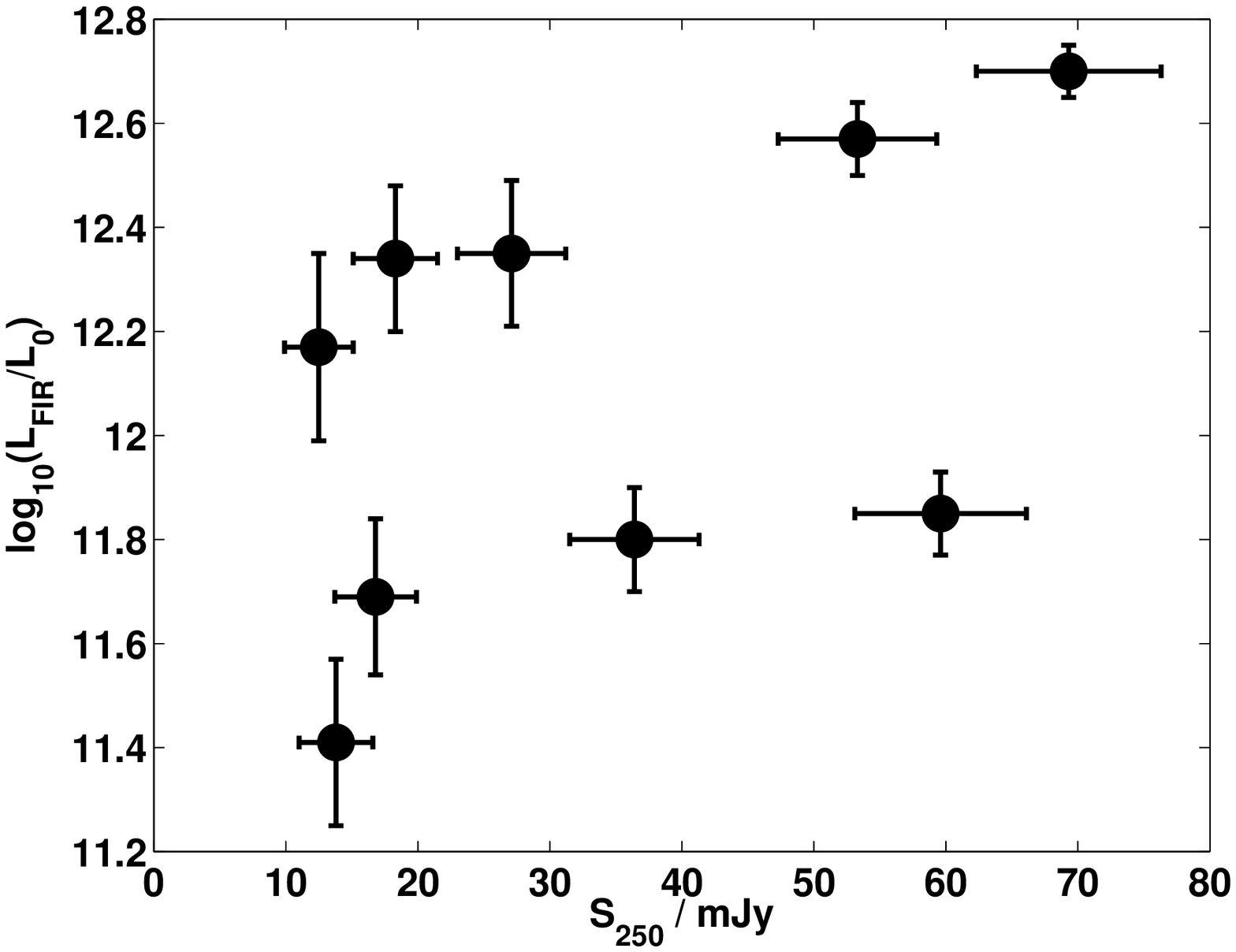} &  \includegraphics[scale=0.45]{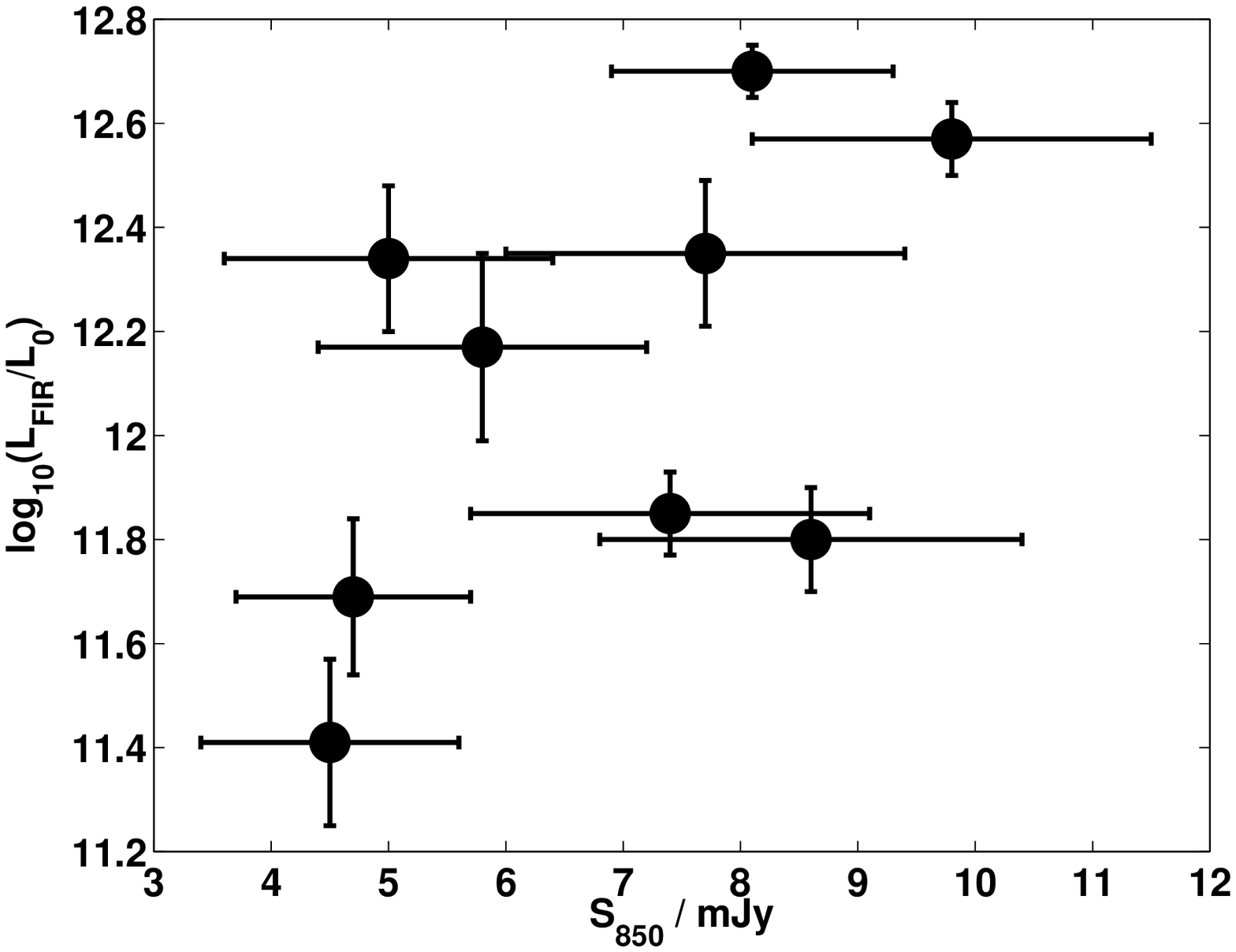} \\
\includegraphics[scale=0.45]{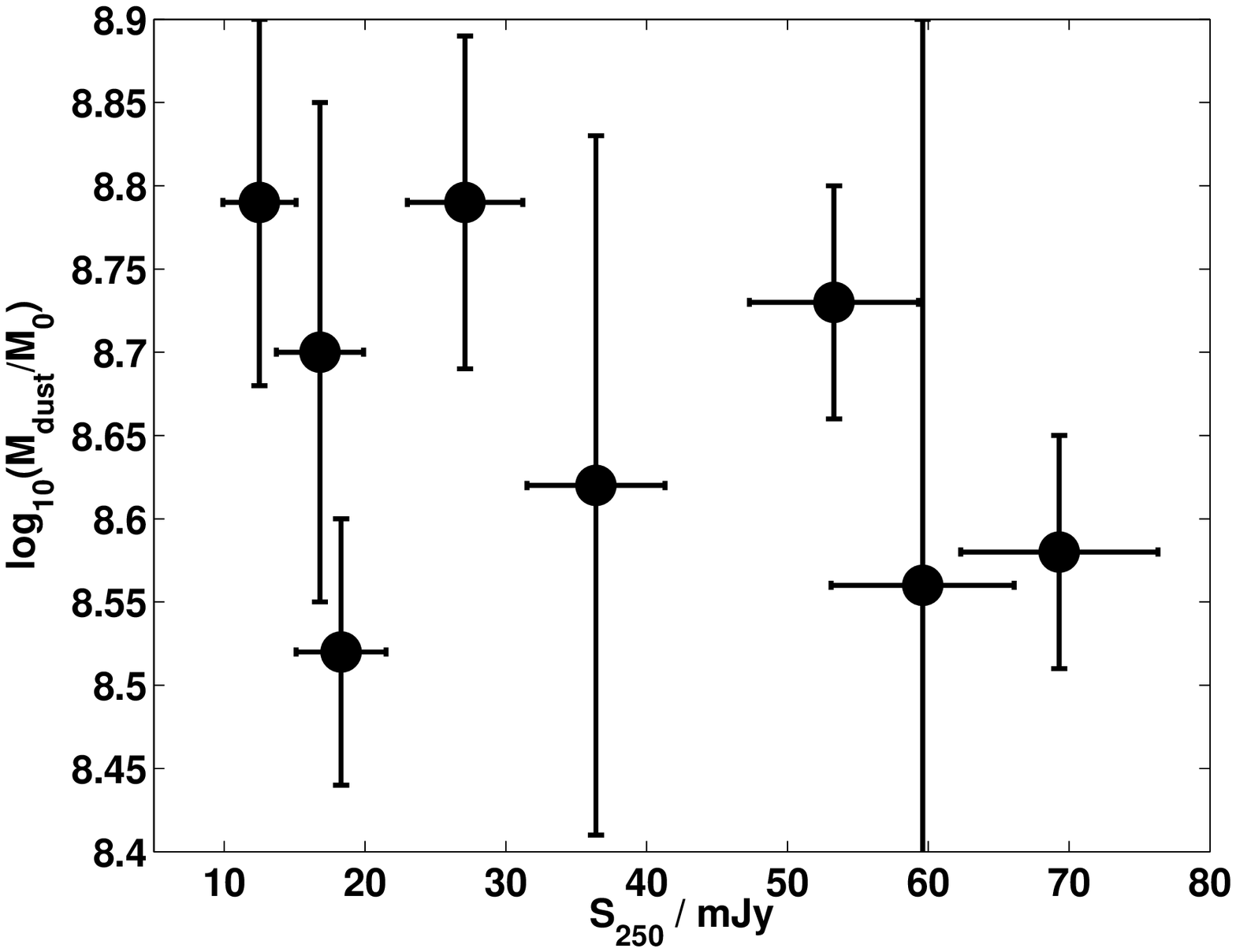} &  \includegraphics[scale=0.45]{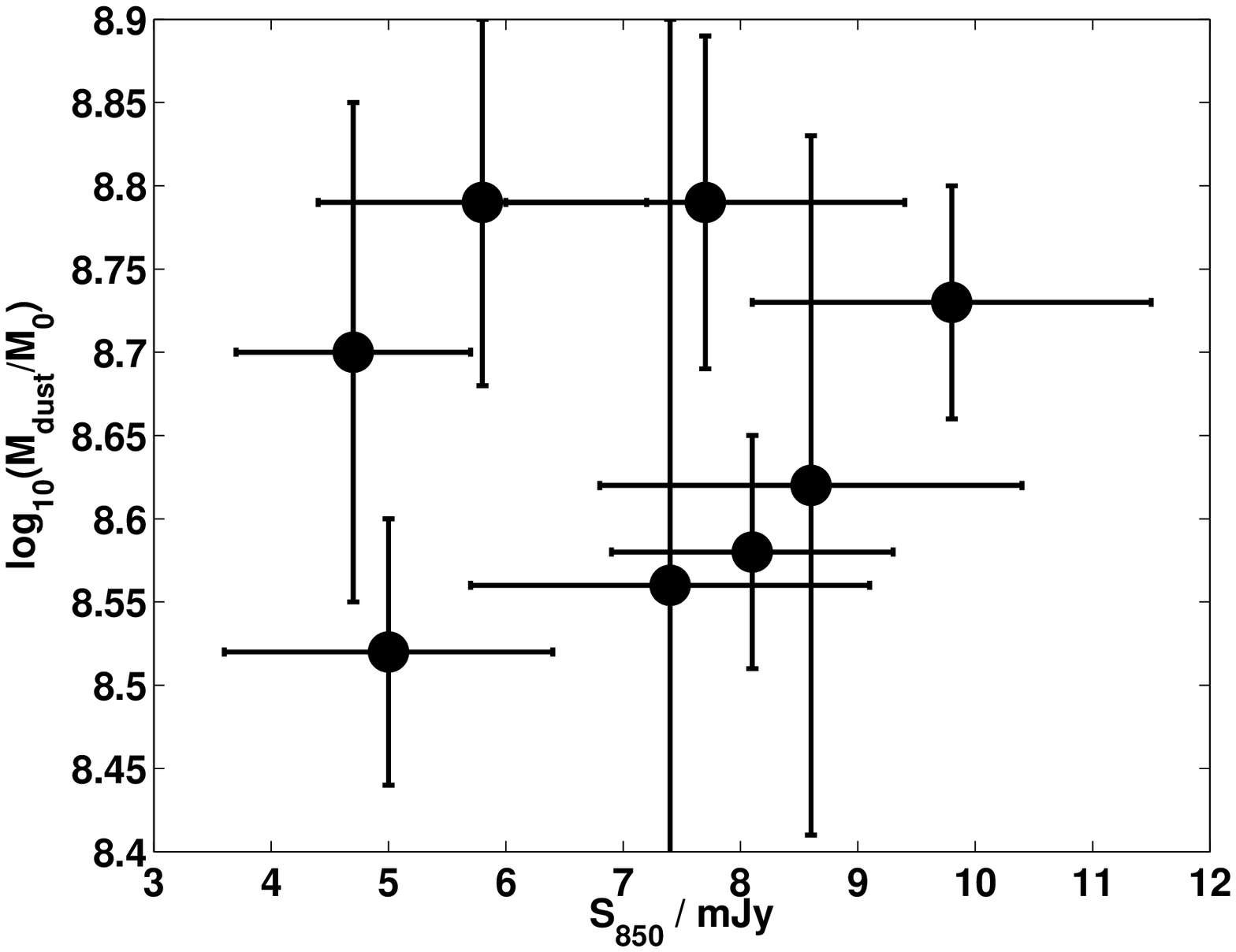} \\
\end{tabular}
 \caption{\textit{Top:} Correlation between the far infrared luminosity estimated from full SED fitting (proxy for star formation) and the 250 and 850$\mu$m fluxes of individually detected AGN. \textit{Bottom:} Correlation between the dust mass estimated from full SED fitting and the 250 and 850$\mu$m fluxes of individually detected AGN}
\label{fig:SLIR}
\end{center}
\end{figure*}

In this section we present the properties of our sample of 19 X-ray AGN that are associated with 850$\mu$m sources in the COSMOS field. Table \ref{tab:properties} presents the stellar masses and various obscuration parameters investigated in this work for each of the 19 AGN. Table \ref{tab:fluxes} presents photometry from \textit{Spitzer} MIPS (24$\mu$m), \textit{Herschel} PACS (100$\mu$m \& 160$\mu$m) and SPIRE (250, 350 and 500$\mu$m). In Fig. \ref{fig:SED} we show the best-fit SEDs for the AGN that are detected in at least two of the \textit{Herschel} bands. We fit a power-law of the form $S_\lambda \propto \lambda^\alpha$ to model the mid infrared emission from the AGN at $\lambda \gtrsim 3 \mu$m, together with a single temperature greybody \citep{Casey:12} with a fixed dust emissivity index of $\beta=2$. The best-fit mid infrared power-law slope, $\alpha$, dust temperature, far infrared luminosity (60-300$\mu$m) and dust mass are given in Table \ref{tab:sedind}. In Fig. \ref{fig:SLIR} we also show the correlation between the far infrared luminosity and the 250 and 850$\mu$m fluxes for these individually detected AGN as well as the correlation between the 250 and 850$\mu$m fluxes and the dust mass.  

\begin{table*}
\begin{center}
\caption{Summary of the properties of the 19 X-ray AGN that are associated with 850$\mu$m sources. Note: the hard X-ray luminosities and redshifts of these sources are presented in Table \ref{tab:IDs}.}
\label{tab:properties}
\begin{tabular}{lccccc}
XID & log$_{10}$(M$_*$/M$_\odot$) & HR & (r-4.5$\mu$m) Vega & E(B-V)$_{\rm{AGN}}$ & E(B-V)$_{\rm{GAL}}$ \\
\hline
13  & 11.26 & $-0.58$ & 5.44 & 0.1 & 0.0 \\
18 & 11.17 & $-0.21$ & 7.68 & 0.6 & 0.0 \\
139  & 10.08 & $-0.37$ & 6.61 & 0.6 & 0.2 \\
160 & 10.61 & $-0.09$ & 5.18 & 0.8 & 0.2 \\
246  & 9.12 & 0.42 & 8.59 & 1.3 & 0.3 \\
250  & 11.11 & -- & 5.02 & 0.1 & 0.5 \\
270 & 11.74 & -- & 8.13 & 0.3 & 0.3 \\
278  & 11.26 & -- & 9.13 & 2.6 & 0.5 \\
353 & 11.24 & $-0.14$ & 6.28 & 0.0 & 0.4 \\
402 & 10.54 & $-0.24$ & 6.57 & 0.3 & 0.4 \\
415  & 9.92 & -- & 6.18 & 0.3 & 0.5 \\
469  & 10.44 & -- & 6.00 & 0.3 & 0.1 \\
10675 & 11.16 & -- & 5.37 & 1.0 & 0.1 \\
10809  & 10.09 & -- & 6.56 & 0.4 & 0.0 \\
30182  & 9.62 & $-0.6$ & 3.82 & 0.3 & 0.0 \\
53922 & 9.25 & $-0.16$ & 6.66 & 2.3 & 0.2 \\
54440 & 11.18 & -- & 9.07 & 0.7 & 0.5 \\
60070  & 11.32 & -- & 7.99 & 0.4 & 0.5 \\
60490 & 10.66 & -- & 6.57 & 0.3 & 0.0 \\
\hline
\end{tabular}
\end{center}
\end{table*}

\begin{table*}
\begin{center}
\caption{Summary of \textit{Spitzer} MIPS and \textit{Herschel} fluxes of the 19 X-ray AGN that are associated with 850$\mu$m sources. Note: the SCUBA-2 850$\mu$m flux densities of these sources are presented in Table \ref{tab:IDs}.}
\label{tab:fluxes}
\begin{tabular}{lcccccc}
XID & S$_{24}$/mJy & S$_{100}$/mJy & S$_{160}$/mJy & S$_{250}$/mJy & S$_{350}$/mJy & S$_{500}$/mJy \\
\hline
13 & 0.91$\pm$0.06 & -- & -- & 13.8$\pm$2.8 & 34.4$\pm$5.2 & 23.9$\pm$4.8 \\
18 & 0.51$\pm$0.02 & -- & -- & $<$9.2 & $<$10.6 & $<$12.2 \\
139 & 2.27$\pm$0.37 & 18.4$\pm$2.9 & 55.7$\pm$3.0 & 59.6$\pm$6.5 & 45.7$\pm$5.9 & 31.3$\pm$5.7 \\
160          &         1.59$\pm$0.12 &                  29.3$\pm$2.1        &          75.2$\pm$3.2 &                 69.3$\pm$7.0    &              51.6$\pm$6.2 &    29.0$\pm$5.4    \\
246 & 0.88$\pm$0.02 & -- & -- & $<$9.0 & $<$12.1 & $<$12.6 \\
250 & 0.28$\pm$0.02 & 4.9$\pm$1.2 & -- & $<$9.2 & $<$10.6 & $<$12.2 \\ 
270          &         0.87$\pm$0.02 &   13.7$\pm$1.6 &  31.3$\pm$3.1 &  53.3$\pm$6.0 &  48.8$\pm$6.0 & 22.5$\pm$5.0 \\
278 & 0.56$\pm$0.02 & -- & 14.9$\pm$3.5 & 18.3$\pm$3.2 & 21.9$\pm$4.1 & 16.2$\pm$3.6 \\  
 353           &        0.40$\pm$0.02 &    -- &   12.5$\pm$2.9 &  16.8$\pm$3.1 & $<$16.1    & $<$16.3  \\
  402 & 0.28$\pm$0.09 & -- & 13.3$\pm$3.3 & $<$14.6 & $<$15.9 & $<$13.6 \\   
   415 & 0.13$\pm$0.02 & -- & -- &$<9.2$ & $<$10.6 & $<$12.2 \\
10675        &       0.09$\pm$0.02 &      -- & -- &          $<$9.2 &       $<$10.3     & $<$12.6    \\
53922      &         1.24$\pm$0.03 &   10.7$\pm$1.8 &  35.2$\pm$3.7 &  36.4$\pm$4.9 &   34.7$\pm$5.2 &   -- \\
54440       &         0.39$\pm$0.02   & -- &    15.7$\pm$3.8 &   27.1$\pm$4.1 &   27.2$\pm$4.7 & $<$16.3     \\
60070        &       0.34$\pm$0.05 & -- & 9.3$\pm$3.0 & 12.5$\pm$2.6 & $<$13.9 & $<$13.6 \\
60490     &              0.08$\pm$0.02 &    -- & -- &   $<$10.0 & $<$15.2   &    $<$12.2 \\
\hline
\end{tabular}
\end{center}
\end{table*}

\begin{table}
\begin{center}
\caption{Best-fit SED parameters for individually detected AGN with at least two detections in the \textit{Herschel} bands.}
\label{tab:sedind}
\begin{tabular}{lcccc}
XID & $\alpha_{\rm{MIR}}$ & T$_{\rm{dust}}$/K & log$_{10}$(L$_{\rm{FIR}}$/L$_\odot$) & log$_{10}$(M$_{\rm{dust}}$/M$_\odot$) \\
\hline
13 & 1.21$\pm$0.07 & 15$\pm$2 & 11.41$\pm$0.16 & 9.08$\pm$2.41 \\
139 & 1.55$\pm$0.09 & 22$\pm$1 & 11.85$\pm$0.09 & 8.56$\pm$0.34 \\
160 & 1.97$\pm$0.06 & 37$\pm$2 & 12.70$\pm$0.05 & 8.58$\pm$0.07 \\
270 & 1.84$\pm$0.04 & 31$\pm$2 & 12.57$\pm$0.07 & 8.73$\pm$0.07 \\
278 & 1.54$\pm$0.08 & 32$\pm$5 & 12.34$\pm$0.14 & 8.52$\pm$0.08 \\
353 & 1.63$\pm$0.07 & 20$\pm$2 & 11.69$\pm$0.15 & 8.70$\pm$0.15 \\
53922 & 1.59$\pm$0.04 & 21$\pm$2 & 11.80$\pm$0.10 & 8.62$\pm$0.21 \\
54440 & 1.79$\pm$0.07 & 27$\pm$4 & 12.35$\pm$0.14 & 8.79$\pm$0.10 \\
60070 & 1.54$\pm$0.10 & 25$\pm$4 & 12.17$\pm$0.18 & 8.79$\pm$0.11 \\
\hline
\end{tabular}
\end{center}
\end{table}

\section{Redshift \& X-Ray Luminosity Distributions}

\begin{figure*}
\begin{center}
\begin{tabular}{cc}
\includegraphics[scale=0.4]{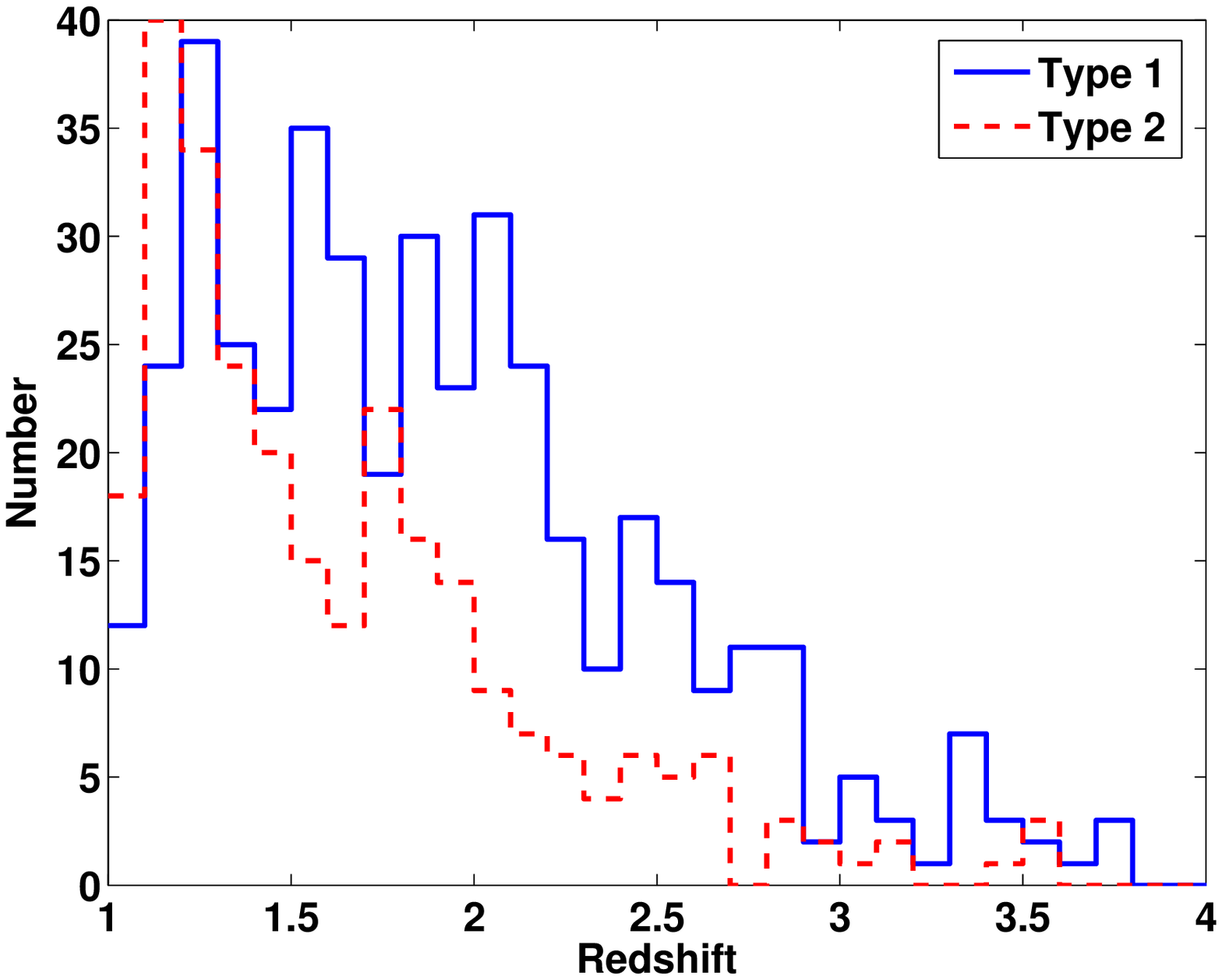} & \includegraphics[scale=0.4]{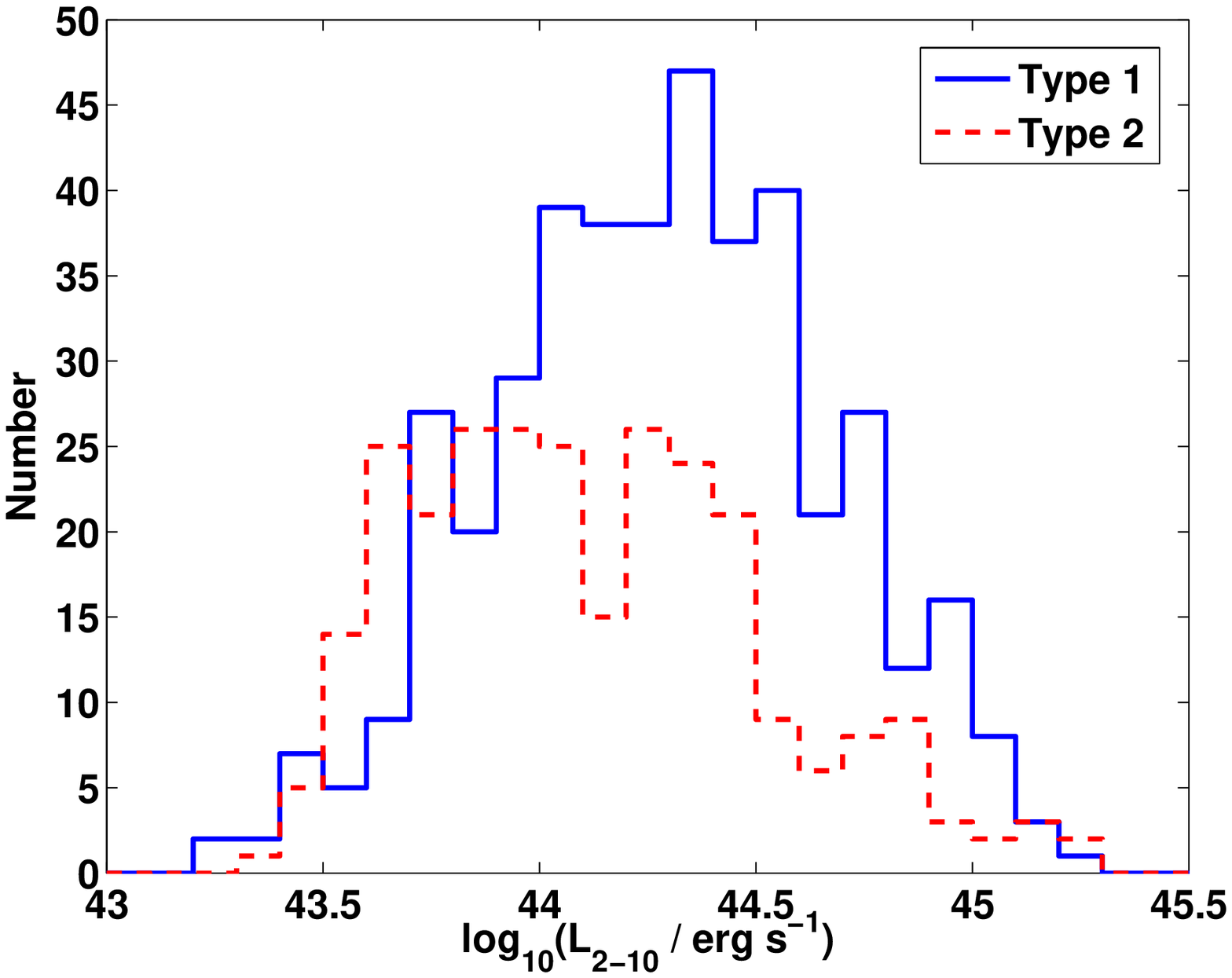} \\
\includegraphics[scale=0.4]{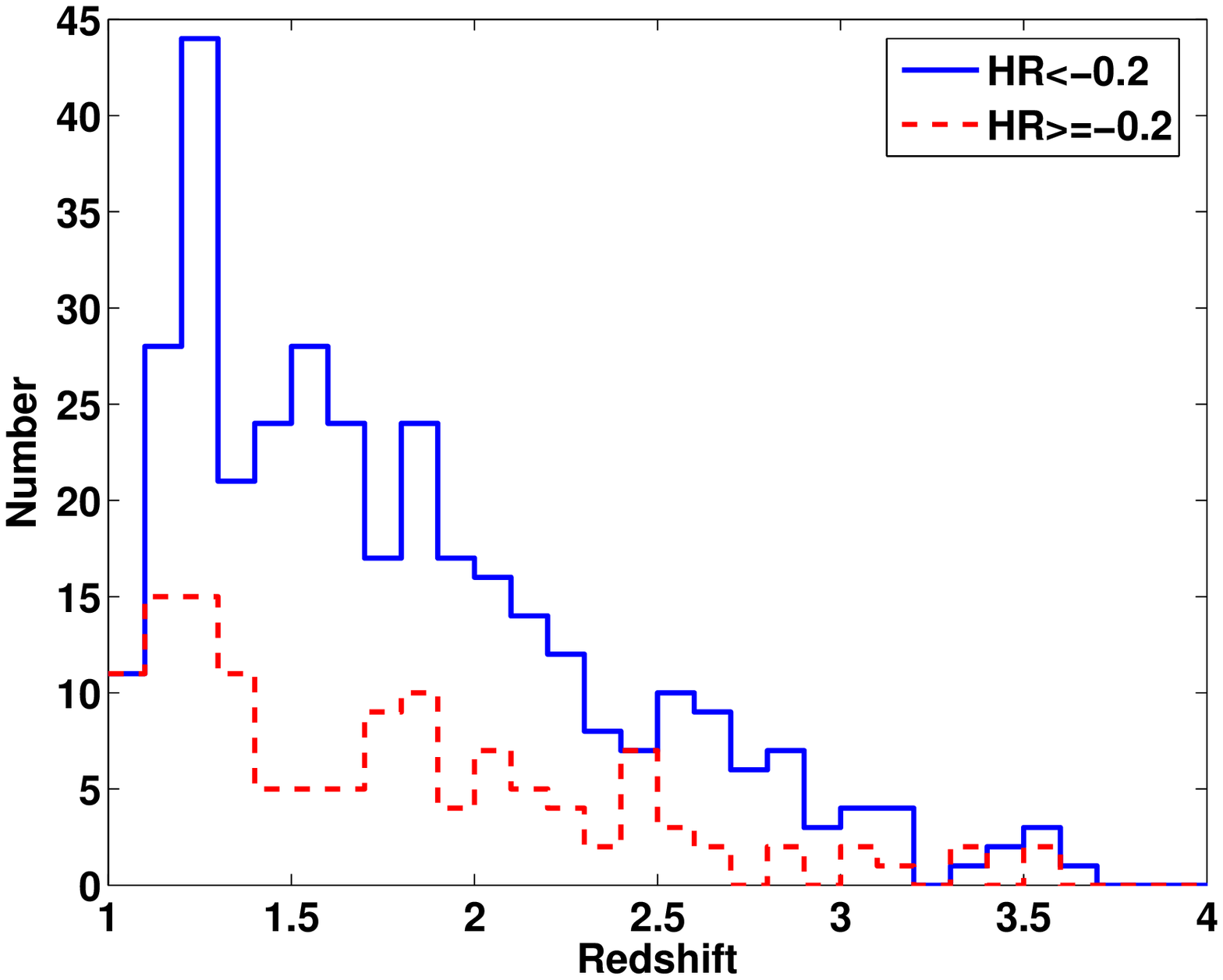} & \includegraphics[scale=0.4]{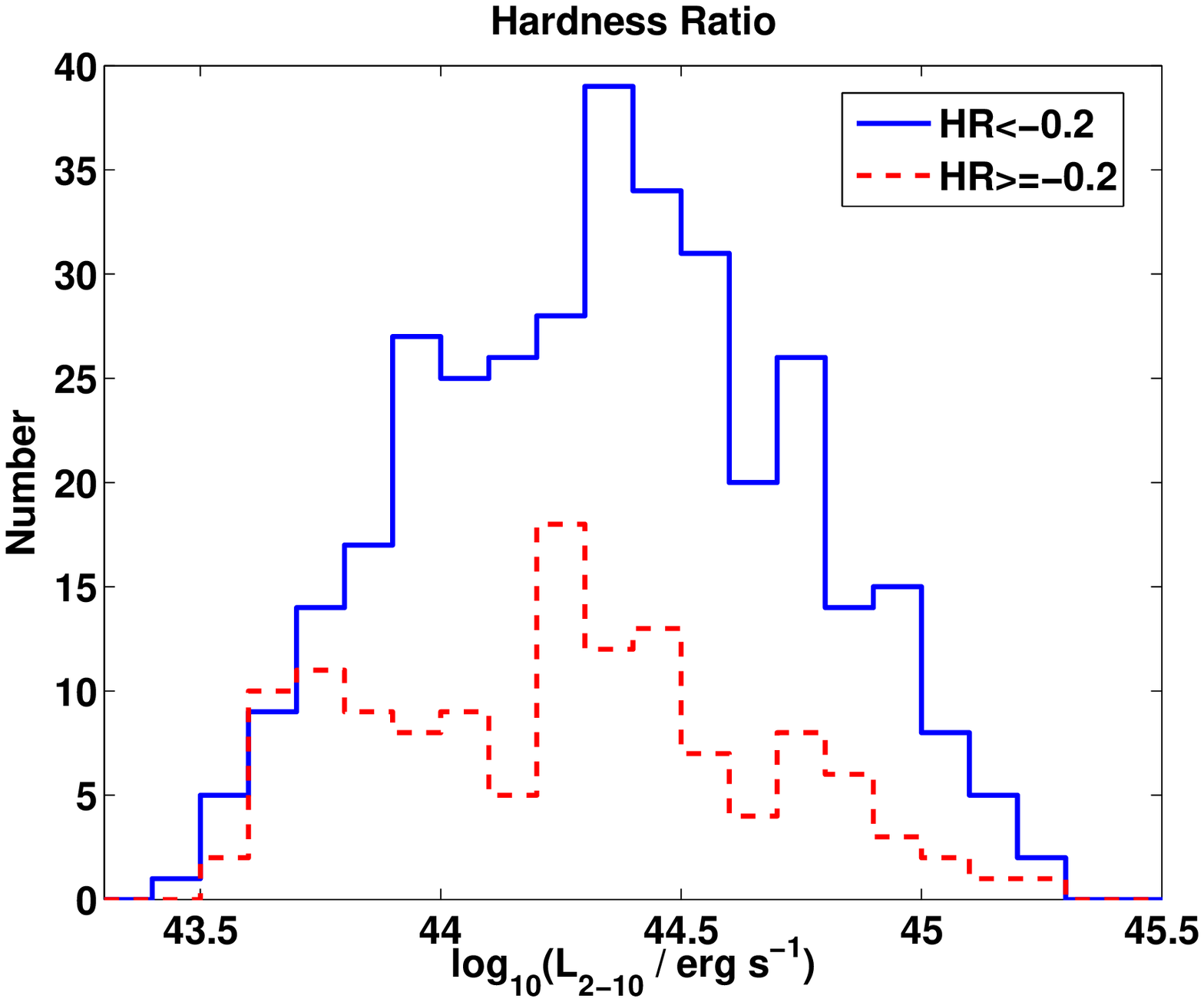} \\
\includegraphics[scale=0.4]{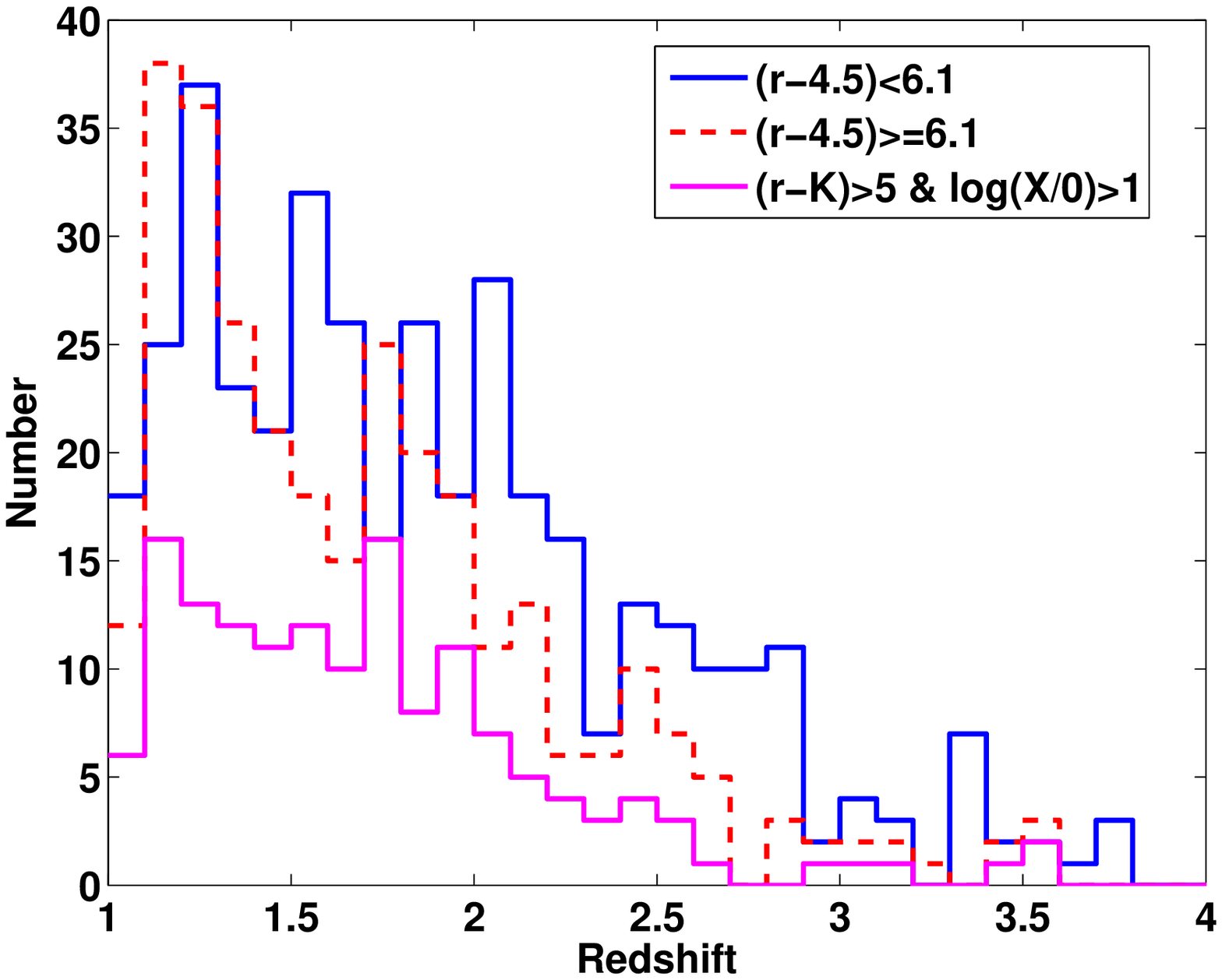} & \includegraphics[scale=0.4]{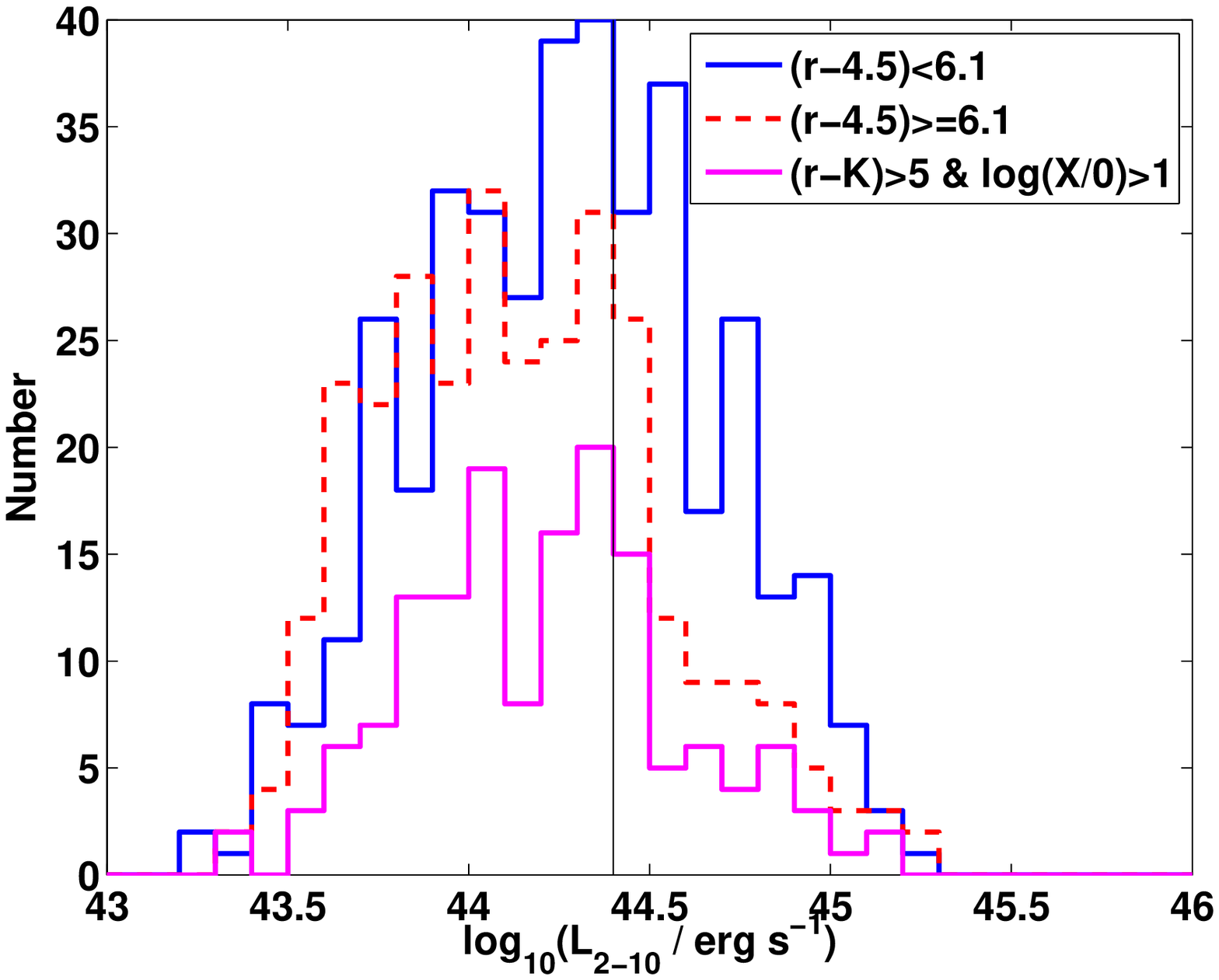} \\
\includegraphics[scale=0.4]{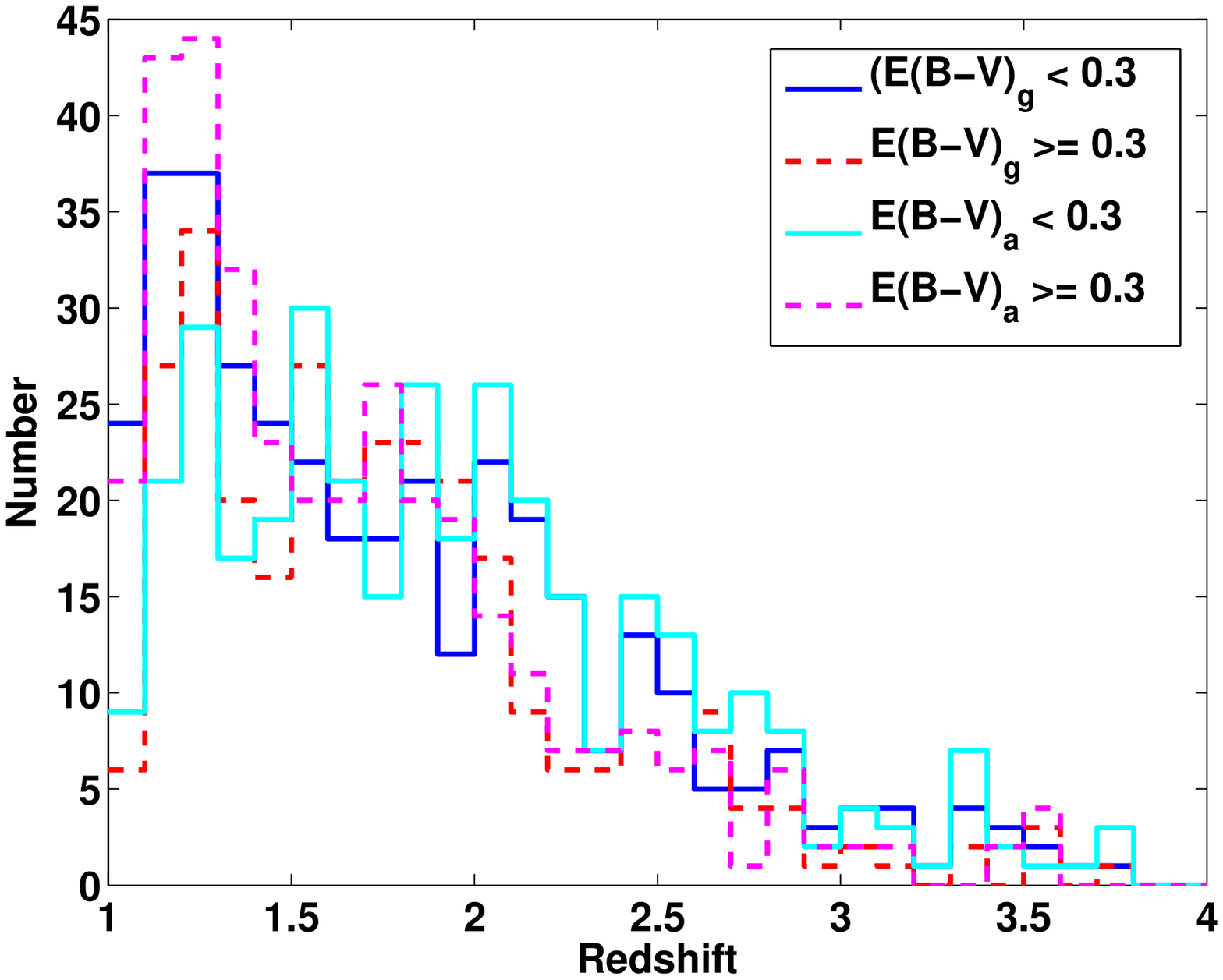} & \includegraphics[scale=0.4]{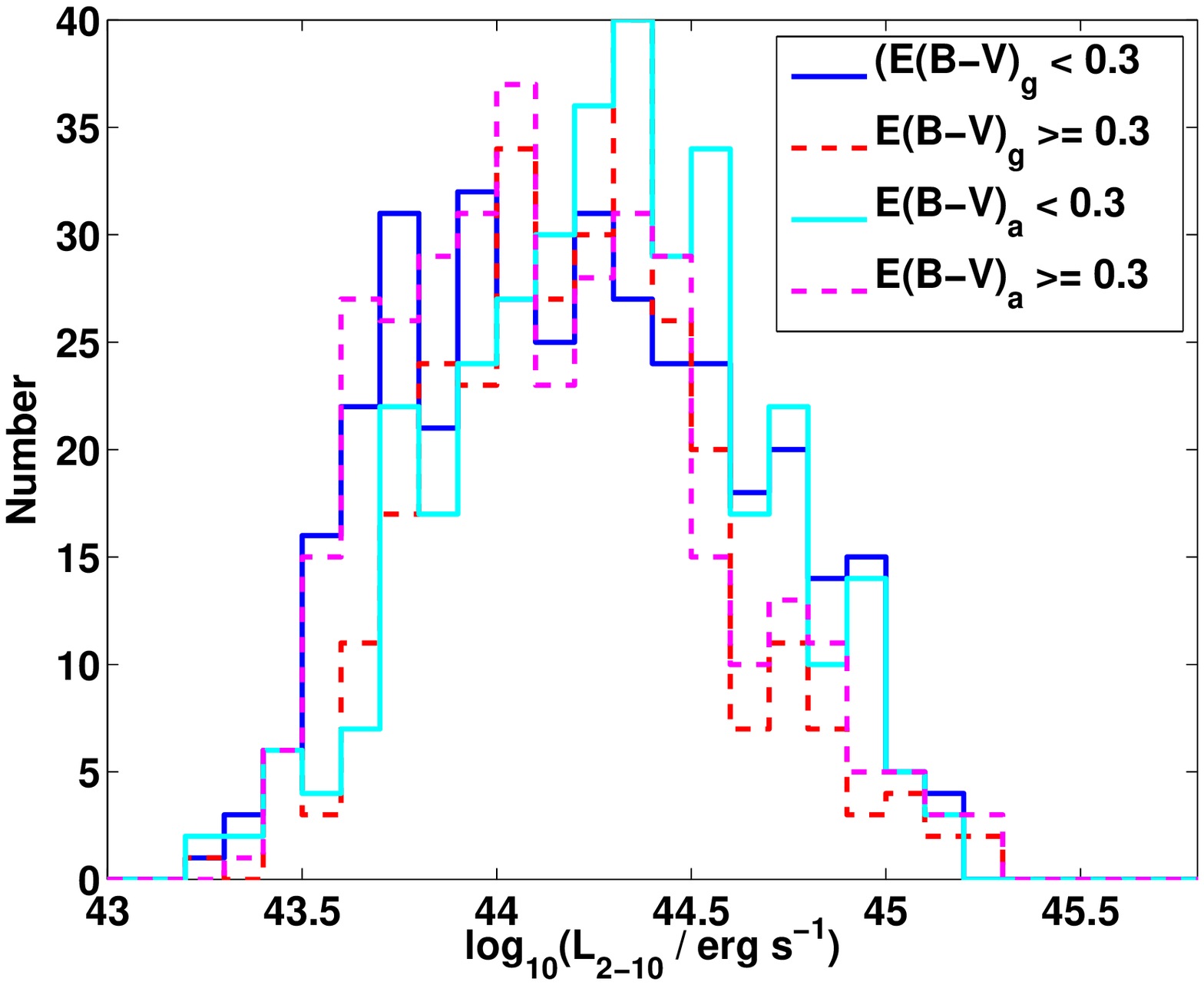} \\
\end{tabular}{cc}
\caption{Redshift (left) and X-ray luminosity (right) distributions for various sub-samples of AGN compared in this work. The top panels show AGN separated as Type 1 and Type 2, the second panels show AGN separated based on hardness ratio, the third panels show AGN separated on the basis of optical-to-infrared colours and the bottom panels show AGN separated on the basis of best-fit dust extinctions. In the third panel, the vertical line marks the X-ray luminosity threshold of log$_{10}$(L$_{2-10})>$44.4 considered in Section \ref{sec:obscuration}.}
\label{fig:dist1}
\end{center}
\end{figure*}

\begin{figure*}
\begin{center}
\begin{tabular}{cc}
\includegraphics[scale=0.4]{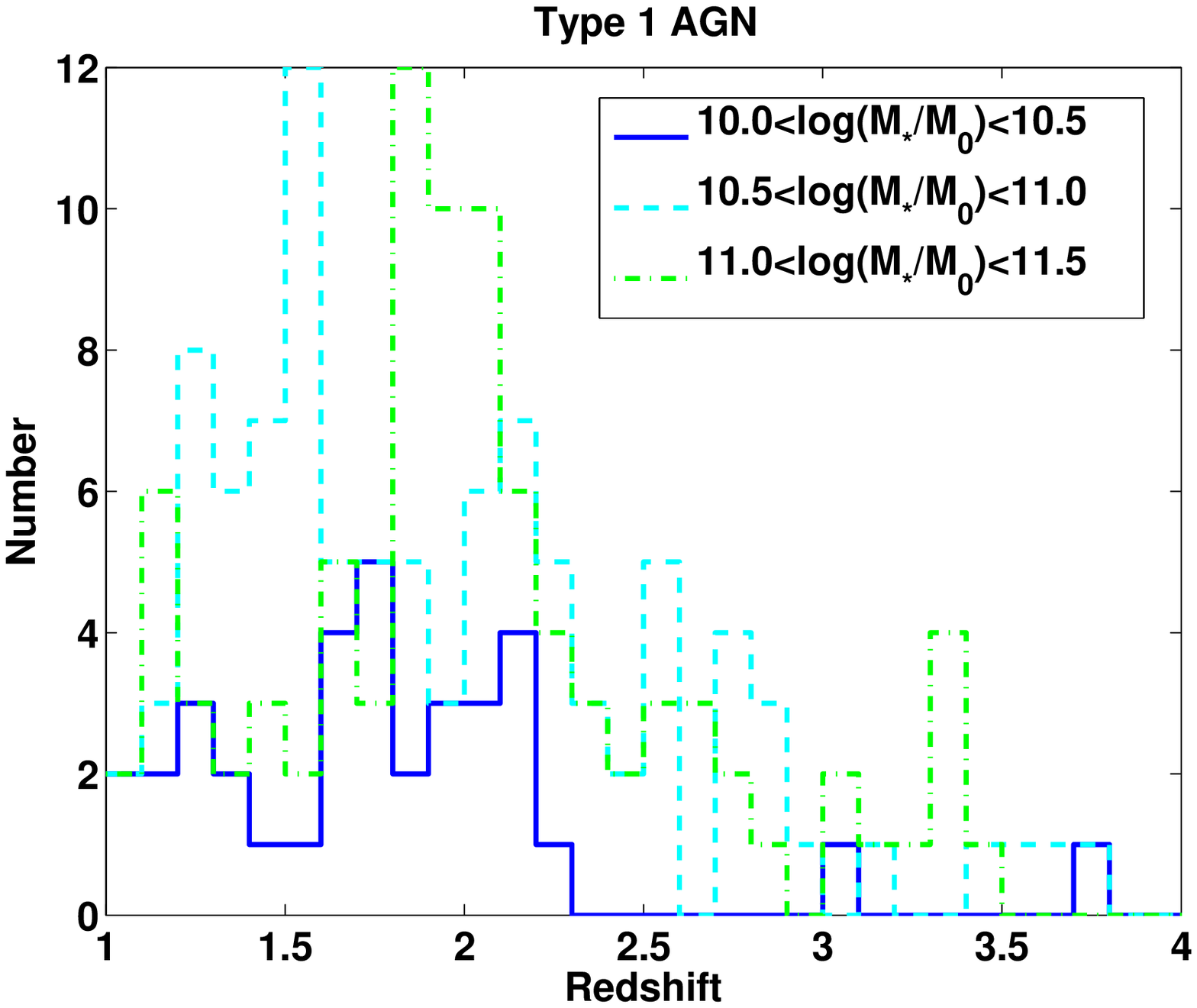} & \includegraphics[scale=0.4]{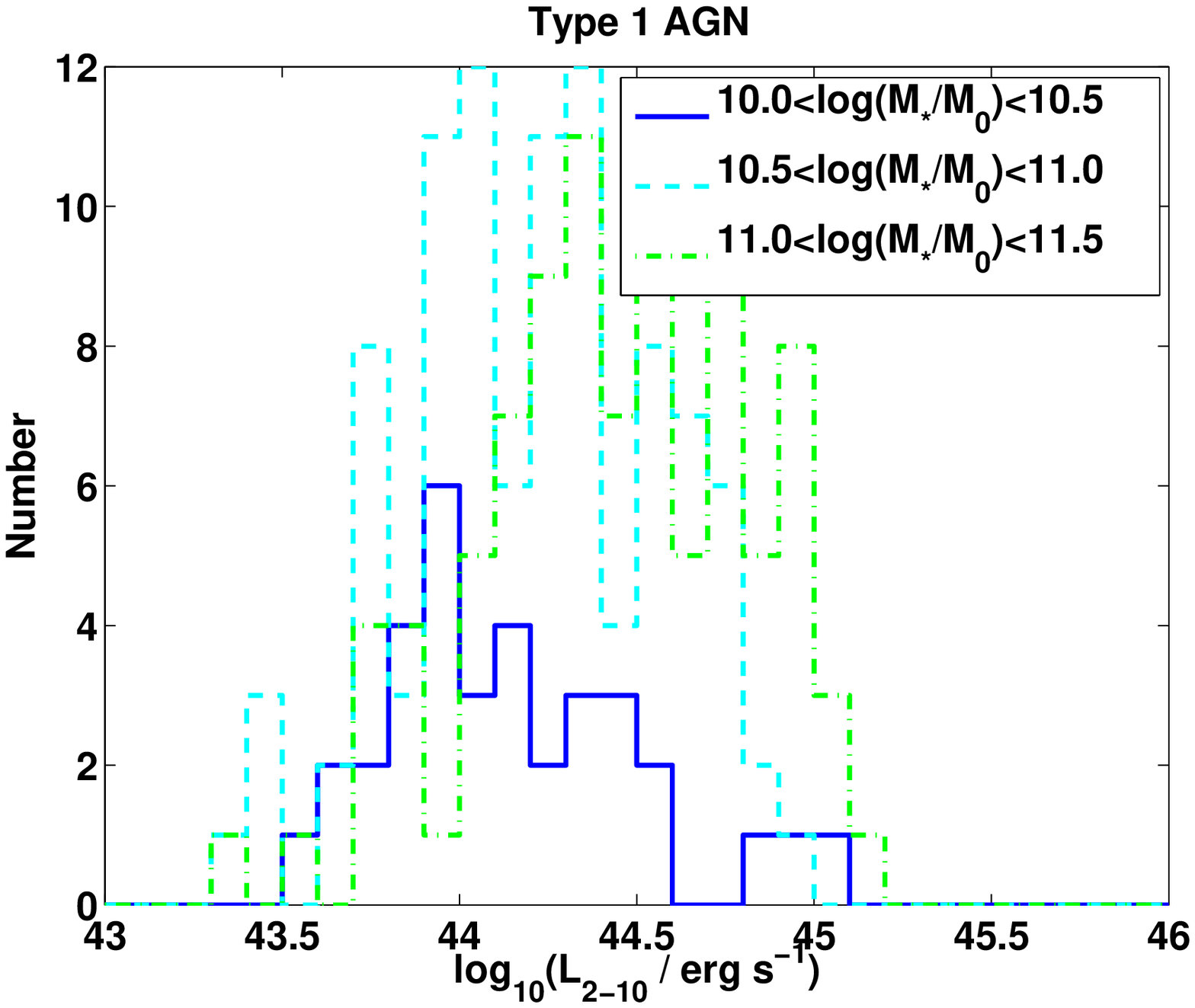} \\
\includegraphics[scale=0.4]{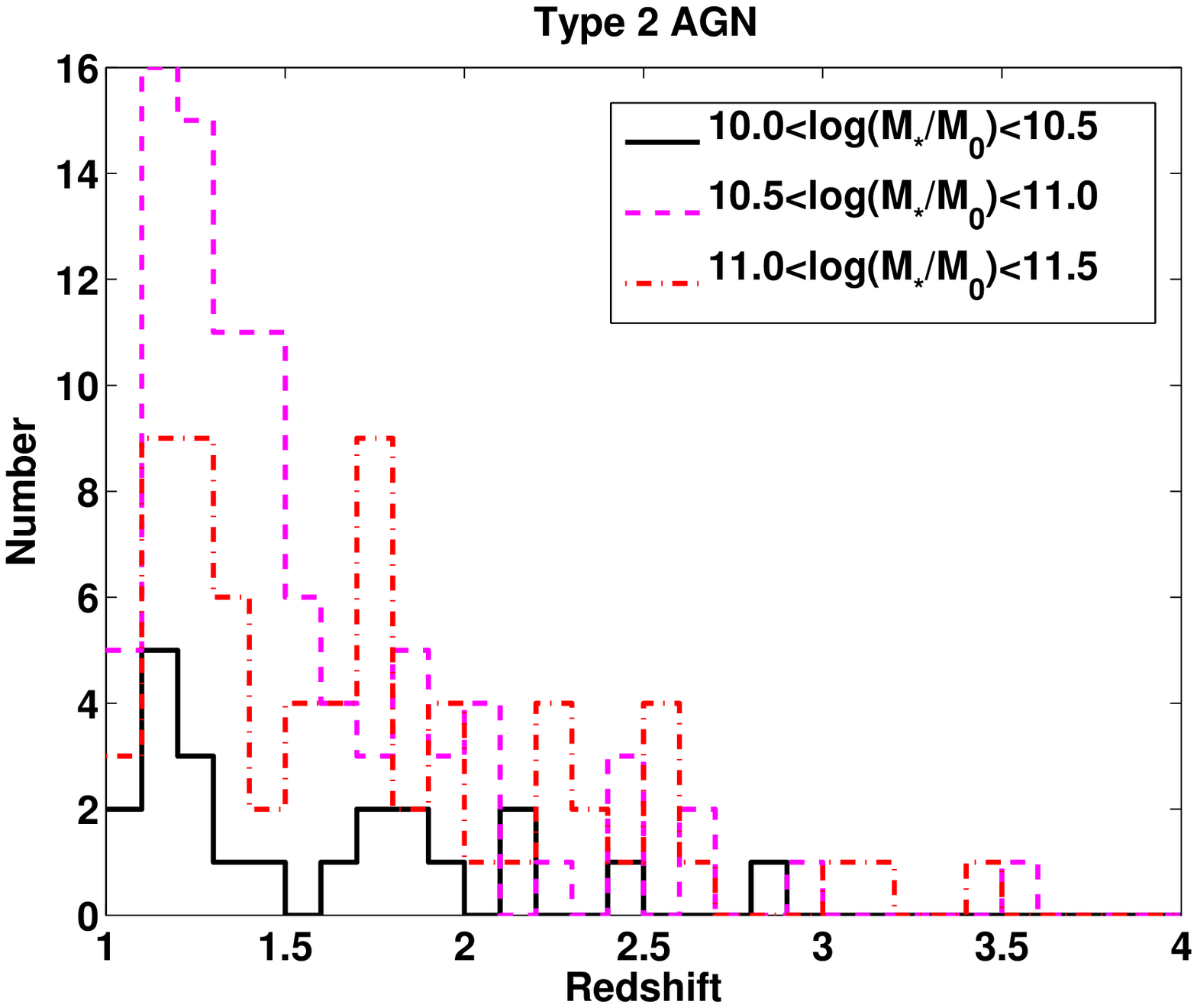} & \includegraphics[scale=0.4]{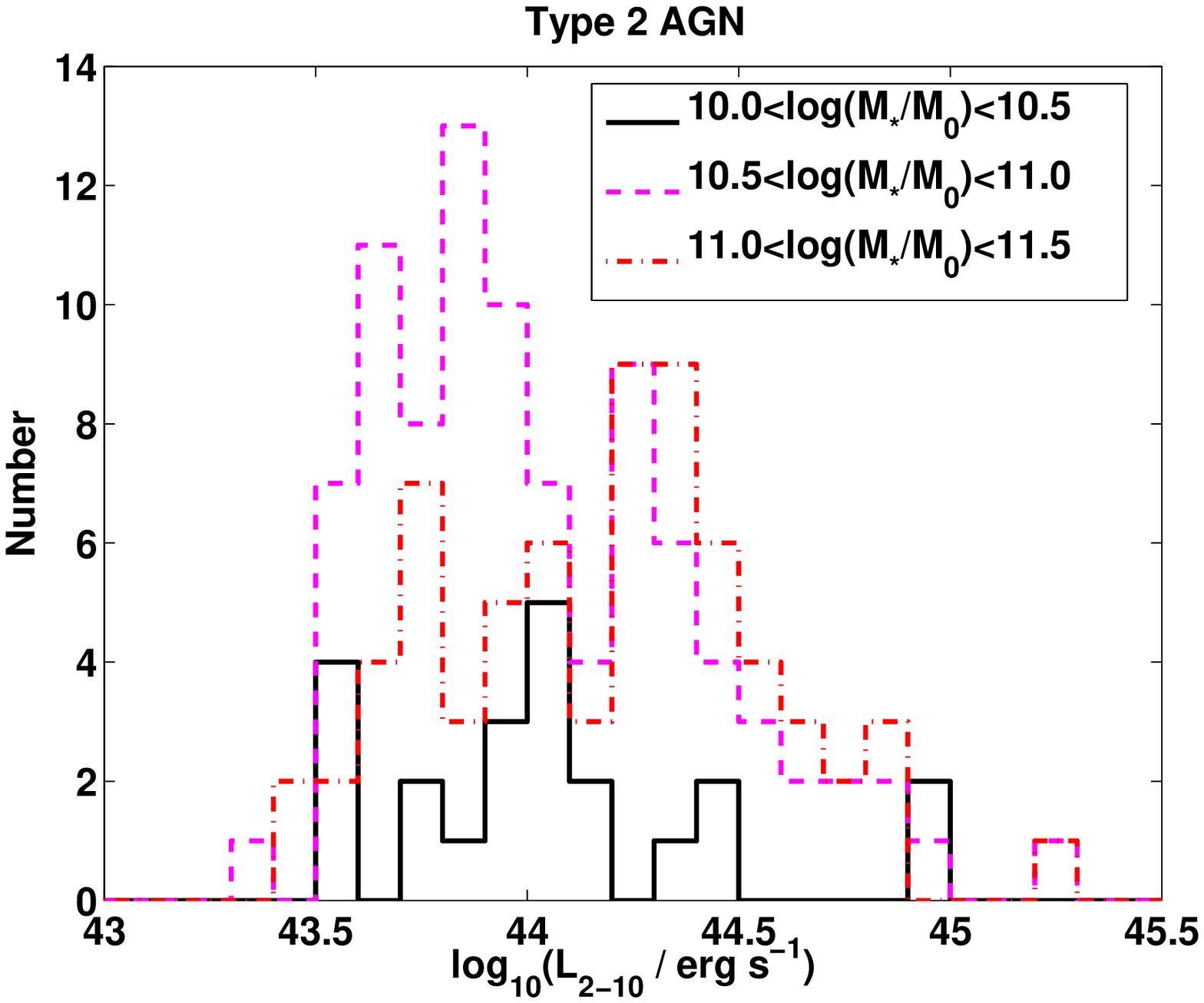} \\
\end{tabular}{cc}
\caption{Redshift (left) and X-ray luminosity (right) distributions for the Type 1 and Type 2 AGN split into the three stellar mass bins considered in Section \ref{sec:gal}.}
\label{fig:dist2}
\end{center}
\end{figure*}

Here we present the redshift and hard X-ray luminosity distributions for the various sub-samples that have been stacked in this analysis. These distributions are shown in Fig. \ref{fig:dist1} and \ref{fig:dist2}. 

\end{document}